\documentclass[lettersize,journal]{IEEEtran}
\usepackage{amssymb,amsfonts}
\usepackage{algorithmic}
\usepackage{graphicx}
\usepackage{textcomp}
\usepackage{xcolor}
\usepackage{amsmath}
\usepackage{bm}
\usepackage{empheq}
\usepackage{caption}
\usepackage{subcaption}
\usepackage{xurl}

\usepackage[normalem]{ulem}
\usepackage{cancel}
\usepackage[font=small,labelfont=bf]{caption}
\usepackage{makecell}
\usepackage{multirow}
\usepackage[sort]{cite}

\begin{document}

\title{Bistatic Passive Sensing via CSI Power}

\author{Zhongqin Wang, \IEEEmembership{Member, IEEE},
        J. Andrew Zhang, \IEEEmembership{Senior Member, IEEE}, 
        Kai Wu, \IEEEmembership{Member, IEEE}, \\
	    Kuangda Chen,
		Min Xu, \IEEEmembership{Member, IEEE},
		Y. Jay Guo, \IEEEmembership{Life Fellow, IEEE}

\IEEEcompsocitemizethanks{
\IEEEcompsocthanksitem Zhongqin Wang, J. Andrew Zhang (Corresponding Author), Kai Wu, Kuangda Chen, Min Xu are with the School of Electrical and Data Engineering and the Global Big Data Technologies Centre, University of Technology Sydney, Sydney 2007, Australia. E-mail:\{zhongqin.wang, andrew.zhang, kai.wu, kuangda.chen, min.xu\}@uts.edu.au
\IEEEcompsocthanksitem Y. Jay Guo is with the Global Big Data Technologies Centre, University of Technology Sydney, Sydney 2007, Australia. E-mail: jay.guo@uts.edu.au
}
}


\maketitle

\begin{abstract}
Passive object sensing with communication signals is a key enabler of perceptive mobile networks and integrated sensing and communication. In practical bistatic deployments, transmitter-receiver asynchrony and hardware impairments introduce time-varying random phase offsets in Channel State Information (CSI). Together with limited bandwidth and small antenna arrays, these effects degrade sensing accuracy. This work proposes a lightweight bistatic passive tracking and sensing framework that operates in the CSI-power domain. CSI power suppresses these offsets without explicit phase calibration, while preserving target-induced sensing cues. We  show that physically admissible constraints in the spatial-frequency domain induced by transmitter-receiver geometry can resolve the mirror ambiguity inherent to real-valued CSI power. Building on these properties, we develop a real-time 3D Fourier-domain processing pipeline that jointly recovers spectral (delay), spatial (angle), and temporal (Doppler) signatures. The resulting features are integrated into an online framework with adaptive motion detection, outlier suppression, and extended Kalman filter tracking with deterministic initialization, followed by position-refined micro-Doppler feature extraction for micro-motion sensing. Extensive experiments, including simulations, a real-world prototype using 3.1~GHz LTE signals, and an open-source gait recognition dataset, demonstrate the effectiveness of the proposed CSI-power-based framework for bistatic passive tracking and sensing.
\end{abstract}
\begin{IEEEkeywords}
Passive tracking, bistatic sensing, CSI power, micro-Doppler, ISAC, PMN.
\end{IEEEkeywords}

\section{Introduction}
\IEEEPARstart{P}{erceptive} Mobile Networks (PMN) aim to endow future cellular infrastructures with native ``radio vision'' by reusing communication signals for environmental sensing \cite{zhang2020perceptive, zhang2021enabling, zhang2022integration}, aligning with Integrated Sensing and Communication (ISAC), where the same spectrum and hardware jointly support data delivery and perception \cite{wu2022joint, 10474422}. A key enabler is Channel State Information (CSI): its spatiotemporal variations encode propagation and motion dynamics, enabling passive sensing without dedicated radar waveforms. Despite this promise, accurate CSI-based passive sensing on communication signals (e.g., IEEE802.11 WiFi, LTE, and 5G NR mmWave) remains challenging in practice due to limited bandwidth, small antenna arrays, and imperfect synchronization. First, communication waveforms often provide only tens of MHz bandwidth (e.g., 1.4--20MHz in LTE), yielding coarse delay resolution and weak multipath separability. Second, many receivers expose only a few antenna elements, which limits the effective aperture and raises sidelobes for Angle of Arrival (AoA) estimation. This challenge is further aggravated by time-varying inter-chain phase offsets caused by Phase Locked Loop (PLL) initial phase differences \cite{zubow2021phase}, making accurate AoA estimation difficult in practice. Third, bistatic deployments with independent transmitter and receiver clocks introduce timing offset (TO) and carrier frequency offset (CFO), causing time-varying CSI phase distortions that degrade Doppler extraction and delay recovery \cite{ma2019wifi, s25020375, wang2025water, 11153070}. These constraints make it difficult to obtain accurate delay-AoA-Doppler signatures from practical CSI measurements.

Most prior work has mainly pursued phase-compensation techniques to recover complex CSI structure \cite{wu2024sensing}. Linear fitting across subcarriers \cite{kotaru2015spotfi, hasanzadeh2025moric} removes an approximately linear phase trend, but cannot fully eliminate TO/CFO effects and is sensitive to model mismatch under rich multipath. Cross-antenna cross-correlation (CACC) \cite{li2017indotrack, qian2018widar2, wang2023single, wang2024passive} and differential CACC (DCACC) \cite{wang2023single, wang2024passive} suppress clock-induced phase distortions, yet introduce mirrored components and typically require at least three antennas for ambiguity mitigation. Cross-antenna signal ratio (CASR) \cite{feng2021lte, feng2022lte, hu2024performance} also alleviates phase offsets and some amplitude distortions, but yields a nonlinear sensing model, often supports only single-target Doppler extraction, and may bias delay and AoA estimation \cite{wang2025towards}. Reference-signal methods \cite{pegoraro2024jump, meneghello2022sharp} construct a reference component to compensate time-varying offsets, but depend on the availability of a stable reference and can degrade under narrowband or single-carrier signaling. Overall, phase-domain compensation is well studied but often calibration-dependent and sensitive to residual TO/CFO, phase jumps, and multipath.

To our knowledge, unlocking sensing directly from CSI magnitude/power remains relatively underexplored. By discarding phase, CSI power is naturally robust to phase rotations induced by TO, CFO, and inter-chain phase inconsistencies, thereby avoiding explicit calibration. This robustness, however, comes with a fundamental challenge: CSI-power observations are real-valued, so the resulting Doppler and spatial-frequency spectra exhibit intrinsic mirror symmetry, which can yield ambiguous delay-AoA-Doppler estimates. For example, the work \cite{karanam2019tracking} exploits CSI power for tracking, but the mirror ambiguity can lead to symmetric solutions, and the sensing region is therefore typically restricted to a half-space (e.g., between the transmitter and receiver) to avoid mirrored tracks. In practical bistatic deployments, an effective and general mechanism to resolve this symmetry remains lacking.

Beyond CSI stabilization, the tracking layer itself faces practical limitations. Many solutions formulate bistatic tracking as nonlinear optimization. Doppler-only methods \cite{li2017indotrack, li2024wifi, 10962318} integrate Doppler over time, but typically require multiple (often at least four) spatially separated receivers and accurate geometry priors to resolve motion direction and stabilize positioning. Multi-dimensional search methods \cite{qian2018widar2, xie2019md} perform joint delay-AoA-Doppler estimation via likelihood-based optimization, but are often initialization sensitive and computationally demanding. Alternatively, Doppler-centric pipelines such as WiDFS \cite{wang2023single} and WiDFS2.0 \cite{wang2024passive} extract dominant Doppler components and then estimate AoA and delay for localization. These methods enable real-time tracking, but their accuracy hinges on Doppler peak stability. Overall, while combining delay, AoA, and Doppler can improve robustness against multipath, practical systems should avoid heavy optimization and initialization-sensitive, high-cost processing to support real-time deployment.

This work presents \emph{WiDFS2.5}, a real-time passive  sensing framework that operates on CSI power to enable unambiguous delay, AoA, and Doppler extraction under clock asynchrony and hardware impairments. Unlike our DCACC-based WiDFS \cite{wang2023single} and WiDFS2.0 \cite{wang2024passive}, this work leverages CSI power and the joint delay-AoA-Doppler structure to localize targets without explicit antenna calibration for AoA estimation. It further provides lightweight, instantaneous position estimates, avoiding the random initialization and heavy optimization commonly required by Doppler-only pipelines. WiDFS2.5 operates with a transmitter and a multi-antenna receiver (AoA requires at least two receiving antennas), and also supports single-input single-output (SISO) delay-Doppler sensing and extension to multiple-input multiple-output (MIMO) settings. Our main contributions are summarized below:

1) We develop a practical bistatic CSI model with hardware and asynchrony distortions, and show that mapping CSI to element-wise power naturally suppresses TO, CFO, and inter-chain phase offsets. We further derive the CSI-power formulation and identify a sensing-relevant static-dynamic cross term that implicitly preserves a structured delay-AoA-Doppler progression, thereby enabling phase-information passive sensing.

2) We propose a lightweight 3D Fourier transform pipeline on the CSI-power cube to jointly estimate delay, AoA, and Doppler. We show that the intrinsic mirror symmetry of power spectra can be suppressed via physically constrained gating, including a non-negative relative-delay constraint (a target-reflected path is typically longer than the Tx-Rx path) and a one-sided relative-AoA constraint (the target moves on one side of the Tx-Rx baseline). We further extend this processing to a delay-AoA-only mode for undersampled scenarios.

3) We integrate the extracted features into a real-time sensing framework with adaptive motion detection, outlier removal across multiple coherent processing intervals (CPIs), and extended Kalman filter (EKF) trajectory tracking with deterministic initialization. Building on the tracked position, we extract position-refined micro-Doppler features and apply a reweighting strategy to obtain a cleaner Doppler signature for downstream sensing applications such as gait recognition.

We validate WiDFS2.5 via simulations against a multi-link Doppler-only baseline~\cite{li2017indotrack, 10962318}, a real-world 1Tx-3Rx LTE prototype at 3.1~GHz/20~MHz achieving real-time sub-meter tracking without antenna calibration~\cite{chen2023development}, and sensing evaluation on the public GaitID dataset~\cite{zhang2020gaitid}. Across these settings, jointly exploiting delay and AoA stabilizes Doppler estimation (especially near zero velocity), provides reliable frame-level position estimates for deterministic EKF initialization, and yields cleaner micro-Doppler features. On GaitID, our position-refined micro-Doppler feature outperforms CASR with more than 90\% identification accuracy.

\section{Bistatic CSI Model}
\label{subsec:bistatic_csi_model}
Consider a bistatic setup with a single-antenna transmitter  and a multi-antenna receiver. We collect CSI across $N_a$ receive antennas and $N_f$ OFDM subcarriers. Let $\mathbf{C}_k\in\mathbb{C}^{N_a\times N_f}$ denote the measured CSI matrix at packet index $k$, whose $(i,j)$-th entry is $CSI_{i,j,k}$, where $i\in\{1,\ldots,N_a\}$ indexes the Rx antenna and $j\in\{1,\ldots,N_f\}$ indexes the subcarrier.

\subsubsection{Non-Ideal CSI Model}
We model the measured CSI as
\begin{equation}
\mathbf{C}_k = \mathbf{D}^{h}_k\,\mathbf{H}_k\,\mathbf{D}^{e}_k,
\label{eq:bistatic_matrix_model}
\end{equation}
where $\mathbf{H}_k\in\mathbb{C}^{N_a\times N_f}$ is the physical channel matrix, decomposed into a static component $\mathbf{H}^{S}$ and a dynamic component $\mathbf{H}^{X}_k$ induced by target motion:
\begin{equation}
\mathbf{H}_k = \mathbf{H}^{S} + \mathbf{H}^{X}_k.
\label{eq:channel_decomposition}
\end{equation}
The diagonal matrices $\mathbf{D}^{h}_k$ and $\mathbf{D}^{e}_k$ capture inter-chain and asynchrony-induced phase distortions, respectively:
\begin{equation}
\mathbf{D}^{h}_k \triangleq \mathrm{diag}\!\left(e^{-{J}\left(\varphi^{h}_{1}+\delta^{h}_{1,k}\right)},\ldots,e^{-{J}\left(\varphi^{h}_{N_a}+\delta^{h}_{N_a,k}\right)}\right),
\label{eq:Dh_def_timevarying}
\end{equation}
\begin{equation}
\mathbf{D}^{e}_k \triangleq \mathrm{diag}\!\left(e^{-{J}\psi_{1,k}},\ldots,e^{-{J}\psi_{N_f,k}}\right),
\label{eq:De_def}
\end{equation}
with
\begin{equation}
\psi_{j,k} \triangleq \varphi_{j,k}^{\mathrm{TO}} + \varphi_{k}^{\mathrm{CFO}}.
\label{eq:to_cfo_phase}
\end{equation}
Here, ${J}$ denotes the imaginary unit. $\varphi^{h}_{i}$ models the constant antenna-chain phase offset, while $\delta^{h}_{i,k}$ captures packet-dependent phase jumps due to PLL-related effects (e.g., per-packet PLL re-locking or initial phase variations). These inter-chain offsets are often piecewise constant, with occasional abrupt $\pi$-radian transitions \cite{zubow2021phase}. In addition, $\varphi^{\mathrm{TO}}_{j,k}$ is the TO-induced phase term that varies across subcarriers and packets, and $\varphi^{\mathrm{CFO}}_{k}$ is the CFO-induced phase term that is common across subcarriers but varies across packets.

\subsubsection{Static and Dynamic Channel Components}
Let $f_j$ denote the $j$-th subcarrier frequency and let $d$ be the Rx antenna spacing of a uniform linear array (ULA). The static component $\mathbf{H}^{S}$ captures the dominant Tx-Rx path and other static reflections, while the dynamic component $\mathbf{H}^{X}_k$ captures the dominant human-induced reflection. We write
\begin{equation}
\mathbf{H}^{S}=\rho^{S}\,\mathbf{a}_r(\theta^{S})\,\mathbf{b}_f^{\mathsf{T}}(\tau^{S})+\widetilde{\mathbf{H}}^{S},
\label{eq:static_rank1}
\end{equation}
\begin{equation}
\mathbf{H}^{X}_k=\rho^{X}\,\mathbf{a}_r(\theta^{X})\,\mathbf{b}_f^{\mathsf{T}}(\tau^{X})\,e^{-{J}2\pi f^{D}(k-1)\Delta t}+\widetilde{\mathbf{H}}^{X}_k,
\label{eq:dynamic_rank1}
\end{equation}
where $\rho^{S}$ and $\rho^{X}$ denote the approximately constant complex amplitudes of the dominant static and human-reflected paths within a coherent processing interval (CPI), respectively. $\tau^{S}$ and $\tau^{X}$ are the corresponding delays (referenced to the first receive antenna), $\theta^{S}$ and $\theta^{X}$ are the AoAs, and $f^{D}$ is the Doppler frequency shift of the dominant human-reflected path within the CPI. Let $\lambda \triangleq c/f_c$ be the carrier wavelength. The receive-array steering vector $\mathbf{a}_r(\theta)\in\mathbb{C}^{N_a}$ and frequency steering vector $\mathbf{b}_f(\tau)\in\mathbb{C}^{N_f}$ are
\begin{equation}
\mathbf{a}_r(\theta) \triangleq
\left[
1,\;
e^{-{J}2\pi \frac{d}{\lambda}\sin\theta},\;
\ldots,\;
e^{-{J}2\pi \frac{(N_a-1)d}{\lambda}\sin\theta}
\right]^{\mathsf{T}},
\label{eq:array_steering}
\end{equation}
\begin{equation}
\mathbf{b}_f(\tau) \triangleq
\left[
e^{-{J}2\pi f_1\tau},\;
e^{-{J}2\pi f_2\tau},\;
\ldots,\;
e^{-{J}2\pi f_{N_f}\tau}
\right]^{\mathsf{T}}.
\label{eq:freq_steering}
\end{equation}
The residual terms $\widetilde{\mathbf{H}}^{S}$ and $\widetilde{\mathbf{H}}^{X}_k$ aggregate the remaining static and dynamic multipath components and are treated as structured interference in subsequent processing.

\subsubsection{Coherent Processing Interval Assumption}
We process CSI in CPIs of length $N_t$ packets. In our implementation, the CSI sampling rate is $1$~kHz and $N_t=128$, corresponding to a CPI duration of $0.128$~s. Over such a CPI, it is reasonable to assume that the dominant path amplitudes remain approximately constant, while its phase varies in a structured manner: linearly across subcarriers (delay), across antennas (AoA), and over packets (Doppler) \cite{zhang2022integration, wang2024passive}. This enables stacking $\{\mathbf{C}_k\}_{k=1}^{N_t}$ into a frequency-space-time cube for joint estimation of $(\tau,\theta,f^{D})$. In some systems with a low CSI sampling rate (e.g., LTE or commodity WiFi under constrained reporting), only a few packets may be available per CPI, making Doppler estimation unreliable. In such cases, we instead exploit the delay-AoA structure for localization, as discussed later.

\section{Unambiguous Sensing Parameter Estimation via CSI Power}
This section presents a lightweight 3D Fourier-domain pipeline that jointly extracts unambiguous delay-AoA-Doppler features from real-valued CSI power.

\subsection{CSI Power for Random Phase Removal}
\label{subsec:power_invariance}
We define the CSI power matrix as the Hadamard product between $\mathbf{C}_k$ and its complex conjugate:
\begin{equation}
\mathbf{P}_k \triangleq \mathbf{C}_k \odot \overline{\mathbf{C}}_k,
\label{eq:Pk_def}
\end{equation}
where $\odot$ denotes the Hadamard product and $\overline{(\cdot)}$ denotes complex conjugation. Equivalently,
\begin{equation}
\left[\mathbf{P}_k\right]_{i,j}=\left|CSI_{i,j,k}\right|^2.
\label{eq:Pk_entry}
\end{equation}
Since Eq. (\ref{eq:Pk_def}) computes the element-wise magnitude-squared, the phase distortions in $\mathbf{D}^{h}_k$ and $\mathbf{D}^{e}_k$ are completely canceled.

Using $\mathbf{H}_k=\mathbf{H}^{S}+\mathbf{H}^{X}_k$ in (\ref{eq:channel_decomposition}), we expand $\mathbf{P}_k$ as
\begin{equation}
\mathbf{P}_k
= \left(\mathbf{H}^{S}+\mathbf{H}^{X}_k\right)\odot \overline{\left(\mathbf{H}^{S}+\mathbf{H}^{X}_k\right)}
= \mathbf{P}^{S} + \mathbf{P}^{X}_k + \mathbf{P}^{C}_k,
\label{eq:Pk_expand}
\end{equation}
where
\begin{equation}
\mathbf{P}^{S}\triangleq \mathbf{H}^{S}\odot \overline{\mathbf{H}^{S}},
\quad
\mathbf{P}^{X}_k\triangleq \mathbf{H}^{X}_k\odot \overline{\mathbf{H}^{X}_k}.
\label{eq:Pk_static_dynamic_power}
\end{equation}
The static-dynamic cross term is
\begin{equation}
\mathbf{P}^{C}_k
\triangleq \mathbf{H}^{S}\odot \overline{\mathbf{H}^{X}_k} \;+\; \overline{\mathbf{H}^{S}}\odot \mathbf{H}^{X}_k
= 2\,\Re\!\left\{\mathbf{H}^{S}\odot \overline{\mathbf{H}^{X}_k}\right\}.
\label{eq:Pk_cross}
\end{equation}
Here, $\mathbf{P}^{S}$ is approximately time-invariant within a CPI, while $\mathbf{P}^{X}_k$ is the self-power of the dynamic component. In typical indoor scenarios, the human-reflected component is much weaker than the dominant static component, so $\mathbf{P}^{X}_k$ is treated as a secondary term. In contrast, $\mathbf{P}^{C}_k$ captures the static-dynamic interaction and carries sensing-relevant delay-AoA-Doppler information.

\subsection{Sensing-Relevant Cross Term}
\label{subsec:sd_interference}
Next, we characterize the sensing-relevant structure of the cross term $\mathbf{P}^{C}_k$ and show how it enables joint parameter estimation in the frequency, spatial, and slow-time domains.

For clarity, we focus on the dominant static path and the dominant human-reflected path in Eq. (\ref{eq:static_rank1}) and Eq. (\ref{eq:dynamic_rank1}), and treat residual multipath as perturbations. From Eq. (\ref{eq:Pk_cross}), the $(i,j)$-th entry of $\mathbf{P}^{C}_k$ is
\begin{equation}
\left[\mathbf{P}^{C}_k\right]_{i,j}
= 2\,\Re\!\left\{ H^{S}_{i,j}\,\overline{H^{X}_{i,j,k}} \right\}.
\label{eq:Pc_entry_re}
\end{equation}

Ignoring residual multipath for exposition, we use
\begin{equation}
H^{S}_{i,j} \approx \rho^{S}
e^{-{J}2\pi f_j\left(\tau^{S}+\frac{(i-1)d}{c}\sin\theta^{S}\right)},
\label{eq:Hs_dom}
\end{equation}
\begin{equation}
H^{X}_{i,j,k} \approx \rho^{X}
e^{-{J}2\pi f_j\left(\tau^{X}+\frac{(i-1)d}{c}\sin\theta^{X}\right)}
e^{-{J}2\pi f^{D}(k-1)\Delta t}.
\label{eq:Hx_dom}
\end{equation}
Substituting Eq. (\ref{eq:Hs_dom}) and Eq. (\ref{eq:Hx_dom}) into Eq. (\ref{eq:Pc_entry_re}) yields
\begin{equation}
\resizebox{\linewidth}{!}{$
\left[\mathbf{P}^{C}_k\right]_{i,j}
\approx 2\,\Re\!\left\{\rho^{S}\overline{\rho^{X}}
e^{-{J}\varphi_{i,j,k}}\right\}
= 2\left|\rho^{S}\rho^{X}\right|\cos\!\left(\varphi_{i,j,k}+\phi_{0}\right),
$}
\label{eq:Pc_cos_form}
\end{equation}
where $\phi_{0}\triangleq \angle(\rho^{S}\overline{\rho^{X}})$ and
\begin{equation}
\begin{aligned}
\varphi_{i,j,k}
\triangleq\;& 2\pi f_j\left(\tau^{X}-\tau^{S}\right)
+2\pi \frac{(i-1)d}{\lambda}\left(\sin\theta^{X}-\sin\theta^{S}\right) \\
&+2\pi f^{D}(k-1)\Delta t.
\end{aligned}
\label{eq:phi_linear}
\end{equation}
Therefore, $\mathbf{P}^{C}_k$ remains real-valued yet retains a structured dependence on the delay difference $\tau_{\Delta}=\tau^{X}-\tau^{S}$ across subcarriers, the AoA difference $s_{\Delta}=\sin\theta^{X}-\sin\theta^{S}$ across antennas, and the Doppler $f^{D}$ across packets. This structure enables joint delay-AoA-Doppler estimation from CSI power.

\subsection{3D Fourier Feature Extraction}
\label{subsec:3dfft_powercube}
We then stack the CSI power measurements within a CPI to form a 3D data cube
\begin{equation}
\mathcal{P}\in\mathbb{R}^{N_a\times N_f\times N_t},
\quad
\left[\mathcal{P}\right]_{i,j,k} \triangleq \left[\mathbf{P}_k\right]_{i,j}.
\label{eq:power_cube_def}
\end{equation}
We apply a cascaded 3D discrete Fourier transform (DFT) to $\mathcal{P}$ along the subcarrier, antenna, and slow-time dimensions. The 3D DFT coefficients are
\begin{equation}
\resizebox{\linewidth}{!}{$
\begin{aligned}
\mathcal{Z}\!\left[\ell,m,n\right]
\triangleq\;&
\sum_{k=1}^{N_t}\sum_{i=1}^{N_a}\sum_{j=1}^{N_f}
\left[\mathcal{P}\right]_{i,j,k}\,
e^{-{J}2\pi\left(
\frac{(j-1)\ell}{N_f}+\frac{(i-1)m}{N_a}+\frac{(k-1)n}{N_t}
\right)}.
\end{aligned}
$}
\label{eq:3d_dft_def}
\end{equation}
for $\ell\in\{0,\ldots,N_f-1\}$, $m\in\{0,\ldots,N_a-1\}$, and $n\in\{0,\ldots,N_t-1\}$. Since the static power $\mathbf{P}^{S}$ is approximately constant within a CPI, it mainly contributes to the DC component along slow time. We suppress background clutter by nulling the DC slice
\begin{equation}
\mathcal{Z}\!\left[\ell,m,0\right] \triangleq 0.
\label{eq:dc_nulling}
\end{equation}
Then $|\mathcal{Z}[\ell,m,n]|$ for $n\neq 0$ forms a delay-AoA-Doppler spectrum, where prominent peaks correspond to target-related reflections. Since $\mathbf{P}^{X}_k$ is secondary, the dominant peak structure is primarily contributed by $\mathbf{P}^{C}_k$, which follows the separable sinusoidal model in Eq. (\ref{eq:Pc_cos_form}) and Eq. (\ref{eq:phi_linear}). Accordingly, the 3D DFT can be interpreted as a bank of separable matched filters over candidate $(\tau_{\Delta}, s_{\Delta}, f^{D})$ triplets, and dominant parameters are obtained by locating the peak $(\ell^*,m^*,n^*)$ of $|\mathcal{Z}[\ell,m,n]|$. However, due to the real-valued nature of CSI power, $|\mathcal{Z}[\ell,m,n]|$ generally exhibits mirrored peaks, and peak selection must be coupled with ambiguity resolution. We address this symmetry-induced ambiguity next.
 
\subsection{Mirror Component Suppression}
\label{subsec:symmetry_mirror}
The CSI-power cross term in Eq. (\ref{eq:Pc_cos_form}) admits an approximately separable cosine model. In particular, the dominant contribution can be written as
\begin{equation}
\left[\mathbf{P}^{C}_k\right]_{i,j}
\approx A\cos\!\left(\varphi_{i,j,k}+\phi_{0}\right),
\label{eq:cos_model_mirror_rewrite}
\end{equation}
where $A\triangleq 2|\rho^{S}\rho^{X}|$ and $\varphi_{i,j,k}$ is a linear progression over subcarrier, antenna, and slow-time indices. Since $\cos(x)=\cos(-x)$, the sign-inverted phase $-\varphi_{i,j,k}$ yields identical observations, leading to mirrored peaks in the 3D DFT magnitude. This induces the mirror mapping
\begin{equation}
(\tau_{\Delta},\,s_{\Delta},\,f^{D})
\;\longleftrightarrow\;
(-\tau_{\Delta},\,-s_{\Delta},\,-f^{D}),
\label{eq:mirror_mapping_rewrite}
\end{equation}
that is, each physical peak has a mirrored counterpart with flipped relative delay, relative sine-AoA, and Doppler. Next, we show that the mirrored branch can be excluded by enforcing physical constraints on delay and AoA.

\begin{figure}
\centering
\includegraphics[width=0.35\textwidth]{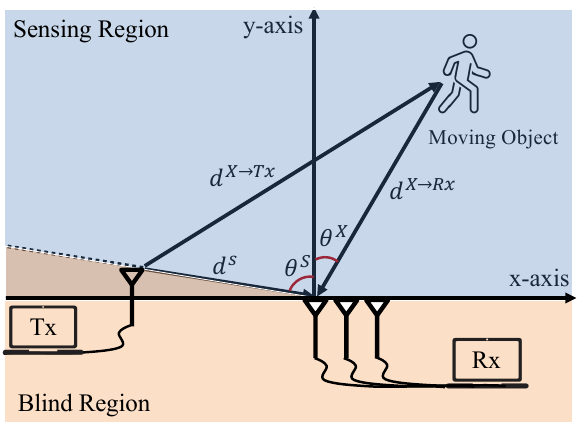}
\caption{Geometry of the bistatic system, where the target moves on one side of the Tx-Rx baseline.}
\label{geometry_tx_rx}
\vspace{-1.5em}
\end{figure}

\textit{Remark 1 (Non-negative Relative Delay).}
Let $\tau^{S}$ correspond to the dominant \emph{static reference path} between Tx and Rx (which may be line-of-sight (LOS) or a strong static reflection), and let $\tau^{X}$ correspond to a target-reflected Tx-target-Rx path. Let $d^{S}$ be the length of the chosen static reference path, and let $d^{\mathrm{Tx}\rightarrow X}$ and $d^{X\rightarrow \mathrm{Rx}}$ be the Tx-target and target-Rx distances, respectively, so that $d^{X}=d^{\mathrm{Tx}\rightarrow X}+d^{X\rightarrow \mathrm{Rx}}$. Under the natural choice that $\tau^{S}$ is taken as the \emph{minimum-delay} dominant static component, the target-reflected path is no shorter than this reference, that is, $d^{X}\ge d^{S}$, which implies $\tau^{X}\ge\tau^{S}$ and
\begin{equation}
\tau_{\Delta}=\tau^{X}-\tau^{S}\ge 0.
\label{eq:tau_delta_nonneg}
\end{equation}

In LoS scenarios, $d^{S}$ coincides with the Tx-Rx distance, and Eq. \eqref{eq:tau_delta_nonneg} follows directly. In NLoS scenarios, however, the strongest static component may not correspond to the geometric Tx-Rx distance. Treating the Tx-Rx distance as $d^{S}$ in such cases can bias the recovered absolute delay and subsequently introduce localization errors (e.g., in Tx position estimation), as will be discussed in Section~\ref{subsubsec:ours_csi_power}.

\textit{Remark 2 (One-Sided AoA).}
Assume the AoA of the static Tx-Rx path, $\theta^{S}$, is known and the target remains in the sensing region without crossing the Tx-Rx baseline (Fig.~\ref{geometry_tx_rx}), that is, it stays on one side of the Tx-Rx line and does not enter the blind region. Under the ULA field-of-view
\begin{equation}
\theta^{S},\theta^{X}\in\left[-\frac{\pi}{2},\frac{\pi}{2}\right],
\label{eq:fov_halfspace}
\end{equation}
the sign of $s_{\Delta}=\sin\theta^{X}-\sin\theta^{S}$ remains fixed, that is, either $s_{\Delta}\ge 0$ or $s_{\Delta}\le 0$. This follows because $\sin(\cdot)$ is strictly increasing over $\left[-\frac{\pi}{2},\frac{\pi}{2}\right]$.

\textit{Remark 3 (Mirrored Triplet Exclusion).}
Let $\Omega$ denote the valid parameter domain induced by Remark 1-2:
\begin{equation}
\Omega \triangleq \Big\{(\tau_{\Delta},s_{\Delta},f^{D}) \,\big|\, \tau_{\Delta}\ge 0,\;
s_{\Delta}\in\mathcal{S}_{\mathrm{one\text{-}side}}\Big\},
\label{eq:admissible_domain}
\end{equation}
where $\mathcal{S}_{\mathrm{one\text{-}side}}$ is the one-sided feasible set determined by $\theta^{S}$. By
Eq. (\ref{eq:mirror_mapping_rewrite}), each triplet $(\tau_{\Delta},s_{\Delta},f^{D})$ is paired with
$(-\tau_{\Delta},-s_{\Delta},-f^{D})$. Since the mirror flips all three signs, any triplet in $\Omega$ has
$\tau_{\Delta}\ge 0$ and $s_{\Delta}\in\mathcal{S}_{\mathrm{one\text{-}side}}$, whereas its mirrored counterpart is excluded from $\Omega$, except when the parameters lie on boundary cases. Therefore, restricting peak selection to $\Omega$ can exclude the mirrored peak, including the Doppler sign-flipped branch.

\subsection{From DFT Bins to Physical Parameters}
We next map the 3D DFT indices $(\ell,m,n)$ to the physical parameters $(\tau_{\Delta},s_{\Delta},f^{D})$ and specify the one-sided conventions used for peak selection. For clarity, we focus on the case $s_{\Delta}\ge 0$ (i.e., $\theta^{X}\ge \theta^{S}$ during the track); the case $s_{\Delta}\le 0$ is handled analogously.

\subsubsection{Delay}
Let $\Delta f$ denote the subcarrier spacing and let $B\triangleq N_f\Delta f$ denote the effective bandwidth. We retain the nonnegative-delay branch by restricting to $\ell\in\{0,1,\ldots,\lfloor N_f/2\rfloor\}$. The delay grid is
\begin{equation}
\tau_{\Delta}[\ell] \triangleq \frac{\ell}{N_f\Delta f}=\frac{\ell}{B},
\quad
\ell\in\{0,1,\ldots,\lfloor N_f/2\rfloor\},
\label{eq:delay_grid_pos}
\end{equation}
with nominal resolution $\Delta\tau\approx 1/B$. Given the static-path delay $\tau^{S}$, the target-related delay is
\begin{equation}
\tau^{X} = \tau^{S}+\tau_{\Delta}.
\label{eq:tauX_recover}
\end{equation}

\subsubsection{AoA}
We map the spatial-DFT index $m$ to a grid on $s_{\Delta}$. Using the centered normalized spatial-frequency grid
\begin{equation}
\mu[m]\triangleq \frac{m}{N_a}-\frac{1}{2},
\label{eq:mu_grid}
\end{equation}
the induced grid is
\begin{equation}
s_{\Delta}[m]\triangleq \frac{\lambda}{d}\,\mu[m],
\label{eq:sdelta_grid}
\end{equation}
so that $s_{\Delta}[m]\in\left[-\frac{\lambda}{2d},\,\frac{\lambda}{2d}\right)$. For half-wavelength spacing
$d=\lambda/2$, Eq. \eqref{eq:sdelta_grid} reduces to $s_{\Delta}[m]=\frac{2m}{N_a}-1$, which samples $s_{\Delta}$ over
$[-1,1)$. Under the one-sided prior $s_{\Delta}\ge 0$, we wrap the negative branch to obtain a one-sided coordinate
$s_{\Delta}^{+}[m]\in[0,2)$ defined as
\begin{equation}
s_{\Delta}^{+}[m]\triangleq
\begin{cases}
\frac{2m}{N_a}, & 0\le m\le \lfloor N_a/2\rfloor,\\[6pt]
2-\frac{2(N_a-m)}{N_a}, & \lfloor N_a/2\rfloor<m\le N_a-1.
\end{cases}
\label{eq:sdelta_pos_map}
\end{equation}
Equivalently, $s_{\Delta}^{+}[m]=(s_{\Delta}[m]+2)\bmod 2$. We then perform peak selection on $s_{\Delta}^{+}[m]$ under
the prior $s_{\Delta}\ge 0$.

Finally, since $\sin\theta^{X}=s_{\Delta}+\sin\theta^{S}$ must satisfy $|\sin\theta^{X}|\le 1$, we have
\begin{equation}
-1-\sin\theta^{S}\ \le\ s_{\Delta}\ \le\ 1-\sin\theta^{S}.
\label{eq:sdelta_feasible}
\end{equation}
For $s_{\Delta}\ge 0$, we keep only bins satisfying $s_{\Delta}^{+}[m]\le 1-\sin\theta^{S}$. Given
$\widehat{s}_{\Delta}$, we recover $\theta^{X}$ via
\begin{equation}
\sin\theta^{X}=\widehat{s}_{\Delta}+\sin\theta^{S},
\label{eq:sin_thetaX_recover}
\end{equation}
followed by the field-of-view constraint in Eq. \eqref{eq:fov_halfspace}.

\subsubsection{Doppler}
The slow-time index $n$ corresponds to the Doppler grid
\begin{equation}
f^{D}[n]\triangleq \frac{n}{N_t\Delta t},
\label{eq:doppler_grid}
\end{equation}
with nominal resolution $\Delta f^{D}\approx 1/(N_t\Delta t)$. We retain all Doppler bins; the $n=0$ component is handled by DC nulling in Eq.~\eqref{eq:dc_nulling}. In indoor human tracking, we further restrict peak selection to a plausible Doppler band assuming $v_{\max}=5$~m/s, i.e., we keep only bins with $|f^{D}[n]|\le f^{D}_{\max}$.

\subsubsection{Practical Grid Refinement}
We refine the grids via zero-padding in one or more dimensions. Specifically, we compute the 3D DFT using lengths
$(N_f^{\mathrm{pad}},N_a^{\mathrm{pad}},N_t^{\mathrm{pad}})$ with $N_f^{\mathrm{pad}}\ge N_f$,
$N_a^{\mathrm{pad}}\ge N_a$, and $N_t^{\mathrm{pad}}\ge N_t$, yielding denser grids and reduced bin quantization. For
example,
\begin{equation}
\resizebox{\linewidth}{!}{$
\begin{aligned}
\tau_{\Delta}[\ell] &\triangleq \frac{\ell}{N_f^{\mathrm{pad}}\Delta f},
\qquad \ell\in\{0,1,\ldots,\lfloor N_f^{\mathrm{pad}}/2\rfloor\},\\
s_{\Delta}^{+}[m] &\triangleq
\begin{cases}
\frac{2m}{N_a^{\mathrm{pad}}}, & 0\le m\le \lfloor N_a^{\mathrm{pad}}/2\rfloor,\\
2-\frac{2(N_a^{\mathrm{pad}}-m)}{N_a^{\mathrm{pad}}}, & \lfloor N_a^{\mathrm{pad}}/2\rfloor<m\le N_a^{\mathrm{pad}}-1,
\end{cases}\\
f^{D}[n] &\triangleq \frac{n}{N_t^{\mathrm{pad}}\Delta t}, \qquad n\in\{0,\ldots,N_t^{\mathrm{pad}}-1\}.
\end{aligned}
$}
\label{eq:grid_refine_examples}
\end{equation}
The zero-padding yields a finer spectral grid to increase the resolution of the extracted delay-AoA-Doppler features.

\subsection{Delay-AoA-Only Processing Under Sparse Sampling}
\label{subsec:2d_wo_doppler}
When only a few CSI packets are available per CPI (e.g., low-rate CSI collection), Doppler processing becomes unreliable. In this case, we drop the Doppler dimension and form a 2D power map by
integrating over slow time:
\begin{equation}
\mathbf{P}_{\mathrm{2D}} \in \mathbb{R}^{N_a\times N_f},
\quad
\left[\mathbf{P}_{\mathrm{2D}}\right]_{i,j}
\triangleq
\sum_{k=1}^{N_t}\left[\mathcal{P}\right]_{i,j,k}.
\label{eq:P2D_def}
\end{equation}
We then apply a 2D DFT across subcarriers and antennas:
\begin{equation}
\begin{aligned}
\mathcal{Z}_{\mathrm{2D}}\!\left[\ell,m\right]
\triangleq\;&
\sum_{i=1}^{N_a}\sum_{j=1}^{N_f}
\left[\mathbf{P}_{\mathrm{2D}}\right]_{i,j}\,
e^{-{J}2\pi\left(
\frac{(j-1)\ell}{N_f}+\frac{(i-1)m}{N_a}
\right)}.
\end{aligned}
\label{eq:2d_dft_def}
\end{equation}

To suppress static background, we initialize a mean spectrum using the first $N_{\mathrm{bg}}$ packets and subtract it
from subsequent estimates. Let $\overline{\mathcal{Z}}_{\mathrm{2D}}[\ell,m]$ be the background mean:
\begin{equation}
\overline{\mathcal{Z}}_{\mathrm{2D}}[\ell,m]
\triangleq
\frac{1}{N_{\mathrm{bg}}}
\sum_{q=1}^{N_{\mathrm{bg}}}
\mathcal{Z}^{(q)}_{\mathrm{2D}}[\ell,m].
\label{eq:Z2D_bg_mean_pkt}
\end{equation}
We then form the background-suppressed spectrum
\begin{equation}
\widetilde{\mathcal{Z}}_{\mathrm{2D}}[\ell,m]
\triangleq
\mathcal{Z}_{\mathrm{2D}}[\ell,m] -\overline{\mathcal{Z}}_{\mathrm{2D}}[\ell,m],
\label{eq:Z2D_bg_subtract_pkt}
\end{equation}
and perform peak selection on $\left|\widetilde{\mathcal{Z}}_{\mathrm{2D}}[\ell,m]\right|$ to obtain $(\tau_{\Delta},s_{\Delta})$.

\section{Passive Object Tracking and Sensing}
This section presents a real-time detection and tracking framework for continuous human tracking.

\subsection{Multi-CPI Fusion for Motion Detection and Filtering}
\label{subsec:robust_multicpi_fusion}

\subsubsection{Adaptive Motion Detection}
For each CPI $k$, we form the background-suppressed 3D spectrum power
\begin{equation}
P_k[\ell,m,n]\triangleq \left|\mathcal{Z}[\ell,m,n]\right|^2,
\label{eq:Pk_spectrum_power}
\end{equation}
and test whether a target-related peak rises above the noise floor. Let $\Omega$ denote the valid index set after mirror-related gating. We locate the strongest candidate
\begin{equation}
(\ell^*,m^*,n^*)=
\arg\max_{(\ell,m,n)\in\Omega} P_k[\ell,m,n],
\label{eq:argmax_peak}
\end{equation}
and compute a local average over a subcube $\mathcal{C}$ centered at $(\ell^*,m^*,n^*)$:
\begin{equation}
T_k \triangleq \frac{1}{|\mathcal{C}|}
\sum_{(\ell,m,n)\in\mathcal{C}} P_k[\ell,m,n],
\label{eq:Tk_subcube_short}
\end{equation}
where $\mathcal{C}$ has size $(2r_\ell\!+\!1)\times(2r_m\!+\!1)\times(2r_n\!+\!1)$. This averaging mitigates spectral leakage and bin-quantization effects.

To estimate the noise floor, we form a noise set $\Omega_{\mathrm{n}}\subseteq\Omega$ by excluding a guard region around $(\ell^*,m^*,n^*)$, and compute a robust median estimator
\begin{equation}
{\mu}_k \triangleq 
\mathrm{median}\Big\{
P_k[\ell,m,n]
\Big\}_{(\ell,m,n)\in\Omega_{\mathrm{n}}}.
\label{eq:mu_hat_short}
\end{equation}
We define the normalized detection statistic
\begin{equation}
\Lambda_k \triangleq \frac{T_k}{{\mu}_k+\varepsilon}.
\label{eq:Lambda_def}
\end{equation}

To improve robustness, we perform \emph{multi-CPI fusion} over a sliding window of $L$ adjacent CPIs by computing
\begin{equation}
\overline{\Lambda}_k \triangleq \mathrm{median}\Big\{\Lambda_{k-q}\Big\}_{q=0}^{L-1}.
\label{eq:Lambda_fuse}
\end{equation}
We declare motion if
\begin{equation}
\overline{\Lambda}_k > \gamma_{\mathrm{cfar}}.
\label{eq:cfar_fuse}
\end{equation}
This temporal fusion suppresses occasional spurious peaks caused by multipath and unstable reflections. 

\subsubsection{Outlier Filtering}
Per-CPI peak extraction may occasionally yield abnormal delay-AoA-Doppler estimates due to unstable reflections and multipath. We therefore apply temporal outlier filtering and reliability-weighted fusion to the per-CPI feature sequence. For a feature $x_k\in\{\widehat{\tau}_{\Delta,k},\,\widehat{s}_{\Delta,k},\,\widehat{f}^{D}_k\}$ within the current fusion window, we compute the Z-score
\begin{equation}
z_k \triangleq \frac{x_k-\bar{x}}{\sigma_x+\varepsilon},
\quad
\bar{x}\triangleq \frac{1}{L}\sum_{k=1}^{L}x_k,
\label{eq:zscore_simple}
\end{equation}
and discard samples with $|z_k|>\zeta$. Over the retained index set $\mathcal{I}$, we fuse the remaining estimates via
\begin{equation}
\widehat{x} \triangleq \sum_{k\in\mathcal{I}} \bar{w}_k\,x_k,
\qquad
\bar{w}_k \triangleq \frac{w_k}{\sum_{q\in\mathcal{I}} w_q+\varepsilon},
\label{eq:weighted_fuse_simple}
\end{equation}
where the weight is set to $w_k=\Lambda_k$ from Eq. \eqref{eq:Lambda_def}.

\subsection{Real-time Object Tracking via EKF}
\label{subsec:ekf_tracking}
We track the target using an EKF driven by fused per-CPI delay, AoA, and Doppler estimates. The Tx location is treated as fixed and provided by deployment measurement.

\subsubsection{State and Motion Model}
As shown in Fig.~\ref{geometry_tx_rx}, we use an Rx-centric Cartesian coordinate system on the horizontal plane. Let
$\mathbf{p}_k=[x_k,y_k]^{\mathsf{T}}$, $\mathbf{v}_k=[\dot{x}_k,\dot{y}_k]^{\mathsf{T}}$, and $\mathbf{a}_k=[\ddot{x}_k,\ddot{y}_k]^{\mathsf{T}}$ denote the target position, velocity, and acceleration at the $k$th fused CPI. The EKF state is
\begin{equation}
\mathbf{x}_k
\triangleq
\big[
\mathbf{p}_k^{\mathsf{T}},\,
\mathbf{v}_k^{\mathsf{T}},\,
\mathbf{a}_k^{\mathsf{T}}
\big]^{\mathsf{T}}
\in\mathbb{R}^{6}.
\label{eq:ekf_state_target}
\end{equation}
We adopt a constant-acceleration model driven by white jerk noise. With sampling interval $\Delta T_k$ between consecutive fused CPIs, define
\begin{equation}
\alpha_k \triangleq \Delta T_k,
\quad
\beta_k \triangleq \tfrac{1}{2}\Delta T_k^2.
\label{eq:alpha_beta_def2}
\end{equation}
The discrete-time dynamics are
\begin{equation}
\mathbf{x}_{k+1}
=
\mathbf{F}_k \mathbf{x}_k
+
\mathbf{w}_k,
\quad
\mathbf{w}_k \sim \mathcal{N}(\mathbf{0},\mathbf{Q}_k),
\label{eq:ekf_state_model_target}
\end{equation}
where
\begin{equation}
\resizebox{\linewidth}{!}{$
\begingroup
\setlength{\arraycolsep}{3pt}
\renewcommand{\arraystretch}{1.05}
\mathbf{F}_k=
\begin{bmatrix}
\mathbf{I}_2 & \alpha_k\mathbf{I}_2 & \beta_k\mathbf{I}_2\\
\mathbf{0}   & \mathbf{I}_2         & \alpha_k\mathbf{I}_2\\
\mathbf{0}   & \mathbf{0}           & \mathbf{I}_2
\end{bmatrix},
\mathbf{Q}_k=
q_{\mathrm{j}}
\begin{bmatrix}
\frac{\Delta T_k^5}{20}\mathbf{I}_2 & \frac{\Delta T_k^4}{8}\mathbf{I}_2 & \frac{\Delta T_k^3}{6}\mathbf{I}_2\\
\frac{\Delta T_k^4}{8}\mathbf{I}_2  & \frac{\Delta T_k^3}{3}\mathbf{I}_2 & \frac{\Delta T_k^2}{2}\mathbf{I}_2\\
\frac{\Delta T_k^3}{6}\mathbf{I}_2  & \frac{\Delta T_k^2}{2}\mathbf{I}_2 & \Delta T_k \mathbf{I}_2
\end{bmatrix}.
\endgroup
$}
\label{eq:ekf_FQ_target}
\end{equation}
Here $q_{\mathrm{j}}$ controls the expected acceleration variability.

\subsubsection{Measurement Model from Range Difference, Relative Sine-AoA, and Doppler}
Let the receiver location be $\mathbf{r}\in\mathbb{R}^2$. For a given Tx location $\mathbf{t}\in\mathbb{R}^2$, the EKF uses the fused per-CPI measurement vector
\begin{equation}
\mathbf{z}_k
\triangleq
\begin{bmatrix}
\widehat{\Delta d}_k\\[1pt]
\widehat{s}_{\Delta,k}\\[1pt]
\widehat{f}^{D}_k
\end{bmatrix}
=
\mathbf{h}\!\left(\mathbf{x}_k;\mathbf{t},\mathbf{r}\right)
+
\mathbf{n}_k,
\quad
\mathbf{n}_k\sim\mathcal{N}(\mathbf{0},\mathbf{R}),
\label{eq:ekf_meas_target}
\end{equation}
where $\mathbf{R}=\mathrm{diag}(\sigma_{\Delta d}^2,\sigma_{s}^2,\sigma_f^2)$.

\paragraph{Range difference}
Define
\begin{equation}
d_{\mathrm{tr}} \triangleq \|\mathbf{t}-\mathbf{r}\|_2,
~
d_{\mathrm{tx}}(\mathbf{p}_k)\triangleq \|\mathbf{p}_k-\mathbf{t}\|_2,
~
d_{\mathrm{rx}}(\mathbf{p}_k)\triangleq \|\mathbf{p}_k-\mathbf{r}\|_2.
\label{eq:dist_defs}
\end{equation}
We model the bistatic range difference as
\begin{equation}
\Delta d(\mathbf{p}_k;\mathbf{t},\mathbf{r})
=
d_{\mathrm{tx}}(\mathbf{p}_k)+d_{\mathrm{rx}}(\mathbf{p}_k)-d_{\mathrm{tr}}.
\label{eq:dd_geom}
\end{equation}

\paragraph{AoA difference}
Let $\theta^{S}$ denote the known AoA of the static reference path. The predicted target AoA is
\begin{equation}
\theta^{X}(\mathbf{p}_k;\mathbf{r})
=
\mathrm{atan2}\!\left(y_k-r_y,\,x_k-r_x\right),
\label{eq:thetaX_geom_global}
\end{equation}
where $\mathbf{p}_k=[x_k,y_k]^{\mathsf{T}}$ and $\mathbf{r}=[r_x,r_y]^{\mathsf{T}}$. The EKF measurement is the relative sine-AoA
\begin{equation}
s_{\Delta}(\mathbf{p}_k;\mathbf{r})
\triangleq
\sin\!\big(\theta^{X}(\mathbf{p}_k;\mathbf{r})\big)-\sin\theta^{S}.
\label{eq:sdelta_meas_model}
\end{equation}

\paragraph{Bistatic Doppler}
Define the unit vectors pointing from the Tx and the Rx to the target:
\begin{equation}
\mathbf{u}_{\mathrm{tx}}
\triangleq
\frac{\mathbf{p}_k-\mathbf{t}}{\|\mathbf{p}_k-\mathbf{t}\|_2},
\quad
\mathbf{u}_{\mathrm{rx}}
\triangleq
\frac{\mathbf{p}_k-\mathbf{r}}{\|\mathbf{p}_k-\mathbf{r}\|_2}.
\label{eq:unitvec_txrx}
\end{equation}
The bistatic Doppler is modeled by 
\begin{equation}
f^{D}(\mathbf{p}_k,\mathbf{v}_k;\mathbf{t},\mathbf{r})
=
\frac{f_c}{c}\left(\mathbf{u}_{\mathrm{tx}}+\mathbf{u}_{\mathrm{rx}}\right)^{\mathsf{T}}\mathbf{v}_k,
\label{eq:fD_geom2}
\end{equation}
where $f_c$ is the carrier frequency and $c$ is the speed of light.

Collecting Eq. \eqref{eq:dd_geom}, Eq. \eqref{eq:sdelta_meas_model}, and Eq. \eqref{eq:fD_geom2}, the measurement function is
\begin{equation}
\mathbf{h}\!\left(\mathbf{x}_k;\mathbf{t},\mathbf{r}\right)
=
\begin{bmatrix}
\Delta d(\mathbf{p}_k;\mathbf{t},\mathbf{r})\\[3pt]
s_{\Delta}(\mathbf{p}_k;\mathbf{r})\\[3pt]
f^{D}(\mathbf{p}_k,\mathbf{v}_k;\mathbf{t},\mathbf{r})
\end{bmatrix}.
\label{eq:ekf_h_target}
\end{equation}

\subsubsection{EKF Recursion and Jacobian Evaluation}
The EKF alternates between prediction using Eq. \eqref{eq:ekf_state_model_target} and update using Eq. \eqref{eq:ekf_meas_target}. Since the AoA measurement is represented in the relative $s_{\Delta}$, no angle wrapping is required in the innovation, avoiding the $2\pi$ discontinuity of angular residuals. The EKF requires the Jacobian $\mathbf{H}_k \triangleq \left.\frac{\partial \mathbf{h}}{\partial \mathbf{x}}\right|_{\widehat{\mathbf{x}}_{k|k-1}}$.
We compute $\mathbf{H}_k$ numerically using a central-difference approximation. In our implementation, the range-difference and $s_{\Delta}$ components depend only on position, while the Doppler component depends on both position and velocity; thus $\partial \mathbf{h}/\partial \mathbf{a}_k=\mathbf{0}$.

\subsubsection{Initialization}
\label{subsubsec:ekf_init}
When motion is detected, we initialize the EKF state as
\begin{equation}
\mathbf{x}_{0}
=
\big[
\widehat{\mathbf{p}}_{0}^{\mathsf{T}},\,
\widehat{\mathbf{v}}_{0}^{\mathsf{T}},\,
\mathbf{0}^{\mathsf{T}}
\big]^{\mathsf{T}},
\label{eq:ekf_init_target}
\end{equation}
where $\widehat{\mathbf{p}}_{0}$ is a geometric Cartesian estimate obtained from the fused range-difference and relative sine-AoA measurements under the given Tx location $\mathbf{t}$. When first two consecutive position estimates are available at initialization, we use
\begin{equation}
\widehat{\mathbf{v}}_{0}=\frac{\widehat{\mathbf{p}}_{1}-\widehat{\mathbf{p}}_{0}}{\Delta T}.
\label{eq:ekf_init_v_target}
\end{equation}
The initial covariance is
\begin{equation}
\mathbf{P}_{0}
=
\mathrm{diag}\!\left(
\sigma_p^2,\sigma_p^2,\sigma_v^2,\sigma_v^2,\sigma_a^2,\sigma_a^2
\right),
\label{eq:ekf_initP_target}
\end{equation}
with $\sigma_v^2,\sigma_a^2\gg\sigma_p^2$. The measurement noise covariance is
$\mathbf{R}=\mathrm{diag}(\sigma_{\Delta d}^2,\sigma_{s}^2,\sigma_f^2)$.

Our pipeline provides per-CPI geometric estimates from range difference and relative sine-AoA, enabling deterministic initialization without random guessing. In contrast, Doppler-only or optimization-based bistatic tracking \cite{li2017indotrack, 10962318, qian2018widar2} often adopts a random or heuristic initialization (e.g., setting $\mathbf{p}_0=\mathbf{0}$), which can lead to slow convergence and make the recovered trajectory sensitive to noise and deployment priors (detailed in Section \ref{subsec:results}).

\subsubsection{Feasibility gating}
To suppress occasional unstable updates, we apply physics-based bounds \emph{after} the EKF update. After obtaining the posterior state $\widehat{\mathbf{x}}_{k|k}$, we compute the implied measurement $\widehat{\mathbf{z}}_{k|k}\triangleq \mathbf{h}(\widehat{\mathbf{x}}_{k|k};\mathbf{t},\mathbf{r})$ and check feasibility. For the bistatic range difference, we enforce $0\le \widehat{\Delta d}_{k|k}\le \Delta d_{\max}$, where $\Delta d_{\max}$ is set by the sensing-region geometry.
For the relative sine-AoA, since $\sin\theta^{X}=\widehat{s}_{\Delta,k|k}+\sin\theta^{S}\in[-1,1]$, we require
$-1-\sin\theta^{S} \le \widehat{s}_{\Delta,k|k} \le 1-\sin\theta^{S}$. For Doppler, we enforce $|\widehat{f}^{D}_{k|k}|\le f^{D}_{\max}$, where $f^{D}_{\max}$ is bounded by the slow-time sampling and is further restricted using a maximum expected target speed (we use $v_{\max}=5$~m/s in indoor tracking). If any bound is violated, we clip the corresponding component of $\widehat{\mathbf{z}}_{k|k}$ to the nearest feasible value.

\subsubsection{Track Management}
\label{subsubsec:ekf_manage}
At each fused CPI, we first perform the EKF prediction. If no motion is detected, we skip the update and propagate by prediction only. Otherwise, we apply an innovation gate to reject abnormal measurements. Let
\begin{equation}
\mathbf{y}_k \triangleq \mathbf{z}_k-\mathbf{h}(\widehat{\mathbf{x}}_{k|k-1}),
\quad
\mathbf{S}_k \triangleq \mathbf{H}_k\mathbf{P}_{k|k-1}\mathbf{H}_k^{\mathsf{T}}+\mathbf{R}.
\label{eq:innov_S_def_tx_full}
\end{equation}
We accept the update if
\begin{equation}
\mathbf{y}_k^{\mathsf{T}}\mathbf{S}_k^{-1}\mathbf{y}_k \le \gamma_g.
\label{eq:ekf_gate_tx_full}
\end{equation}
If the gate fails, we treat it as a missed association and propagate by prediction only. Let $N_{\mathrm{term}}$ denote the maximum allowed number of consecutive missed updates. We terminate the track if misses persist for $N_{\mathrm{term}}$ consecutive CPIs, or if its visibility ratio $\eta \triangleq N_{\mathrm{vis}}/N_{\mathrm{age}}$ drops below $\eta_{\min}$, where $N_{\mathrm{vis}}$ is the number of CPIs with accepted updates and $N_{\mathrm{age}}$ is the track age (in CPIs). A track is confirmed only after it survives for at least $N_{\mathrm{conf}}$ CPIs; otherwise, it remains tentative.



\subsection{Position-Refined Micro-Doppler}
\label{subsec:position_based_sensing}
Beyond trajectory tracking, we leverage the estimated target location to extract cleaner micro-Doppler signatures.
The key idea is to spatially focus the 3D delay-AoA-Doppler spectrum around the target-associated delay-AoA region so
that each CPI yields a target-dominated complex observation. The resulting complex sequence across CPIs exhibits
periodic modulations induced by gait micro-motions (e.g., cadence and limb swings), which are informative for sensing applications.

\subsubsection{Delay-AoA focusing}
For each CPI $k$, we obtain the 3D spectrum $\mathcal{Z}_k[\ell,m,n]$ and its magnitude
$S_k[\ell,m,n]\triangleq|\mathcal{Z}_k[\ell,m,n]|$. We first locate the dominant bin
\begin{equation}
(\ell_k,m_k,n_k)=\arg\max_{\ell,m,n}~ S_k[\ell,m,n].
\label{eq:peak_argmax_lmn}
\end{equation}
To reduce bin quantization and local leakage, we apply a 2D Gaussian focusing window in the delay-AoA plane centered at
$(\ell_k,m_k)$:
\begin{equation}
G_k[\ell,m]\triangleq
\exp\!\Big(
-\tfrac{1}{2}\big(\tfrac{\ell-\ell_k}{\sigma_\ell}\big)^2
-\tfrac{1}{2}\big(\tfrac{m-m_k}{\sigma_m}\big)^2
\Big),
\label{eq:gaussian_focus}
\end{equation}
and form the focused complex spectrum
\begin{equation}
\mathcal{Z}^{\mathrm{foc}}_k[\ell,m,n]\triangleq G_k[\ell,m]\;\mathcal{Z}_k[\ell,m,n].
\label{eq:Z_foc_def}
\end{equation}

\subsubsection{Doppler reweighting}
Due to finite CPI length and spectral leakage, a sign-flipped Doppler counterpart may still remain even after delay-AoA
gating. We suppress it by comparing each Doppler bin with its mirrored index
\begin{equation}
n^{\dagger}\triangleq (N_t-n)\bmod N_t.
\label{eq:doppler_mirror_index}
\end{equation}
Let $S^{\mathrm{foc}}_k[\ell,m,n]\triangleq|\mathcal{Z}^{\mathrm{foc}}_k[\ell,m,n]|$. We compute the mirror-contrast ratio
\begin{equation}
r_{D,k}[\ell,m,n]\triangleq
\frac{S^{\mathrm{foc}}_k[\ell,m,n]}{S^{\mathrm{foc}}_k[\ell,m,n^{\dagger}]+\varepsilon},
\label{eq:doppler_mirror_ratio}
\end{equation}
and apply a soft reweighting
\begin{equation}
\mathcal{Z}^{\mathrm{rw}}_k[\ell,m,n]\triangleq
\big(r_{D,k}[\ell,m,n]\big)^{\kappa}\,
\mathcal{Z}^{\mathrm{foc}}_k[\ell,m,n],
\quad \kappa\in(0,1].
\label{eq:doppler_mirror_reweight}
\end{equation}
This favors the dominant Doppler side while down-weighting bins dominated by their sign-flipped counterparts.
 
\begin{figure}[t]
    \centering
    \begin{subfigure}[t]{0.49\linewidth}
        \centering
        \includegraphics[width=\textwidth]{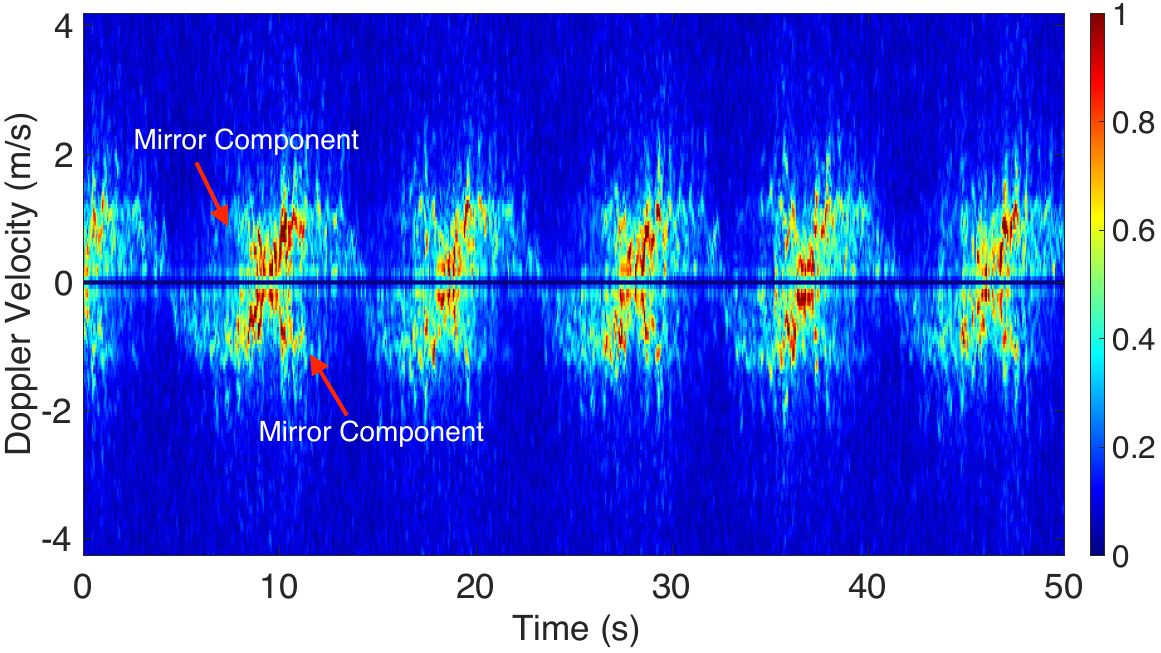}
        \subcaption{3D DFT-based micro-Doppler}
        \label{fig:traj_2d:range}
    \end{subfigure}
    \begin{subfigure}[t]{0.49\linewidth}
        \centering
        \includegraphics[width=\textwidth]{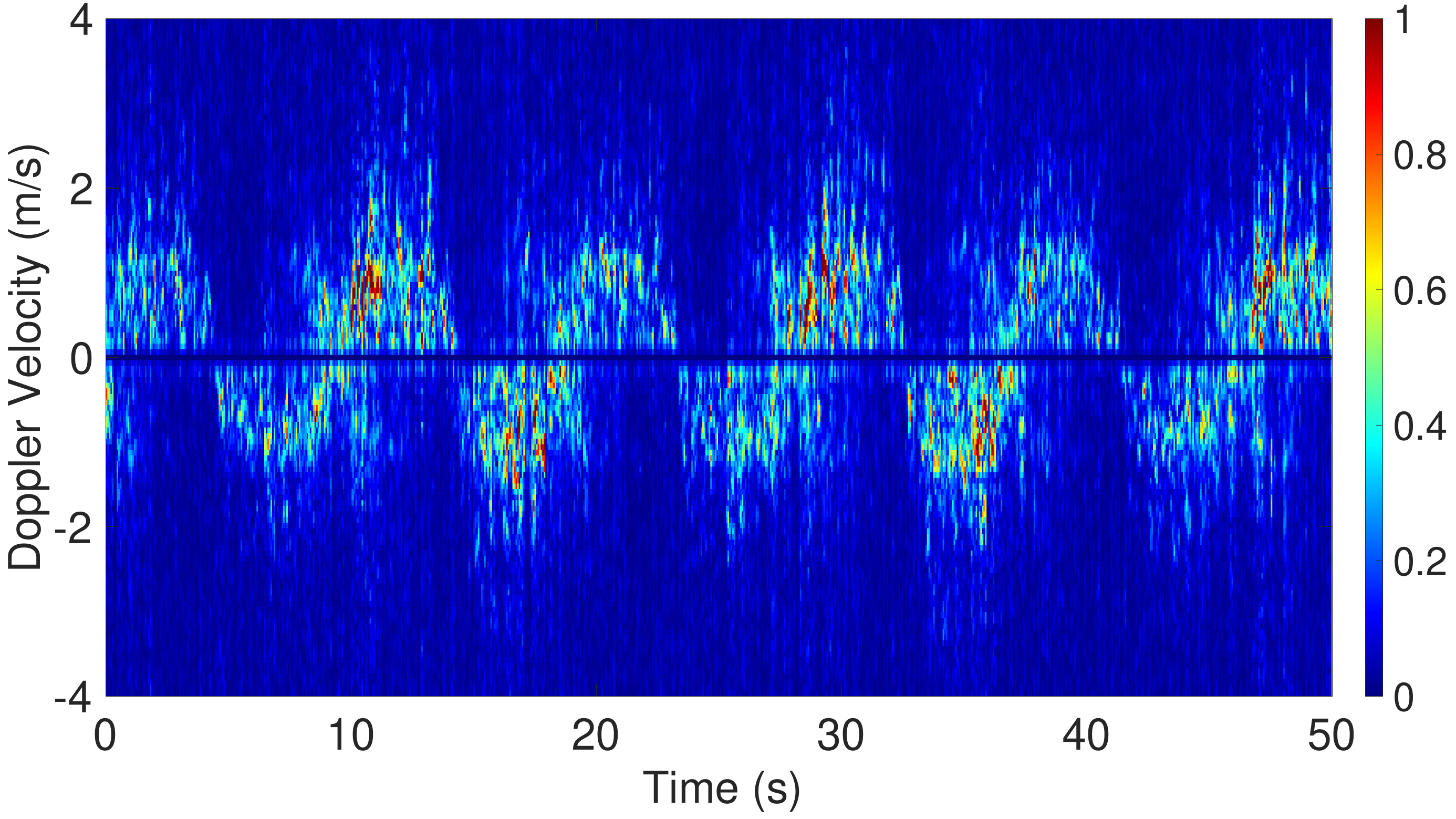}
        \subcaption{Position-based micro-Doppler}
        \label{fig:traj_2d:aoa}
    \end{subfigure}
    \caption{Micro-Doppler heatmaps: (a) baseline without spatial focusing, and (b) position-refined micro-Doppler via delay-AoA focusing.}
    \label{fig:microdoppler_comp}
    \vspace{-1.5em}
\end{figure}

\subsubsection{Across-CPI micro-Doppler spectrogram}
For each CPI, we form a 1D Doppler profile by averaging the magnitude across the delay-AoA plane:
\begin{equation}
S^{\mathrm{MD}}_k[n]\triangleq
\frac{1}{N_\ell N_m}\sum_{\ell}\sum_{m}
\left|\mathcal{Z}^{\mathrm{rw}}_k[\ell,m,n]\right|,
\label{eq:micro_doppler_line}
\end{equation}
where the averaging region is implicitly defined by the support of $G_k[\ell,m]$.
Stacking $\{S^{\mathrm{MD}}_k[n]\}$ over successive CPIs (with CPI step $T_{\mathrm{step}}$) yields a position-refined
micro-Doppler spectrogram. For reference, we also compute a baseline line without focusing and reweighting:
\begin{equation}
S^{\mathrm{base}}_k[n]\triangleq
\frac{1}{N_\ell N_m}\sum_{\ell}\sum_{m}
\left|\mathcal{Z}_k[\ell,m,n]\right|.
\label{eq:micro_doppler_line_baseline}
\end{equation}

\subsubsection{Single-Antenna Case}
If only one Rx antenna is available, AoA is unavailable. We therefore apply delay-only focusing: for each CPI,
we select the dominant delay bin $\ell_k$ and optionally apply a 1D Gaussian window around $\ell_k$ on
$\mathcal{Z}_k[\ell,n]$. The per-CPI micro-Doppler profile is then formed as $S_k[n]=|\mathcal{Z}_k[\ell,n]|$ after
delay focusing, and stacking $\{S_k[n]\}$ over CPIs yields the  micro-Doppler.

\subsubsection{Example and Visualization}
Fig.~\ref{fig:microdoppler_comp} compares two time-Doppler micro-Doppler heatmaps from an indoor bistatic WiFi setup (5~GHz, 20~MHz bandwidth, 30 subcarriers) for a linear walking trajectory. Fig.~\ref{fig:traj_2d:range} shows a baseline map obtained by directly averaging $|\mathcal{Z}_k[\ell,m,n]|$ over the delay-AoA plane. Fig.~\ref{fig:traj_2d:aoa} shows the proposed position refined map, where a delay-AoA Gaussian focusing window (with Doppler-mirror reweighting) is applied before averaging. Overall, spatial focusing concentrates energy on the target return and suppresses background leakage and residual mirror contamination, yielding clearer periodic motion patterns.

\subsection{Sparse-Sampling Case}
When Doppler processing is unavailable, we operate on the 2D spectrum $P^{\mathrm{2D}}_k[\ell,m]\triangleq\left|\widetilde{\mathcal{Z}}_{\mathrm{2D}}[\ell,m]\right|^2$ using a 2D neighborhood in place of $\mathcal{C}$. We compute the per-CPI statistic $\Lambda_k$ analogously and smooth it over a sliding window of $L$ CPIs via Eq. \eqref{eq:Lambda_fuse} and Eq. \eqref{eq:cfar_fuse} to suppress occasional spurious peaks.
For the per-CPI estimates $x_k\in\{\widehat{\tau}_{\Delta,k},\,\widehat{s}_{\Delta,k}\}$, we apply the window-based
outlier filtering and then fuse the retained samples by calculating a  average, yielding more stable delay-AoA estimates and smoother trajectories.

\begin{table}[t]
\centering
\caption{Simulation parameters.}
\label{tab:sim_params}
\begin{tabular}{l c}
\hline
\textbf{Parameter} & \textbf{Value} \\
\hline
Carrier frequency $f_c$ & 5~GHz \\
Total bandwidth $B$ & 20~MHz \\
CSI subcarriers $N_f$ & 30 \\
Receivers $R$ & 4 (baseline); our method uses $R=1$ \\
Rx antennas per receiver $M$ & 3 \\
Antenna spacing $d$ & $\lambda/2$~m \\
CSI sampling rate $f_s$ & 1~kHz \\
Sampling interval $\Delta t$ & 1~ms \\
Static amplitude $\rho^{S}$ & 1.0 (reference) \\
Dynamic amplitude scale $\rho^{X}_0$ & 0.15 (reference, distance-scaled) \\
TO $\mathrm{std}(\tau^{\mathrm{TO}}_k)$ & 20~ns \\
Packet phase $\phi^{\mathrm{CFO}}_k$ & i.i.d.\ uniform on $[-\pi,\pi]$ \\
Inter-chain phase offsets & enabled (ours) \\
Additive noise & AWGN, SNR $\in\{15,30\}$~dB (or none) \\
Ellipse center $(x_c,y_c)$ & (0, 4)~m \\
Ellipse axes $(a_e,b_e)$ & (4, 3)~m \\
Lap period $T_0$ & 25~s \\
Number of laps & 3 \\
\hline
\end{tabular}
\vspace{-1.5em}
\end{table}

\section{Implementation}
\label{sec:exp_setup}

\subsection{Simulation Data Generation}
\label{subsec:sim_setup}
To benchmark against multi-receiver-based Doppler tracking pipelines~\cite{li2017indotrack, li2024wifi, 10962318}, we generate controlled bistatic CSI using the signal model in Section~\ref{subsec:bistatic_csi_model}. The setup uses one transmitter and $R=4$ receivers, each with a $M=3$-element ULA, placed around the motion region to provide diverse viewing angles. Key parameters are summarized in Table~\ref{tab:sim_params}.

\subsubsection{Target trajectory}
The target follows an elliptical trajectory with center $(x_c,y_c)$, axes $(a_e,b_e)$, and lap period $T_0$. Its position $\mathbf{p}_k$ is generated analytically, and the velocity $\mathbf{v}_k$ is obtained from the corresponding derivative.

\subsubsection{CSI synthesis}
For each receiver, CSI is synthesized by superposing a dominant static component (Tx-Rx) and a target-induced dynamic component (Tx-target-Rx), where the phase on each antenna/subcarrier is determined by geometric delays. We set the static amplitude to $\rho^{S}=1$ as a reference. The dynamic amplitude is distance-dependent and scaled with the bistatic path length $d^{X}_{r,k}$ (with the Tx-Rx length as the reference path), using an attenuation exponent $\alpha$ (we use $\alpha=1$). Practical impairments are included by injecting packet-dependent timing offset and random CFO phase, together with receiver-dependent inter-chain phase offsets.

\subsection{LTE CSI Acquisition in Real-World Experiments}
\label{subsec:csi_acq}
We evaluate the tracking performance of WiDFS2.5 using an over-the-air 3.1~GHz LTE CSI acquisition platform \cite{chen2023development}.

\begin{figure}[t]
\centering
\begin{subfigure}[t]{0.466\linewidth}
    \centering
    \includegraphics[width=\linewidth]{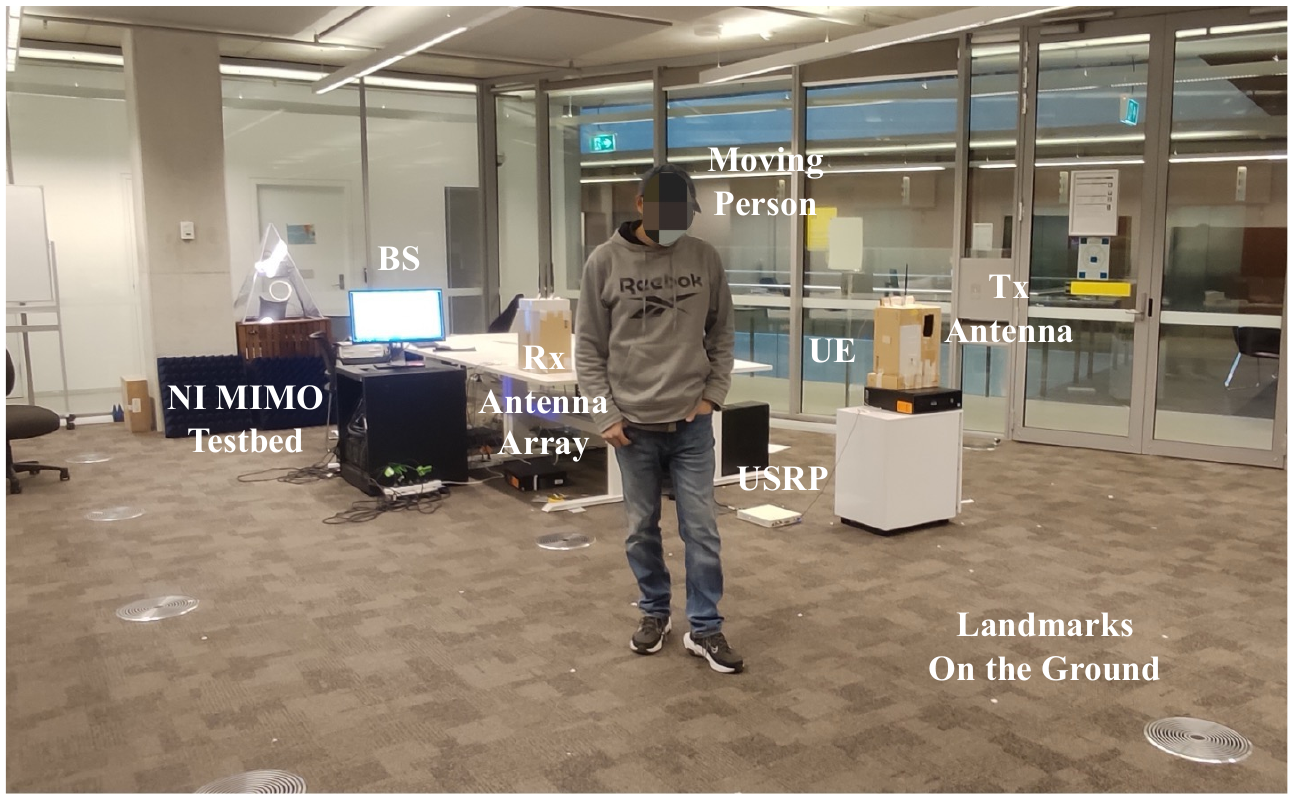}
    \subcaption{Experiment scenario}
    \label{fig:exp:scenario}
\end{subfigure}
\hfill
\begin{subfigure}[t]{0.52\linewidth}
    \centering
    \includegraphics[width=\linewidth]{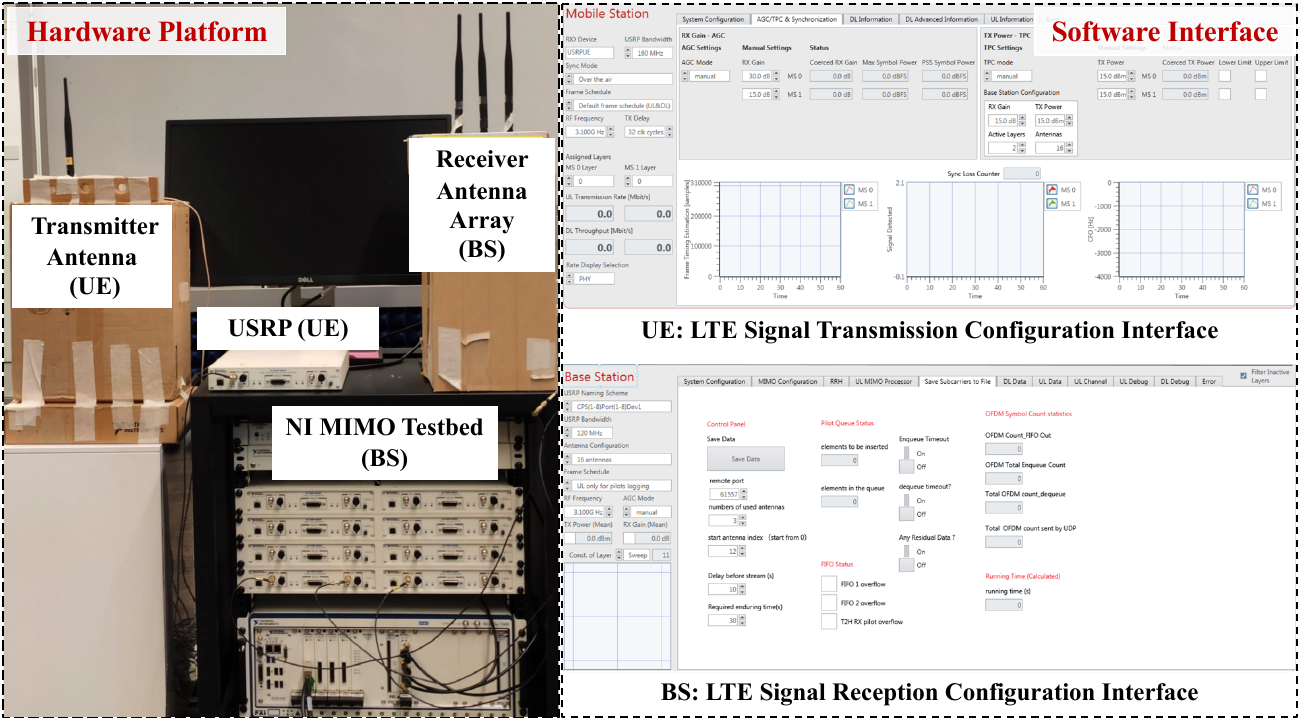}
    \subcaption{Testbed and software interface}
    \label{fig:exp:testbed}
\end{subfigure}
\caption{Experimental setup using a 3.1~GHz LTE signal.}
\label{fig:lte_exp_setup}
\vspace{-1em}
\end{figure}

\textit{1) System setup.}
As shown in Fig.~\ref{fig:lte_exp_setup}, the platform comprises a NI massive-MIMO prototyping base station (BS) and a software-defined-radio user equipment (UE). We use a 1Tx-3Rx bistatic configuration: the UE transmits uplink pilots and the BS estimates CSI from three receive antennas. No clock synchronization or prior antenna calibration is assumed.

\textit{2) Pilot structure.}
The system follows the 3GPP LTE frame structure (10~ms frames with 10 subframes). Uplink pilots are transmitted once per slot. The uplink occupies a 20~MHz channel with 1200 nominal subcarriers at 15~kHz spacing. Pilots are extracted at the resource-block (RB) level, i.e., one pilot-bearing tone per RB, yielding 100 active pilot subcarriers per slot. This corresponds to an effective frequency spacing of 180~kHz, and CSI is obtained on these 100 tones.

\textit{3) CSI extraction and streaming.}
We implement a real-time CSI streaming interface using LabVIEW Communications integrated with the massive-MIMO framework. Only pilot tones are retained to reduce streaming throughput. To mitigate UDP packet loss, the CSI stream is downsampled to 1~kHz before transmission to the host computer, where subsequent processing is performed in real time.

\begin{figure}
\centering
    \centering
    \includegraphics[width=0.6\linewidth]{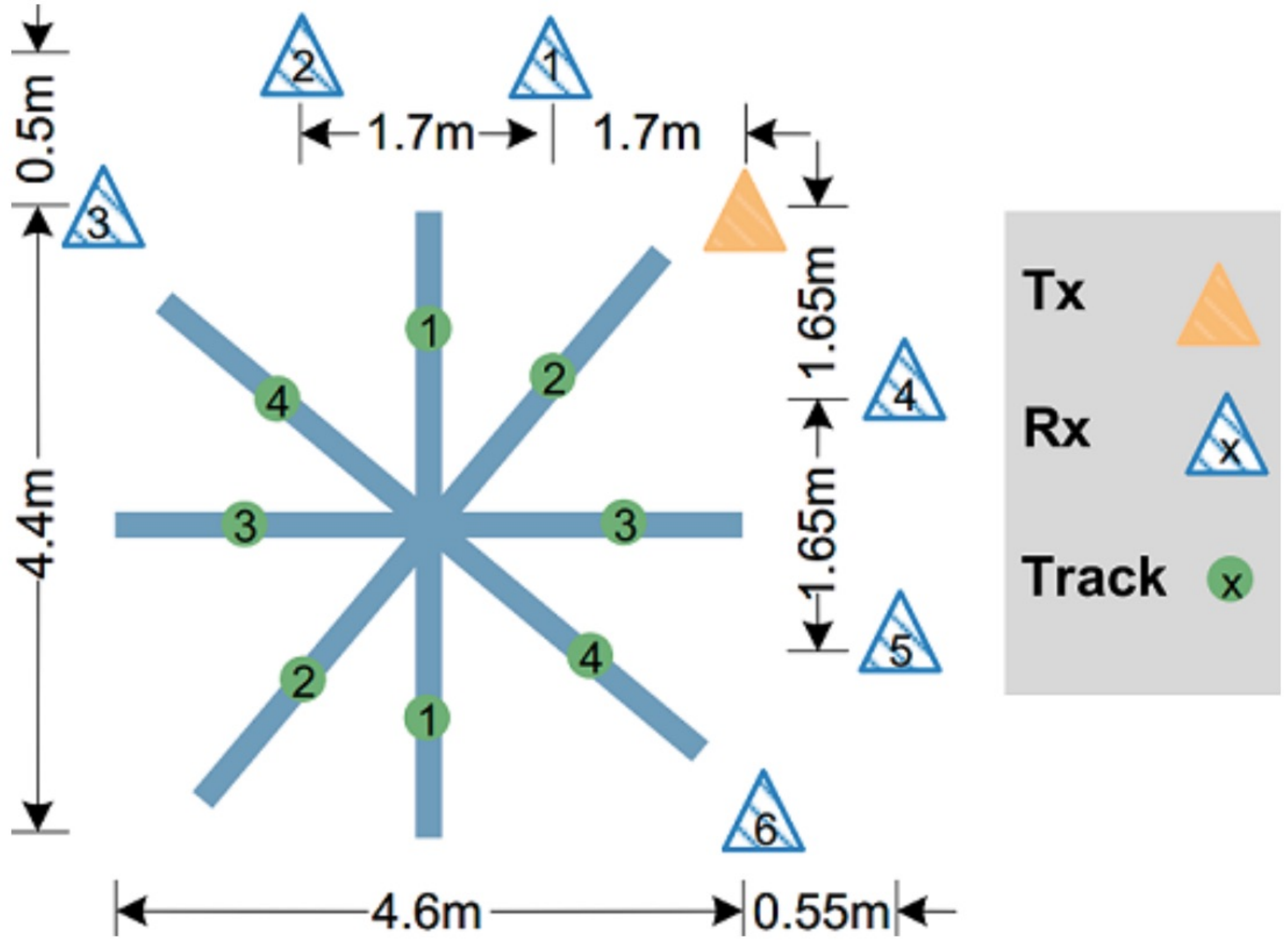}
    \caption{WiFi CSI measurement geometry of GaitID dataset \cite{zhang2020gaitid}.}
    \label{fig:gaitid_setup}
\vspace{-1.5em}
\end{figure}

\subsection{WiFi GaitID Dataset for Sensing Validation}
\label{subsec:gaitid}
We evaluate sensing performance using the public \emph{GaitID} dataset \cite{zhang2020gaitid}, collected in an indoor bistatic WiFi-CSI setup at 5~GHz using commodity Intel~5300 NICs. As illustrated in Fig.~\ref{fig:gaitid_setup}, the dataset deploys one transmitter and six receivers at different locations, and records CSI while volunteers walk along four predefined tracks in both directions under diverse indoor environments. Each sample corresponds to a short walking segment containing multiple steps and is labeled by subject identity. The dataset contains 11 subjects and 22,497 CSI
sequences in total. In this work, we use GaitID to validate the sensing quality of our micro-Doppler features.

\subsection{Algorithm Configuration}
\label{subsec:algo_params}

\subsubsection{DFT configuration}
Within each CPI, we form the CSI-power cube $\mathcal{P}\in\mathbb{R}^{N_a\times N_f\times N_t}$ and compute a 3D DFT
with zero-padding. Unless otherwise stated, we use DFT lengths
$(N_f^{\mathrm{pad}},N_a^{\mathrm{pad}},N_t^{\mathrm{pad}})=(128,32,32)$. We apply mirror-contrast reweighting with
$\kappa=0.5$. 

\subsubsection{Motion detection and multi-CPI fusion}
We compute the motion score in Eq. \eqref{eq:Lambda_def} using a $3\times3\times3$ local neighborhood. The per-CPI score $\Lambda_k$ is temporally fused by a sliding-window median over $L=128$ consecutive CPIs via Eq. \eqref{eq:Lambda_fuse}. Consecutive windows are shifted by 64 CPIs (50\% overlap). In our implementation, the fusion window spans approximately 1.5~s. Motion is declared when the fused score satisfies $\overline{\Lambda}_k>\gamma_{\mathrm{cfar}}$, where $\gamma_{\mathrm{cfar}}=5$ is chosen empirically from our subsequent experiment (Section \ref{subsubsec:motion_detection}).

\subsubsection{EKF Tracker and Track Management}
We run an EKF at the fused-CPI rate. Before each update, we apply an innovation gate by comparing the fused measurement with its EKF prediction and computing a Mahalanobis-type cost using the predicted covariance and measurement noise. The update is accepted only if the cost is below $\gamma_g=9$; otherwise, we skip the update and propagate by prediction only. A track is reported once it survives for $N_{\mathrm{age}}=5$ consecutive frames. A track is terminated if it has $N_{\mathrm{term}}=20$ consecutive rejected/missed updates, or if its visibility ratio $\eta \triangleq N_{\mathrm{vis}}/N_{\mathrm{age}}$ falls below $\eta_{\min}=0.6$. Tracks with $N_{\mathrm{age}}<20$ are tentative.

\begin{figure*}
\centering
\begin{subfigure}[t]{0.329\linewidth}
    \centering
    \includegraphics[width=\textwidth]{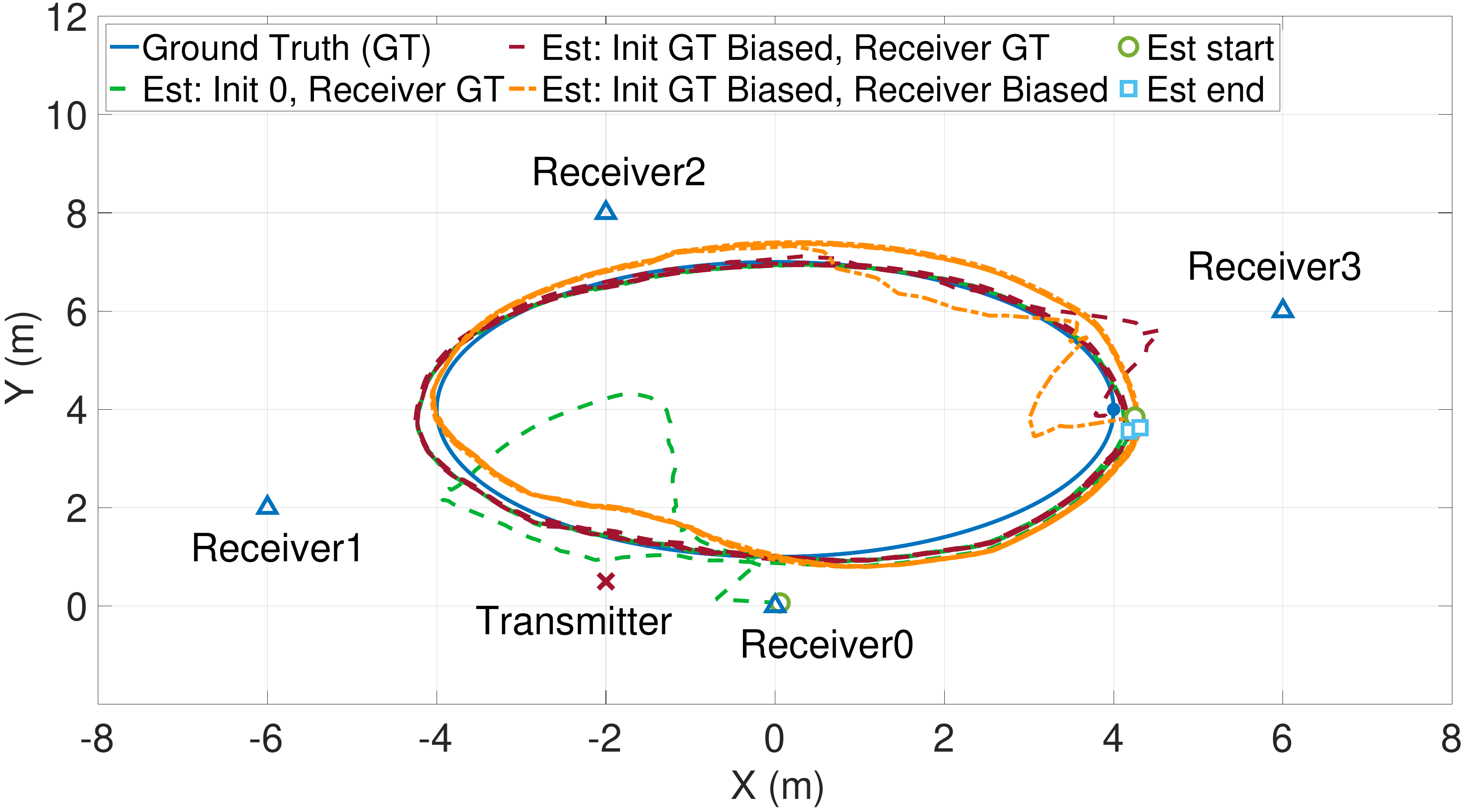}\\
    \includegraphics[width=\textwidth]{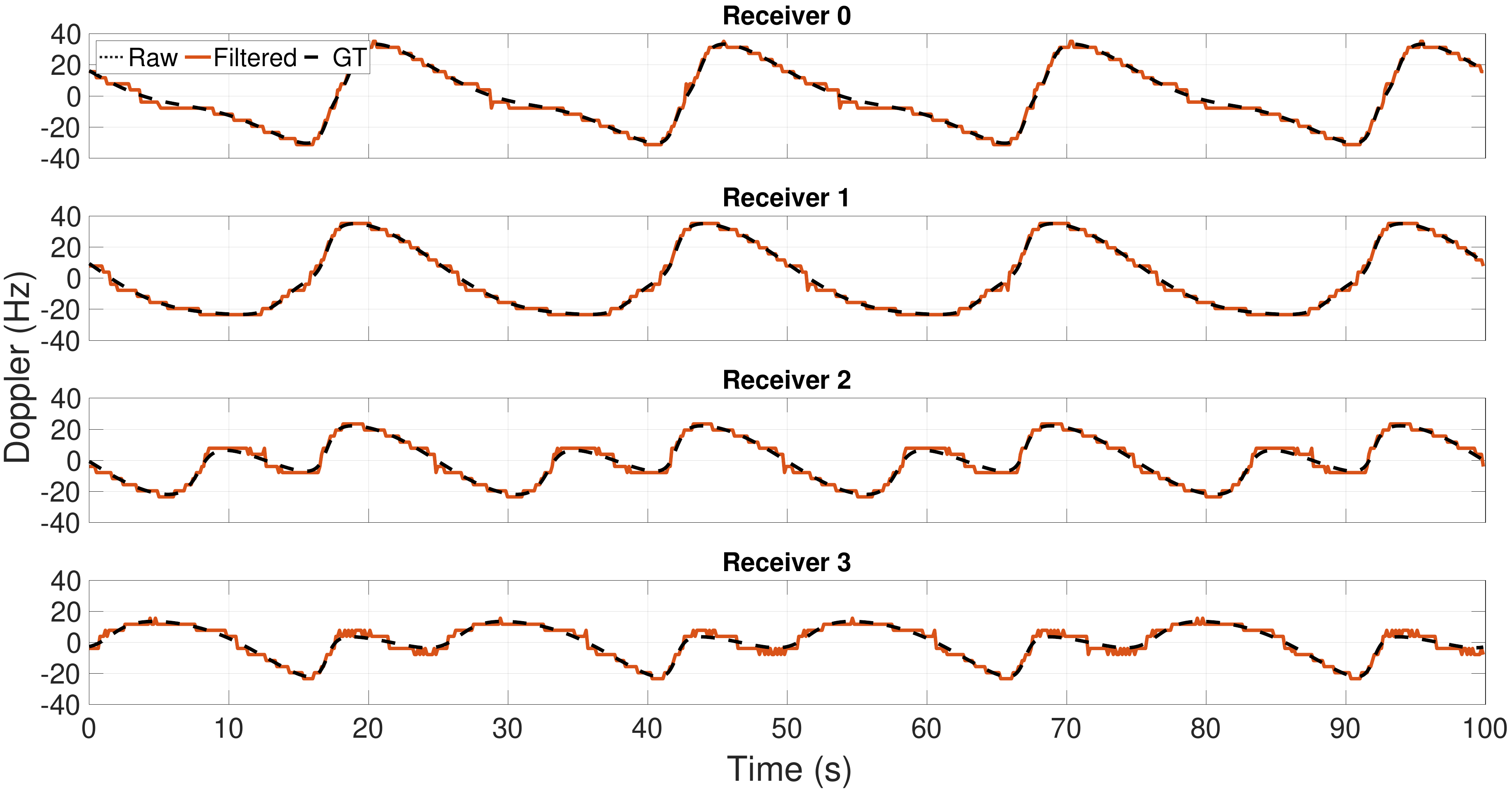}\\
    \subcaption{Impact of initialization and geometry bias (noise-free, CASR)}
    \label{fig:traj_all:init_bias}
\end{subfigure}
\hfill
\begin{subfigure}[t]{0.329\linewidth}
    \centering
    \includegraphics[width=0.97\textwidth]{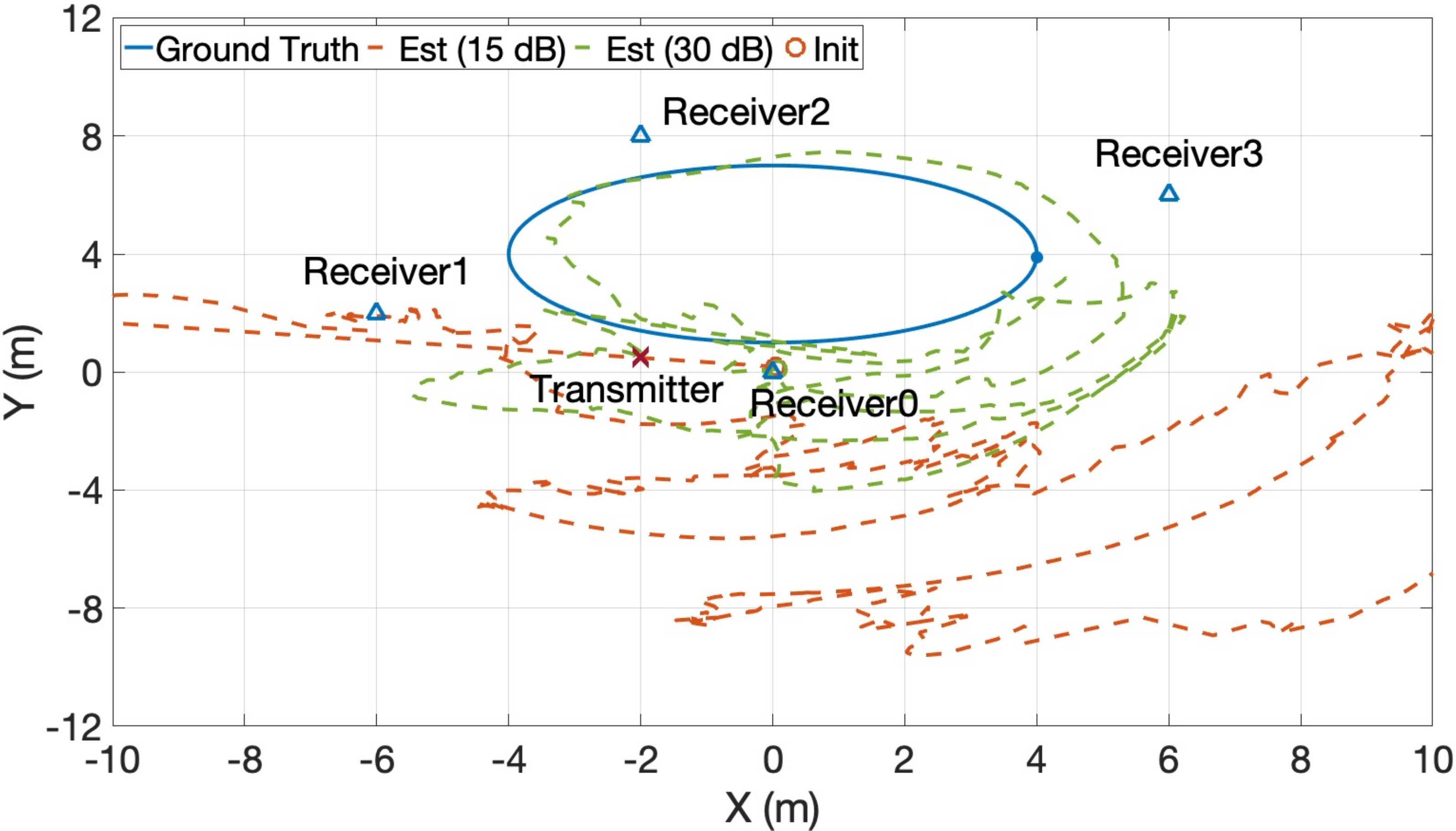}\\
    \includegraphics[width=\textwidth]{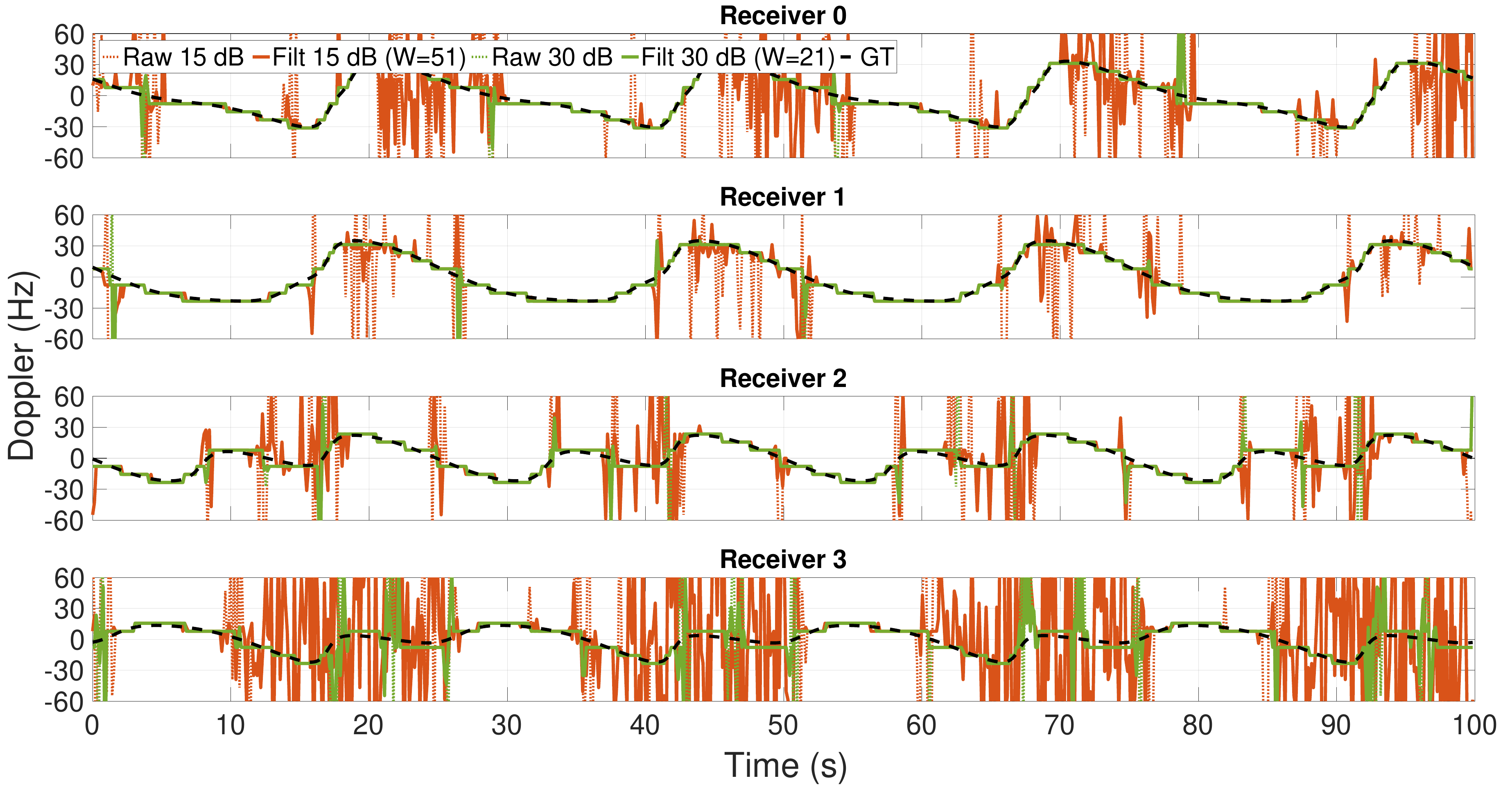}\\
    \subcaption{Impact of measurement noise (15 dB vs 30 dB,  CASR)}
    \label{fig:traj_all:snr}
\end{subfigure}
\hfill
\begin{subfigure}[t]{0.329\linewidth}
    \centering
    \includegraphics[width=\textwidth]{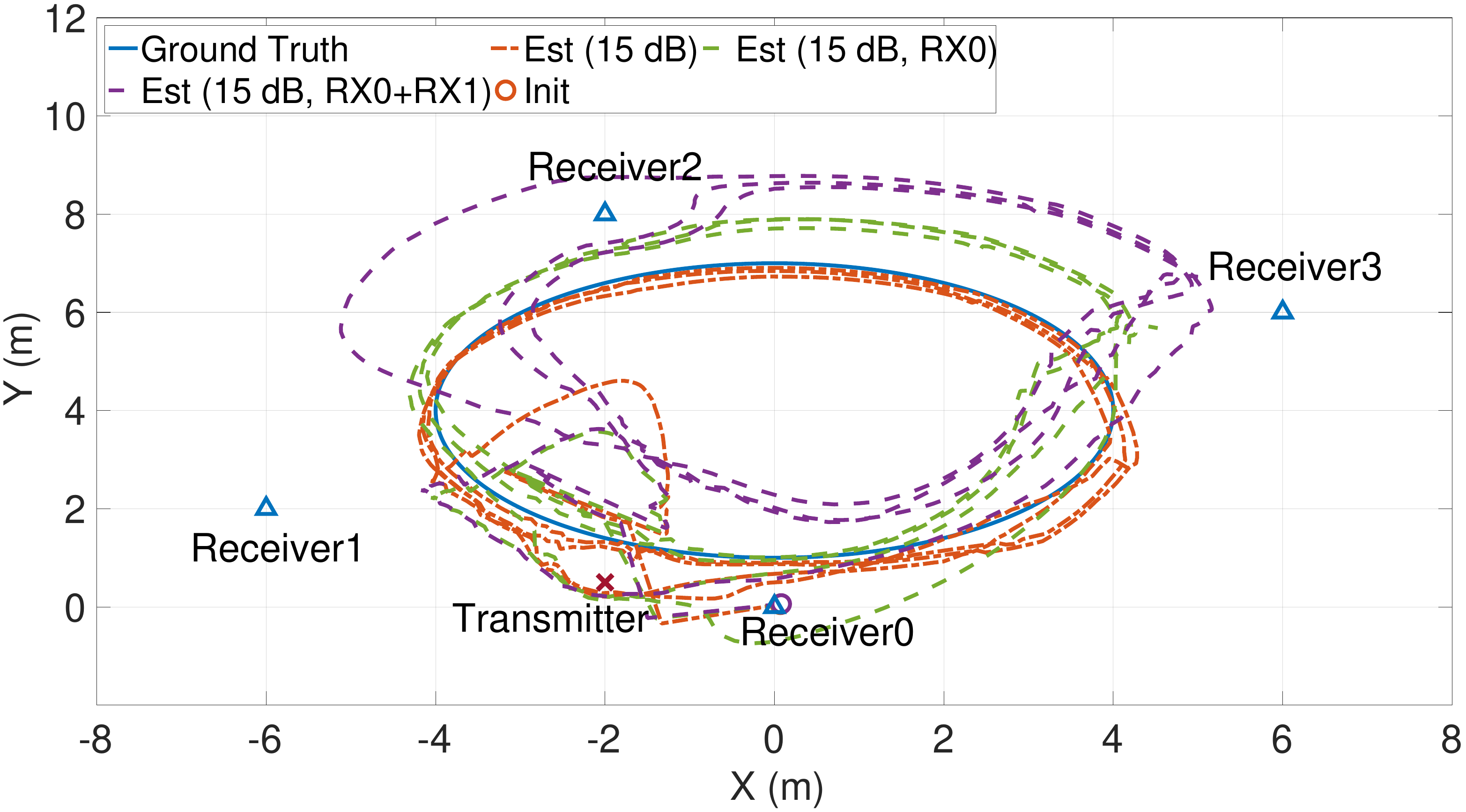}\\
    \includegraphics[width=\textwidth]{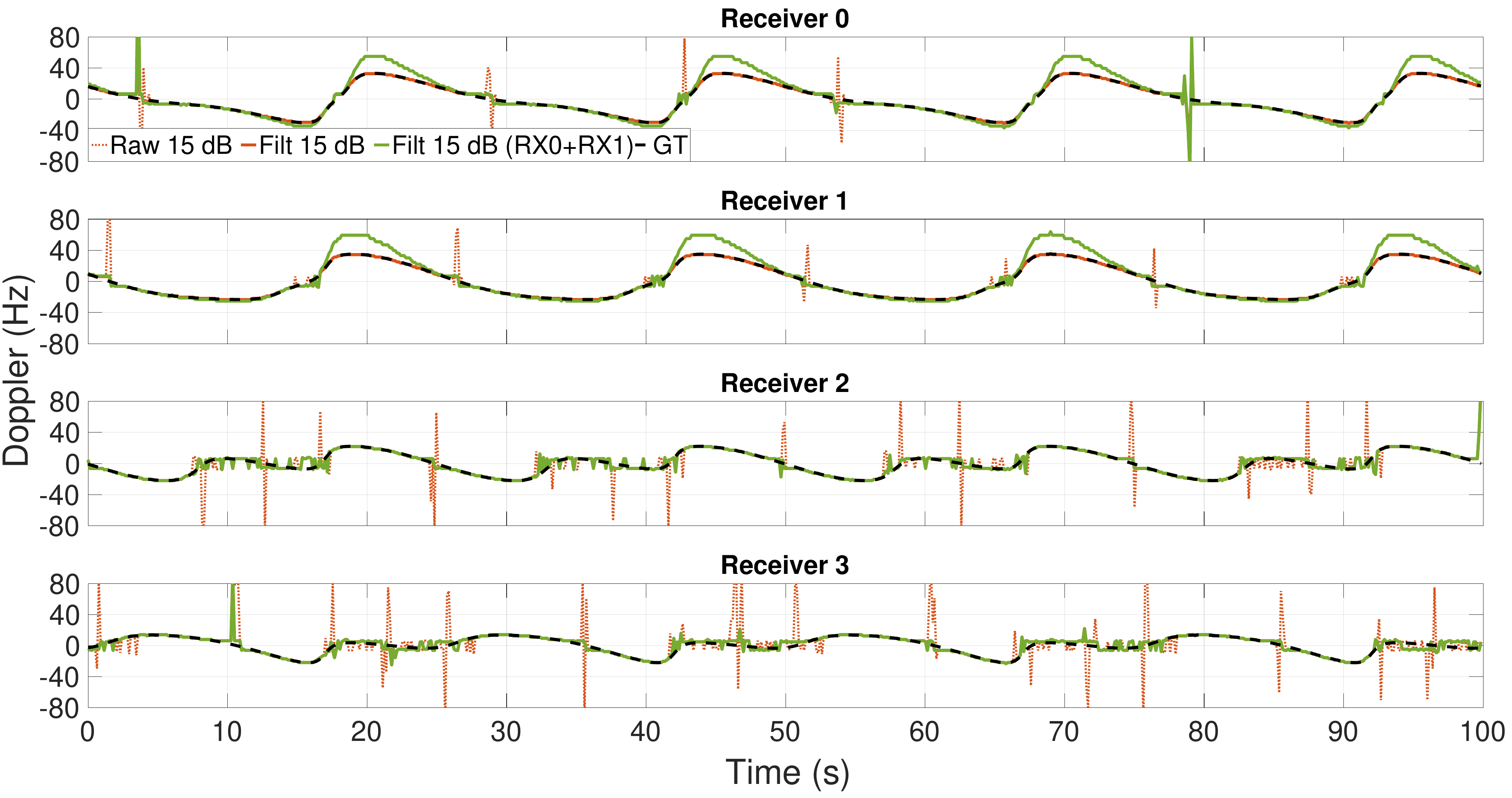}\\
    \subcaption{Impact of Doppler bias under a single link and multiple links (15 dB, CSI power)}
    \label{fig:traj_all:rxdiv}
\end{subfigure}
\caption{Doppler-only tracking performance under different conditions (multi-receiver, CASR for Doppler estimation).}
\label{fig:traj_all}
\vspace{-1em}
\end{figure*}

\begin{figure}
    \centering
    \begin{subfigure}[t]{\linewidth}
        \centering
        \includegraphics[width=0.75\textwidth]{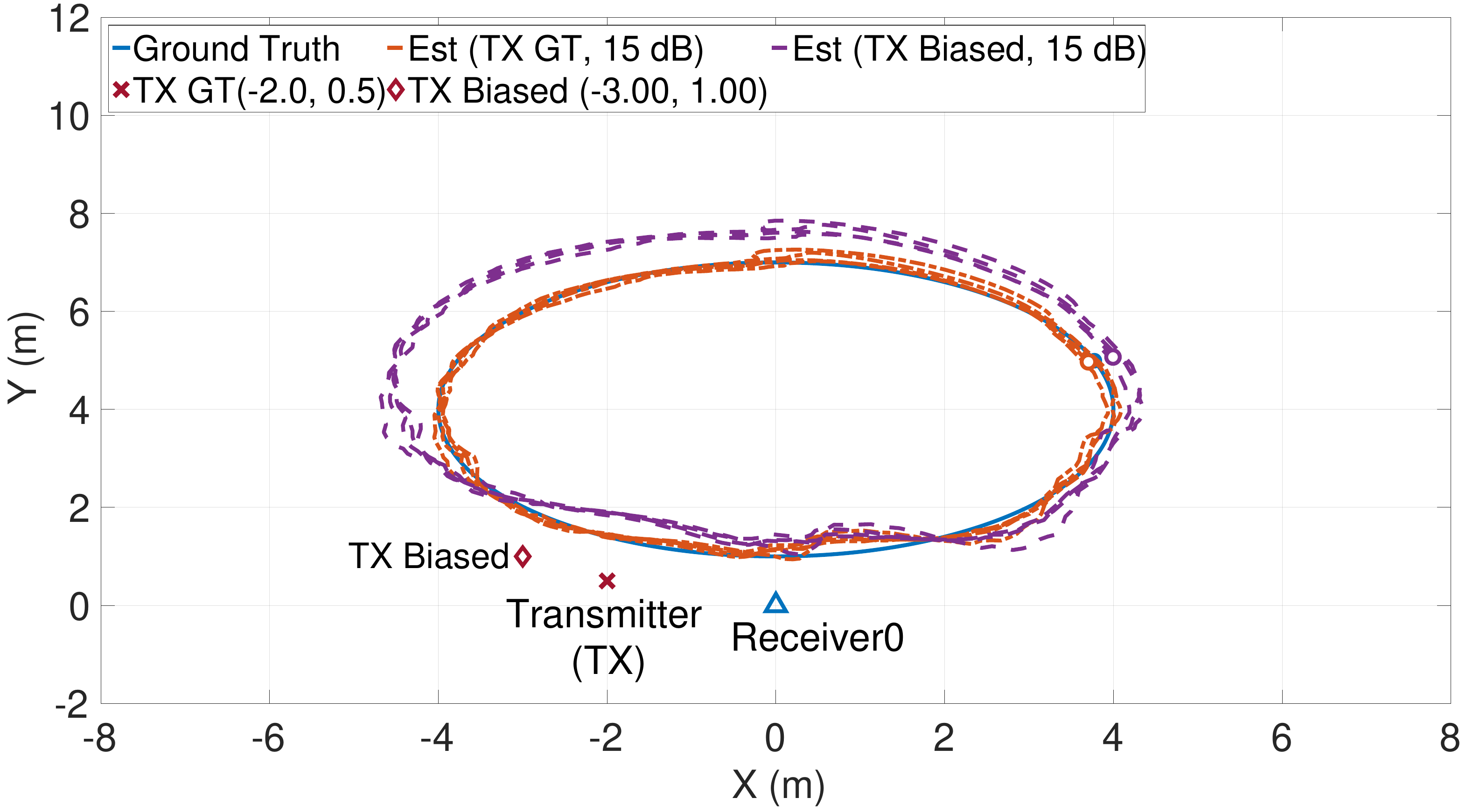}
        \subcaption{Impact of Tx bias (15dB)}
        \label{fig:txbias:a}
    \end{subfigure}
    \vspace{0.4em}
    \begin{subfigure}[t]{\linewidth}
        \centering
        \includegraphics[width=0.75\textwidth]{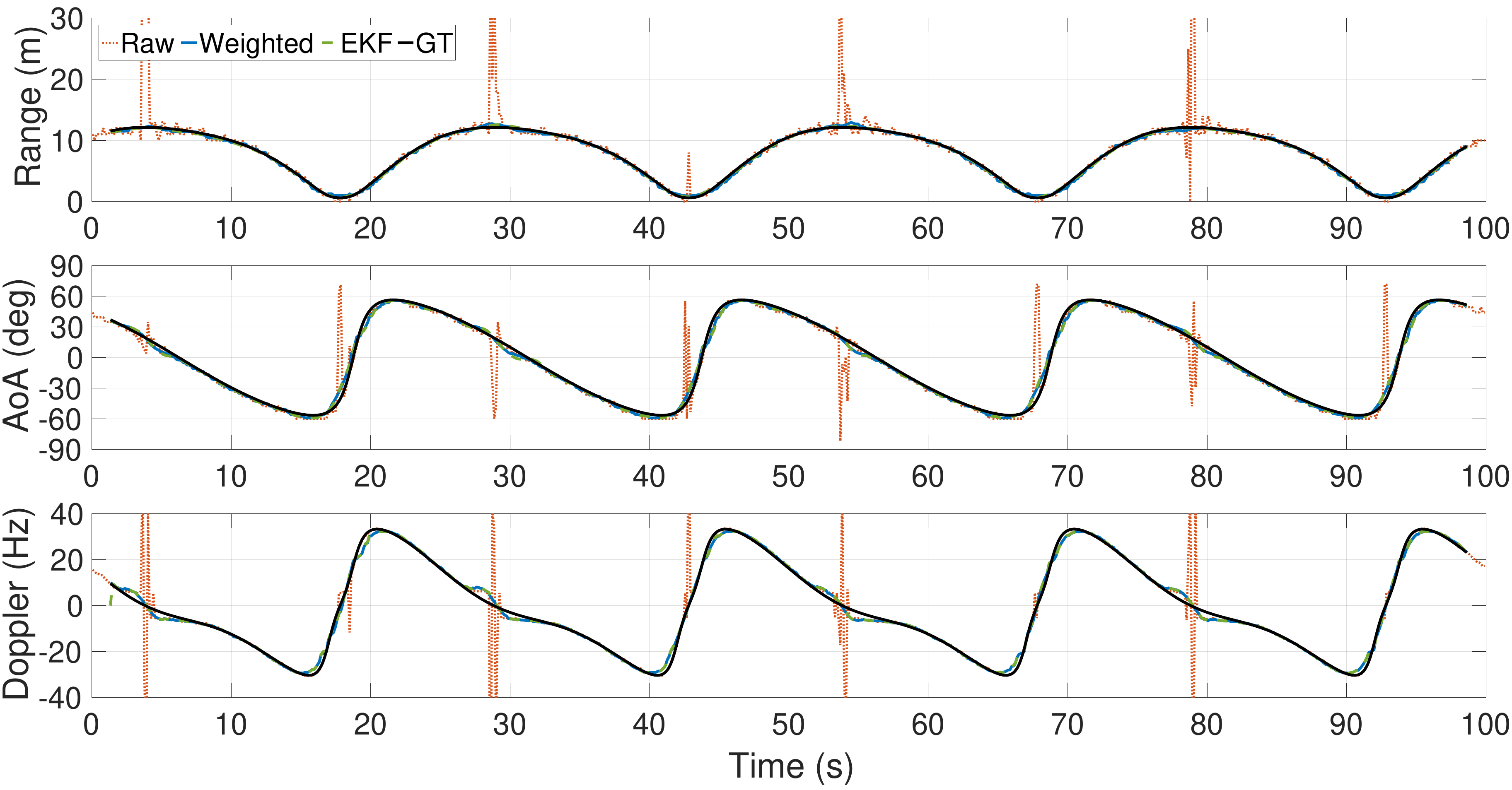}
        \subcaption{Extracted range, AoA, and Doppler measurements}
        \label{fig:txbias:b}
    \end{subfigure}
    \caption{Ours range-AoA-Doppler tracking performance (single-receiver, CSI power for multi-dimension parameter extraction).}
    \label{fig:txbias}
    \vspace{-1em}
\end{figure}

\subsection{Baselines}
\label{subsec:baselines}
\subsubsection{Tracking}
We consider a Doppler-only multi-receiver tracking pipeline~\cite{li2017indotrack, 10962318} as a representative baseline. The method extracts Doppler measurements from multiple bistatic Tx-Rx links. At each update, the target position $\mathbf{p}=[x,y]$ and velocity $\mathbf{v}=[v_x,v_y]$ are inferred from a nonlinear Doppler measurement model, which can be interpreted as solving a local nonlinear least-squares problem around the predicted state. Since Doppler measurements do not provide an absolute position anchor, the baseline uses a random initialization for $(\mathbf{p}_0,\mathbf{v}_0)$ and relies on subsequent updates to converge. Here, we implement this Doppler-only baseline using an EKF with a constant-velocity motion model. At each CPI, the EKF linearizes the bistatic Doppler measurement function around the predicted state and updates the target position and velocity using the multi-link Doppler observations.

We also considered a CSI-magnitude-based tracker~\cite{karanam2019tracking} that jointly estimates AoA and motion parameters and performs tracking with a particle filter and joint probabilistic data association filter. However, because it does not resolve the mirror ambiguity of power spectra, it typically assumes a restricted feasible region to avoid mirrored solutions. As our evaluation targets unconstrained tracking without such region assumptions, we do not include it as a baseline.

\subsubsection{Sensing}
For gait identification in sensing evaluation, we compare (a) CASR-based micro-Doppler extracted from the CSI-magnitude pipeline, (b) a raw 3D-DFT baseline without focusing, (c) the proposed position-refined micro-Doppler with delay-AoA focusing (plus Doppler-mirror reweighting), and (d) a range-refined delay-only focusing variant for the single-antenna (no-AoA) case. Each representation is used as input features to a lightweight deep network for classification. We report accuracy, precision, recall, and F1-score. Notably, GaitID was collected with Intel~5300 NICs, whose CSI phase is relatively stable with negligible time-varying inter-antenna $\pi$ jumps, making CASR applicable. In contrast, other chipsets~\cite{zubow2021phase} may exhibit inter-antenna phase discontinuities, which motivates CSI-power-based processing for improved cross-platform robustness.

\section{Results}
\label{subsec:results}

\subsection{Simulation Results}
\label{subsec:sim_results}
We compare our method with a Doppler-only baseline under different conditions. Fig.~\ref{fig:traj_all} shows a Doppler-only multi-receiver baseline with one transmitter and four receivers deployed around the motion region to provide favorable bistatic geometry (Receiver0 is the coordinate origin). The Tx/Rx locations are $\mathbf{t}=[-2.0,0.5]$ and $\mathbf{r}_0=[0.0,0.0]$, $\mathbf{r}_1=[-6.0,2.0]$, $\mathbf{r}_2=[-2.0,8.0]$, $\mathbf{r}_3=[6.0,6.0]$. The baseline assumes synchronous multi-link Doppler observations and estimates the target state from these measurements. In contrast, Fig.~\ref{fig:txbias} shows our single-receiver solution, which extracts range (delay), AoA, and Doppler from the CSI-power spectrum and tracks the target using an EKF.

\subsubsection{Impact of initialization and geometry bias}
We first study the Doppler-only multi-receiver baseline under an ideal noise-free setting. Fig.~\ref{fig:traj_all:init_bias} shows the trajectories estimated from CASR-extracted Doppler measurements collected by four receivers. We compare two initialization strategies while keeping receiver locations accurate: ``Init GT-biased, Rx GT'' initializes the EKF near the true initial state with a small perturbation, whereas ``Init 0-biased, Rx GT'' starts from $\mathbf{p}_0=\mathbf{0}$ with a biased initial state. As expected, an initialization closer to the true starting point leads to faster convergence and fewer transient errors. We further evaluate geometry bias by introducing random perturbations (within 1~m) to the receiver locations, denoted as Init GT-biased, Rx biased''. This mismatch in receiver geometry produces a systematic trajectory shift relative to the ground truth.

\subsubsection{Impact of measurement noise}
Fig.~\ref{fig:traj_all:snr} evaluates the Doppler-only baseline under AWGN at 15~dB and 30~dB, with a zero initialization $\mathbf{p}_0=\mathbf{0}$. When relying on CASR-extracted Doppler, the pipeline is most vulnerable when the true Doppler is near zero: the Doppler peak becomes weak, and residual DC/clutter leakage plus AWGN can dominate the spectrum even after suppression. As a result, CASR may select spurious peaks, producing large Doppler outliers. At higher SNR (30~dB vs.\ 15~dB), both the frequency and magnitude of outliers decrease. To mitigate these artifacts, we apply a Hampel outlier filter with sliding windows $\texttt{win}=51$ and $21$: a sample is flagged if its deviation from the local median exceeds $3$ scaled MADs, and is then replaced by the local median. While filtering removes isolated outliers, residual bias can still accumulate through the EKF recursion, causing gradual trajectory drift, especially at lower SNR.

\subsubsection{Impact of Doppler bias}
Fig.~\ref{fig:traj_all:rxdiv} evaluates Doppler bias effects under 15~dB. We extract Doppler from the CSI-power
delay-AoA-Doppler tensor and use delay/AoA to guide peak selection, producing a relatively stable Doppler sequence;
the few remaining outliers can be removed by a Hampel filter, after which the trajectory matches the ground truth well.
We then examine bias induced by non-rigid human motion, where different body parts yield inconsistent Doppler returns.
We emulate this by applying a nonlinear distortion to the Doppler measurements: ``Rx0'' distorts only Rx0, while
``Rx0+Rx1'' distorts both Rx0 and Rx1 (with $\mathbf{p}_0=\mathbf{0}$). The Doppler-only tracker exhibits significant
trajectory bias, which is larger for ``Rx0+Rx1''. This highlights a key limitation of Doppler-only tracking: it assumes a consistent Doppler from a single effective scattering center across links, which may not always hold for non-rigid targets in realistic environments.

\subsubsection{Ours CSI-power solution}
\label{subsubsec:ours_csi_power}
Fig.\ref{fig:txbias} reports our single-receiver tracking results using CSI-power multi-dimensional feature extraction. We jointly estimate range (delay), AoA, and Doppler and feed the fused measurements to the EKF. Occasional outliers mainly occur when the true Doppler approaches zero (slow motion). Under 15dB AWGN, we compare two transmitter-location settings: Tx GT'' uses the true Tx position, whereas Tx biased'' introduces a deployment bias. With Tx GT'', the EKF converges quickly and the recovered trajectory closely matches the ground truth; with Tx biased'', a systematic shift appears due to geometric mismatch. Compared with Doppler-only tracking, our approach is less sensitive to non-rigid human scattering: while Doppler may vary across body parts, delay and AoA remain tied to the target location and provide stable geometric constraints at each CPI, reducing drift. Moreover, incorporating delay and AoA relaxes the need for favorable multi-receiver geometry, enabling accurate tracking with a single receiver without random initialization.

\begin{figure*}
\centering
\begin{subfigure}[t]{0.329\linewidth}
    \centering
    \includegraphics[width=\textwidth]{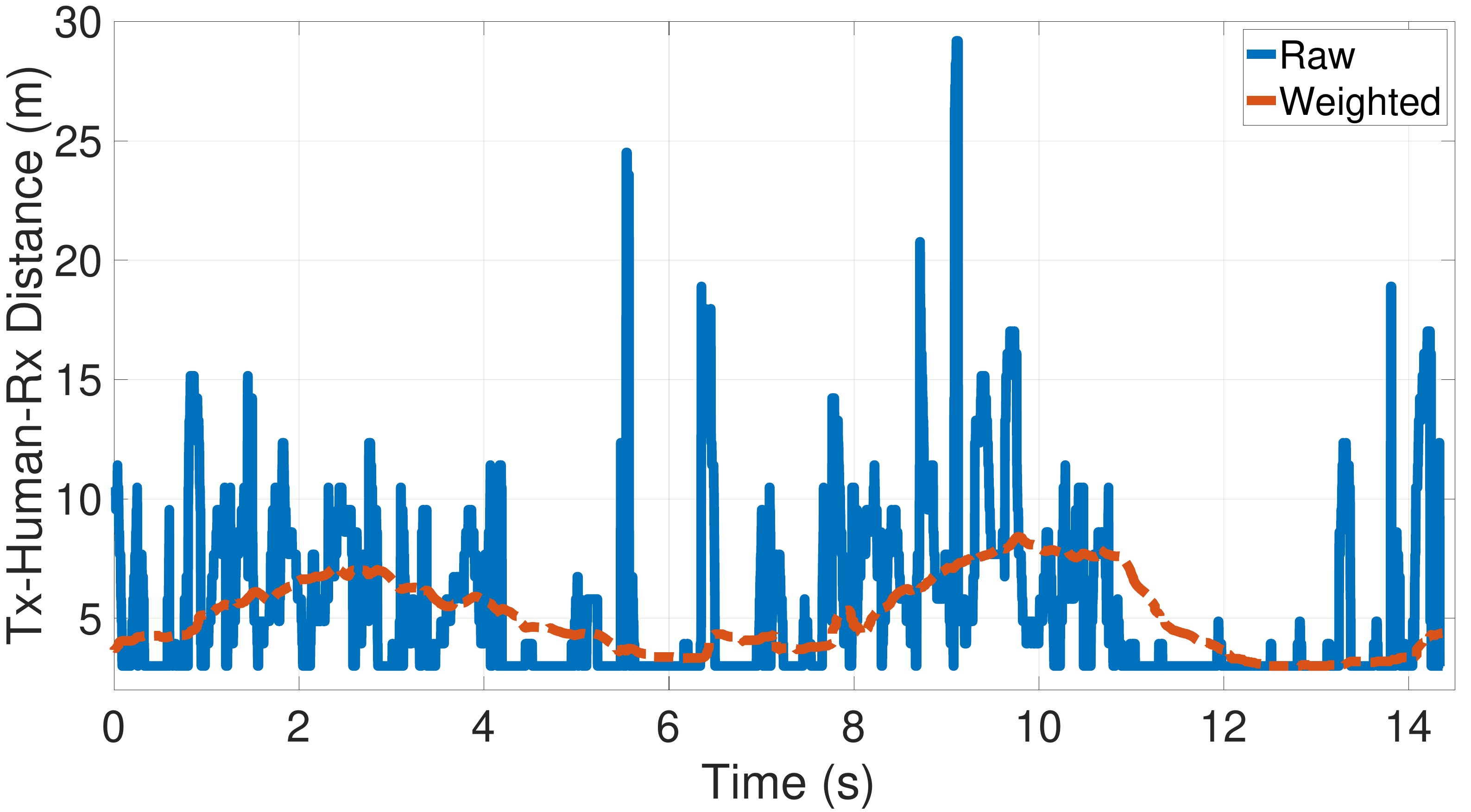}\\
    \includegraphics[width=\textwidth]{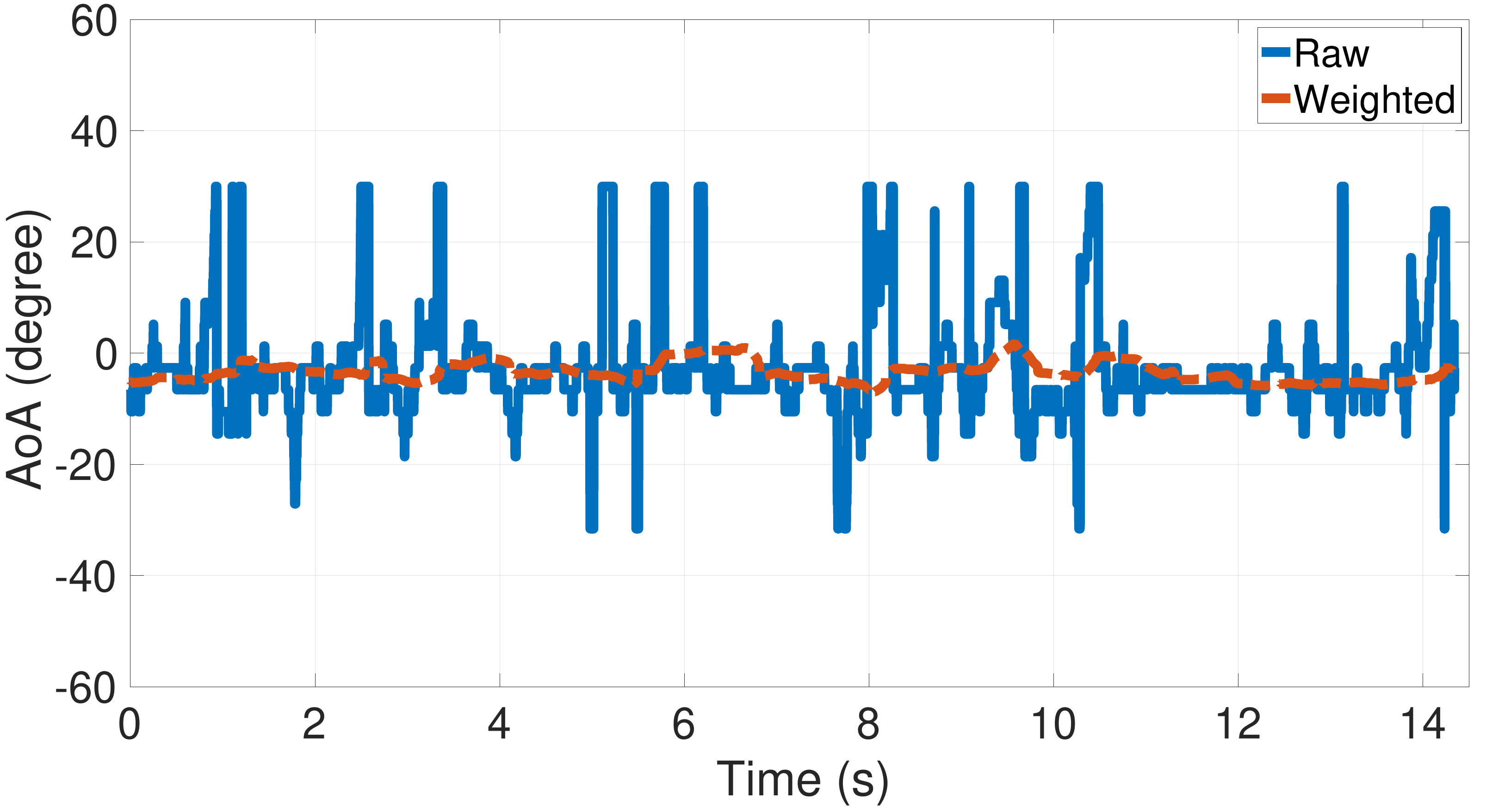}\\
    \includegraphics[width=\textwidth]{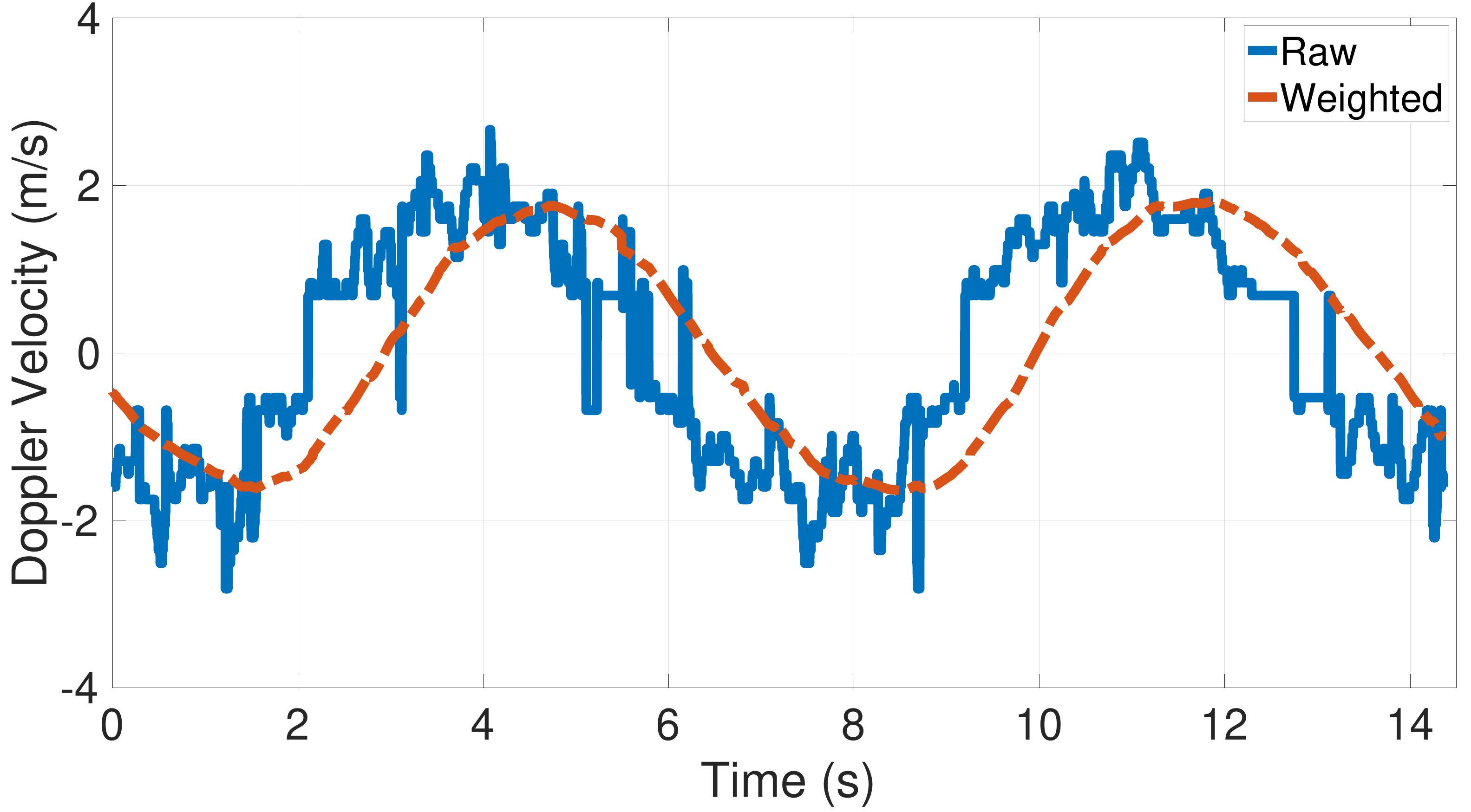}\\
    \includegraphics[width=\textwidth]{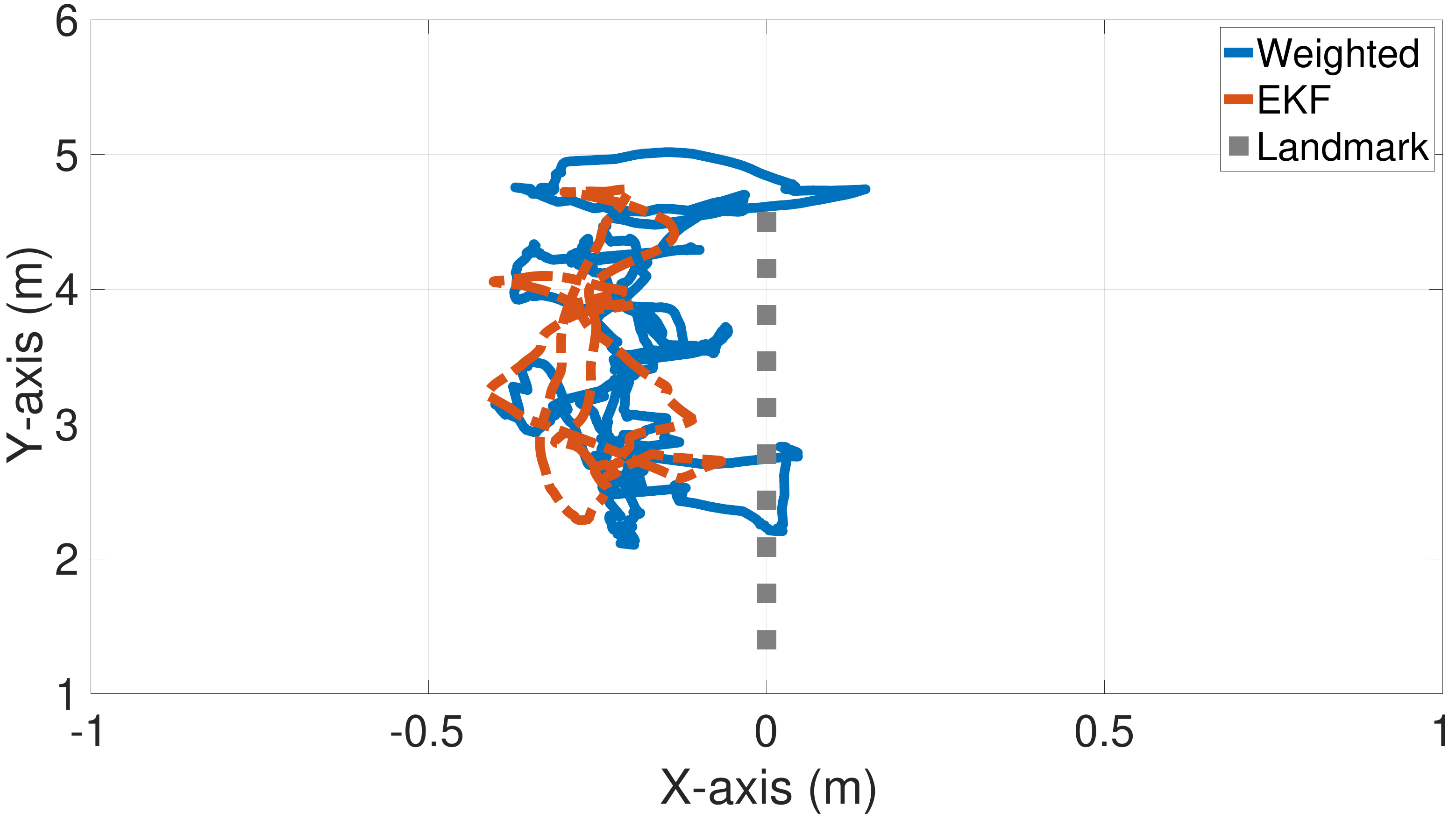}\\
    \includegraphics[width=\textwidth]{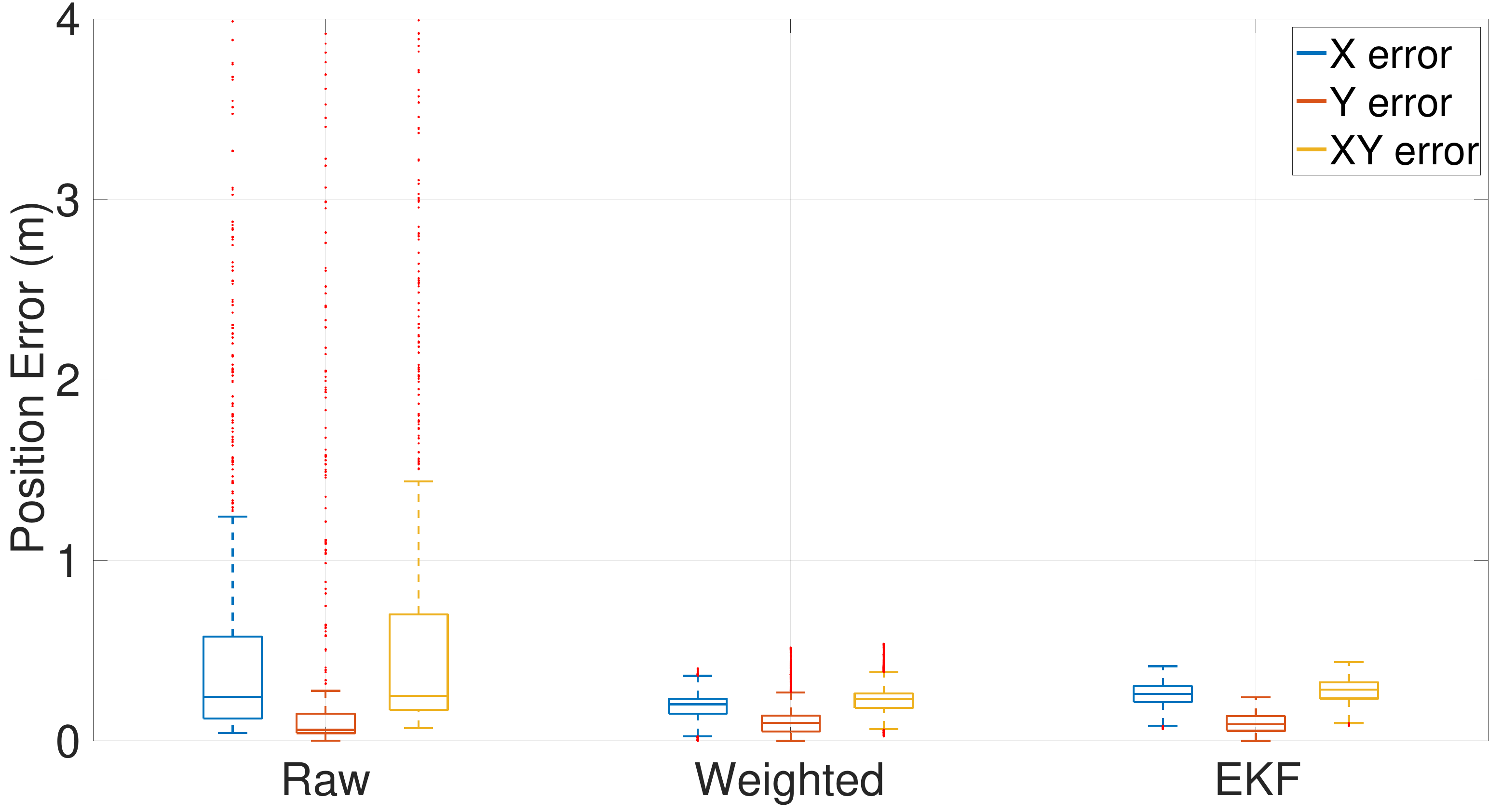}
    \subcaption{Linear trajectory.}
    \label{fig:traj_all:lin}
\end{subfigure}
\hfill
\begin{subfigure}[t]{0.329\linewidth}
    \centering
    \includegraphics[width=\textwidth]{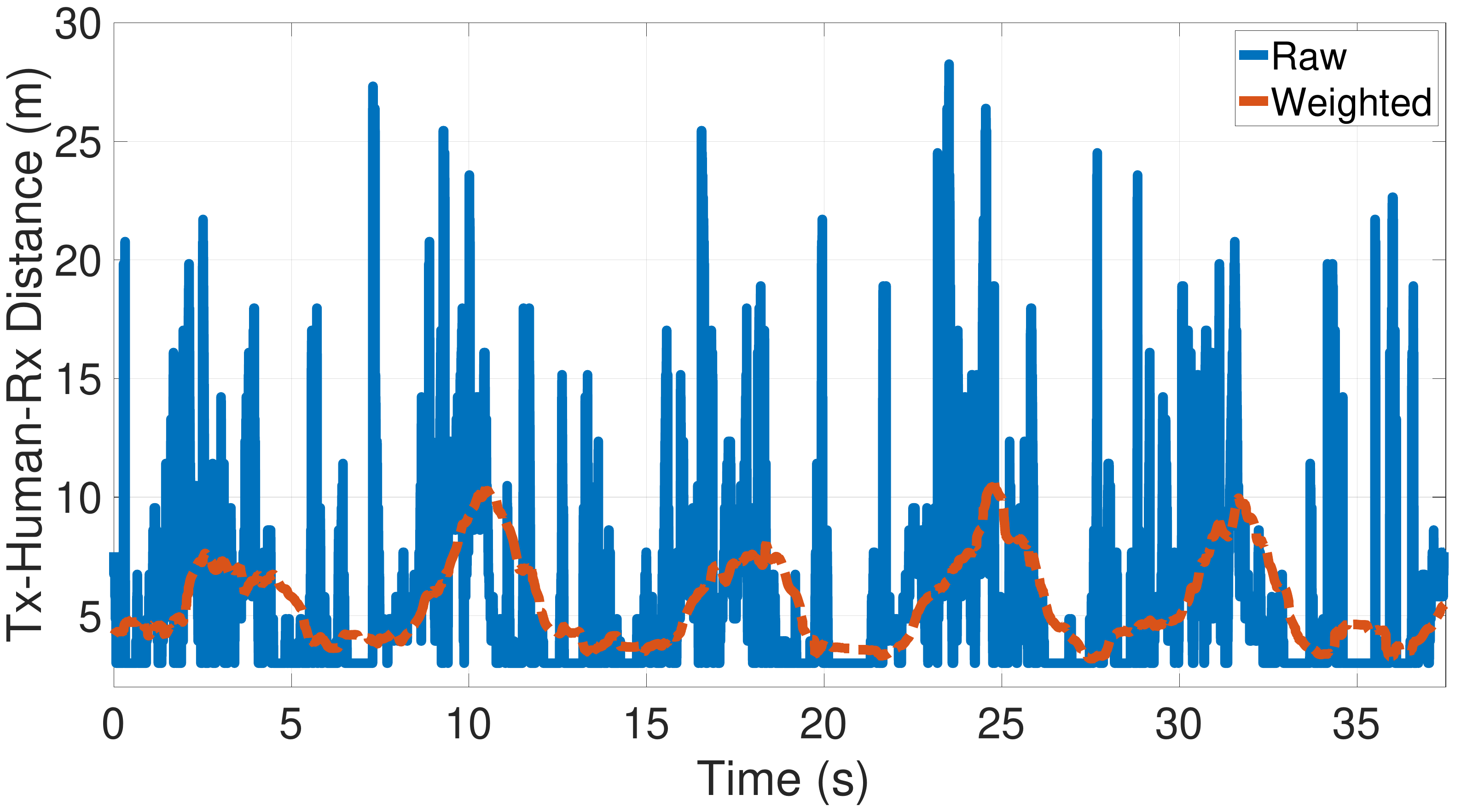}\\
    \includegraphics[width=\textwidth]{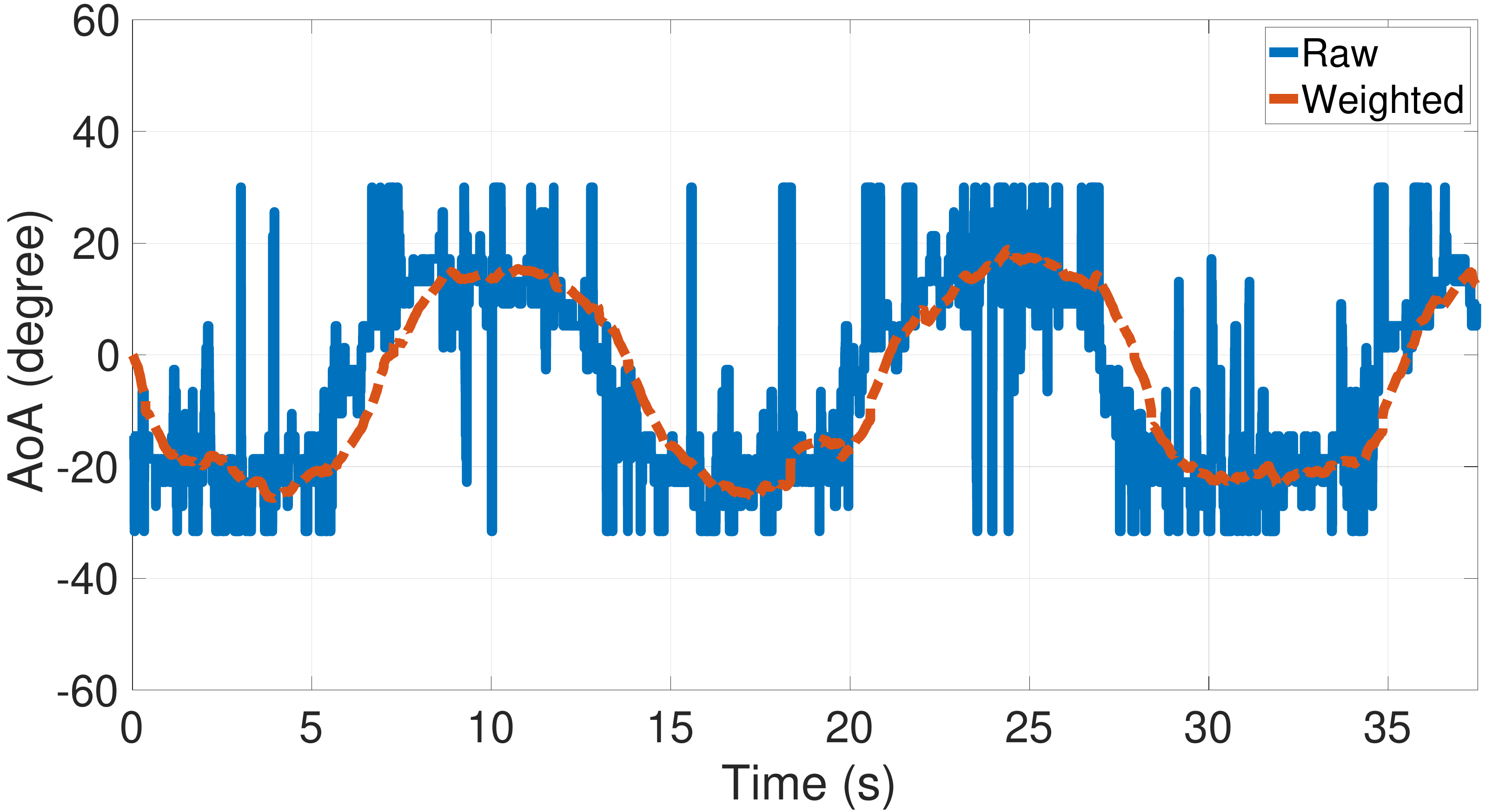}\\
    \includegraphics[width=\textwidth]{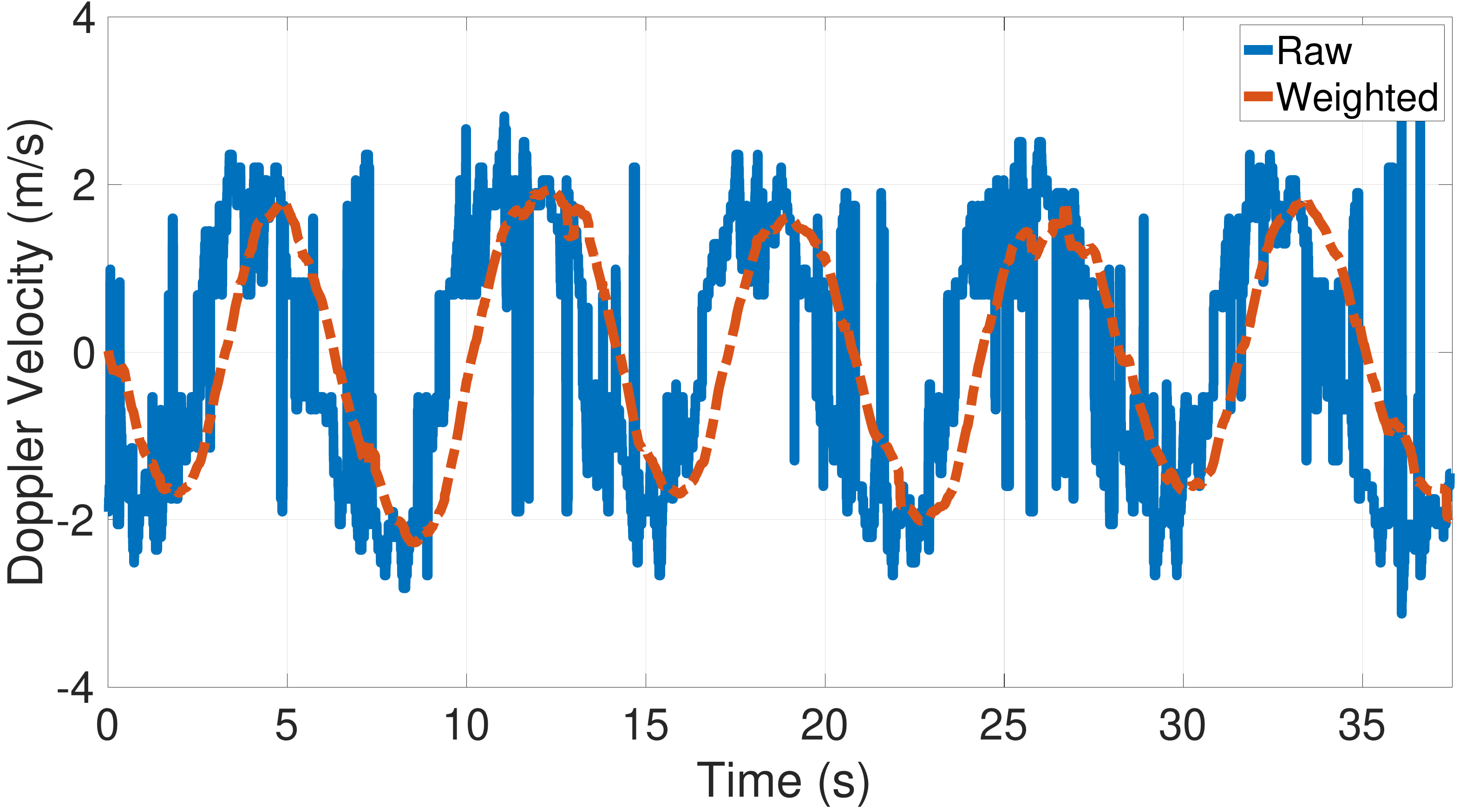}\\
    \includegraphics[width=\textwidth]{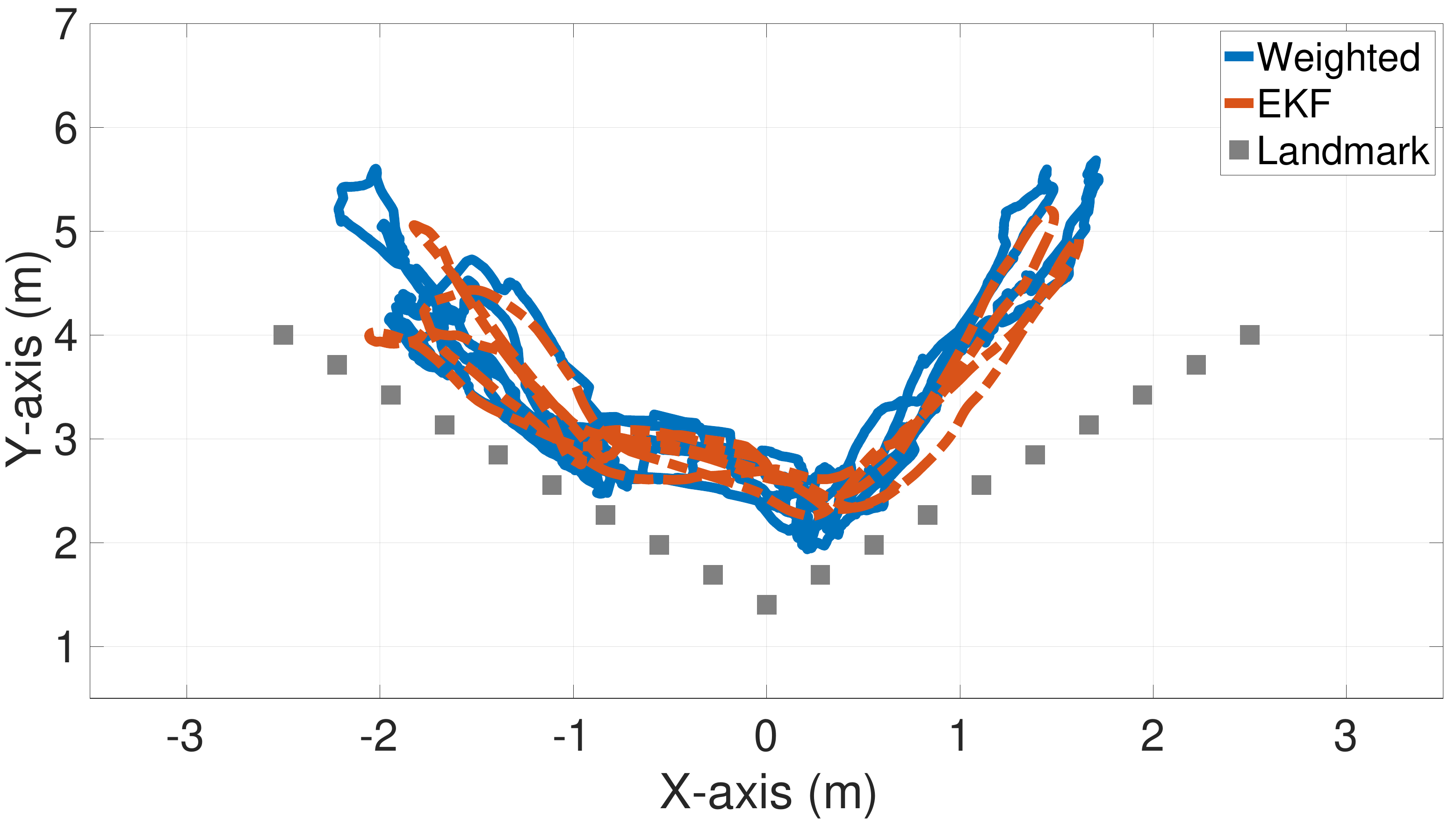}\\
    \includegraphics[width=\textwidth]{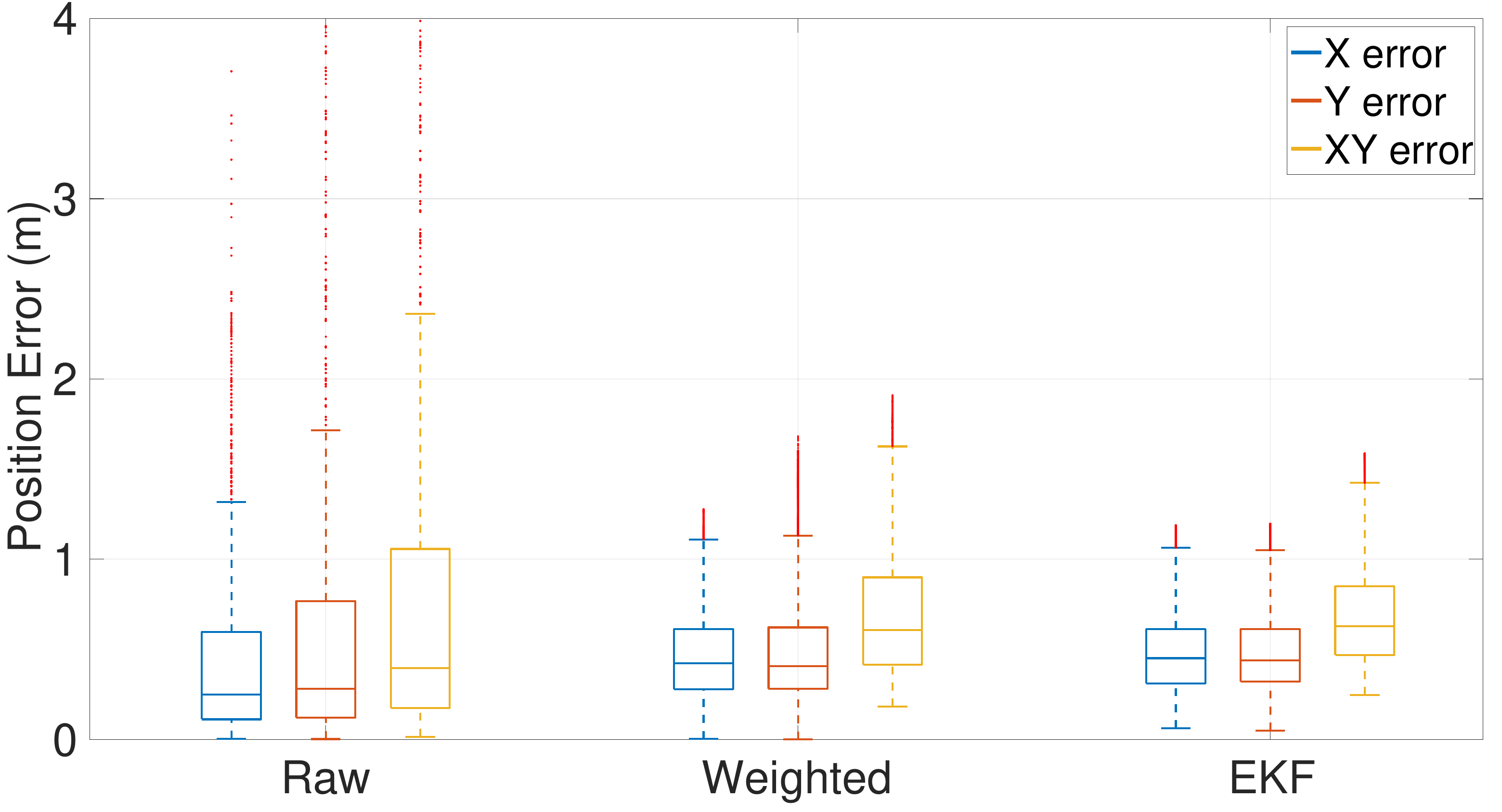}
    \subcaption{V-shaped trajectory.}
    \label{fig:traj_all:v}
\end{subfigure}
\hfill
\begin{subfigure}[t]{0.329\linewidth}
    \centering
    \includegraphics[width=\textwidth]{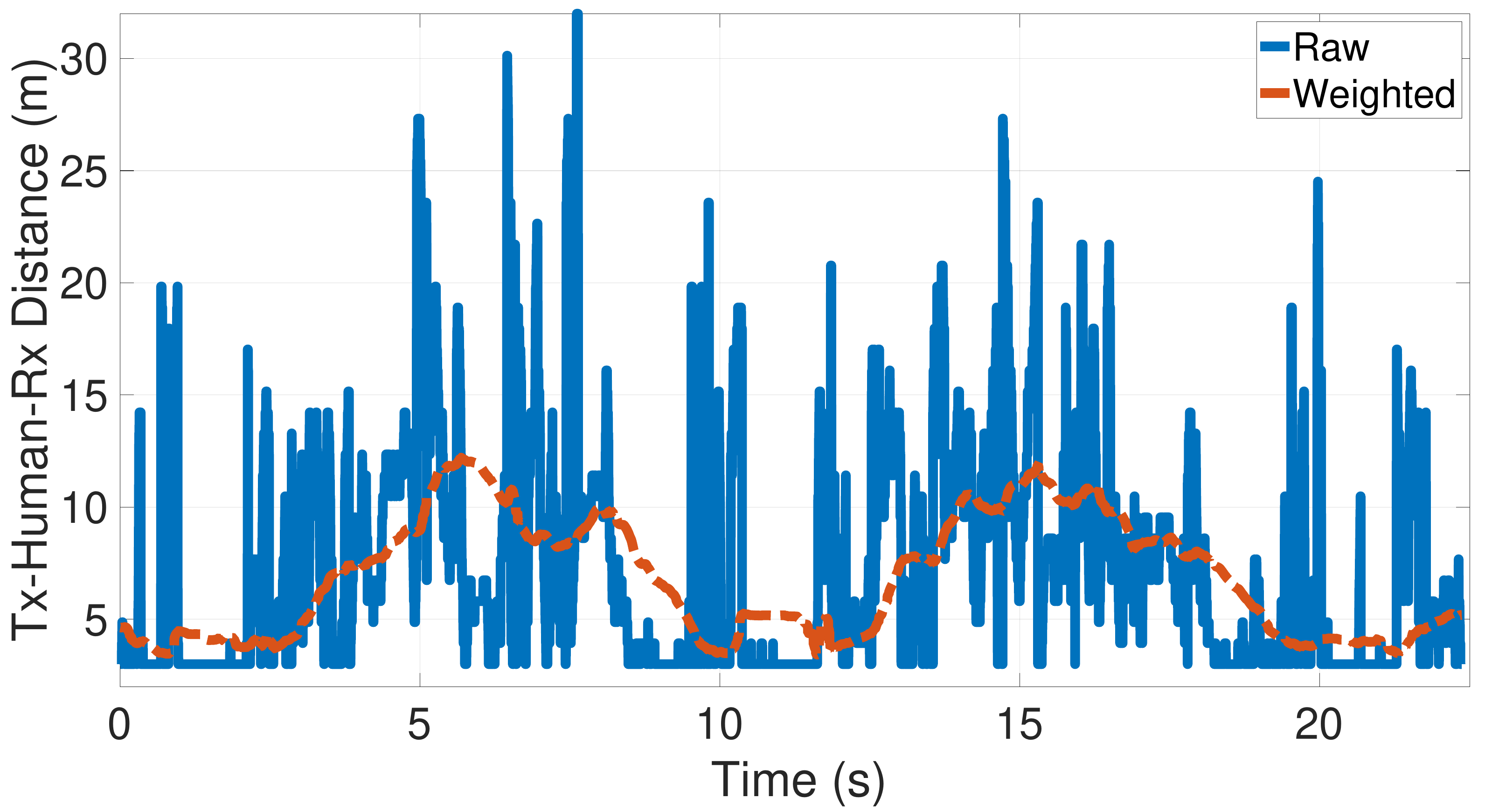}\\
    \includegraphics[width=\textwidth]{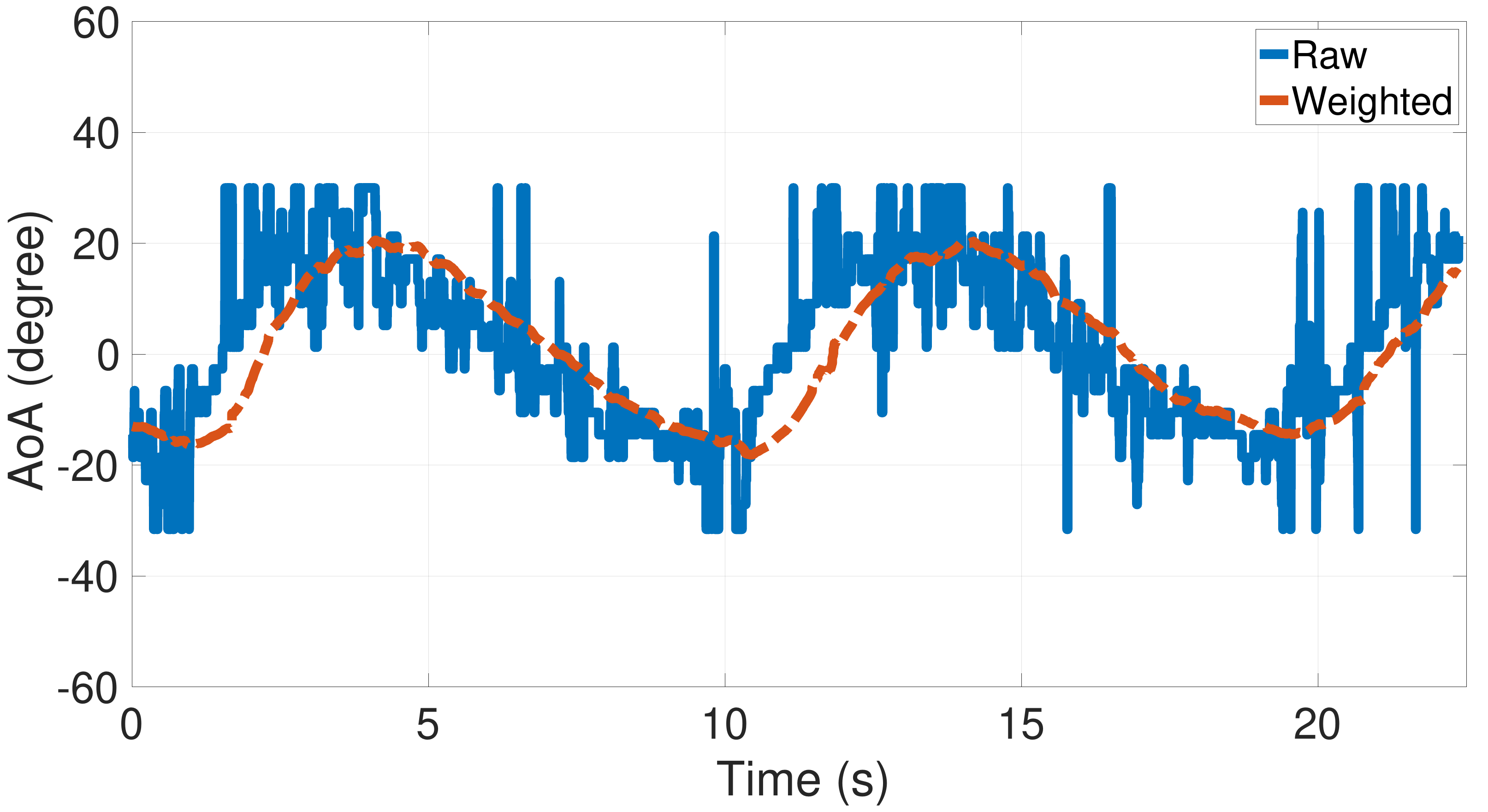}\\
    \includegraphics[width=\textwidth]{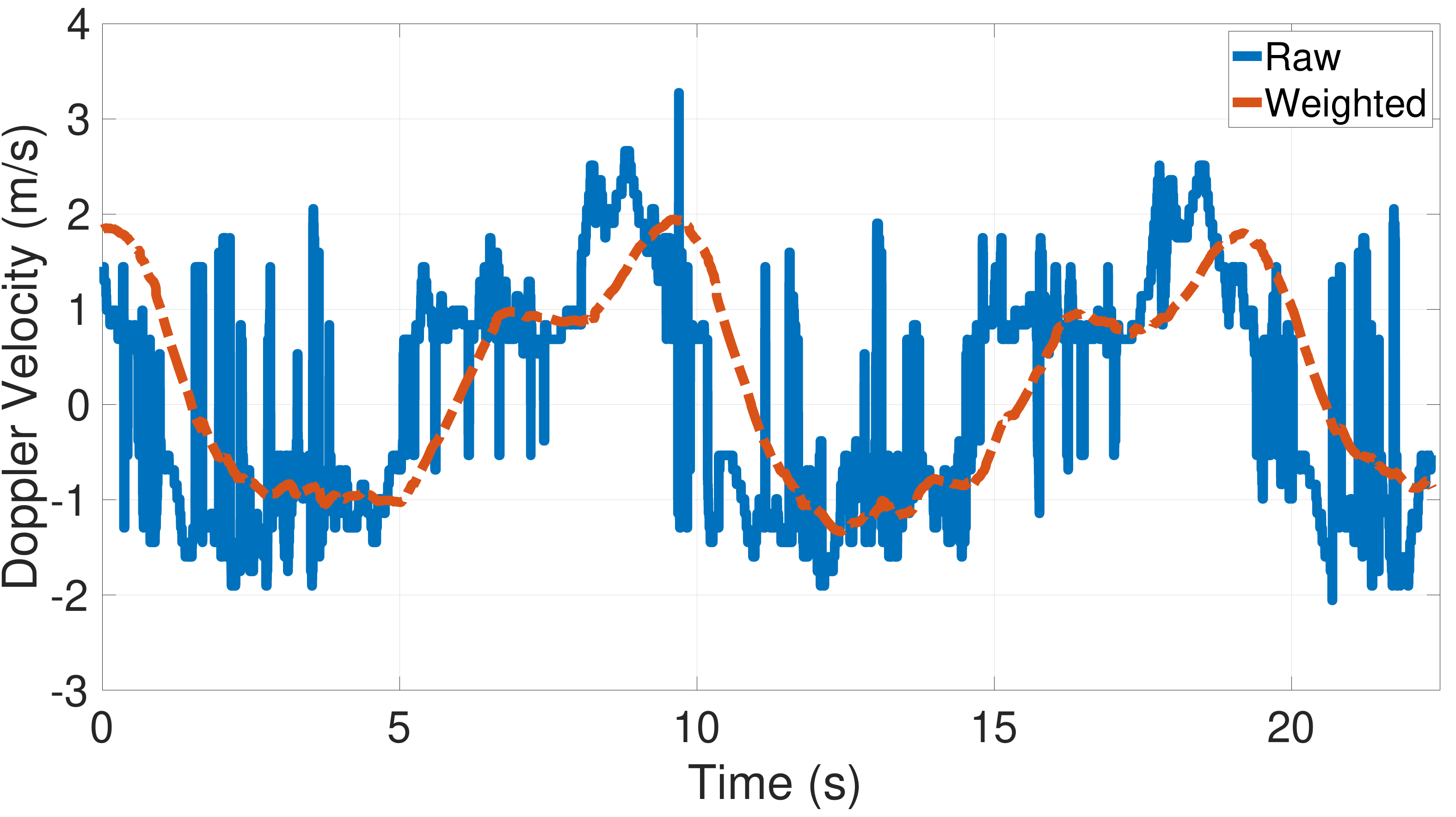}\\
    \includegraphics[width=\textwidth]{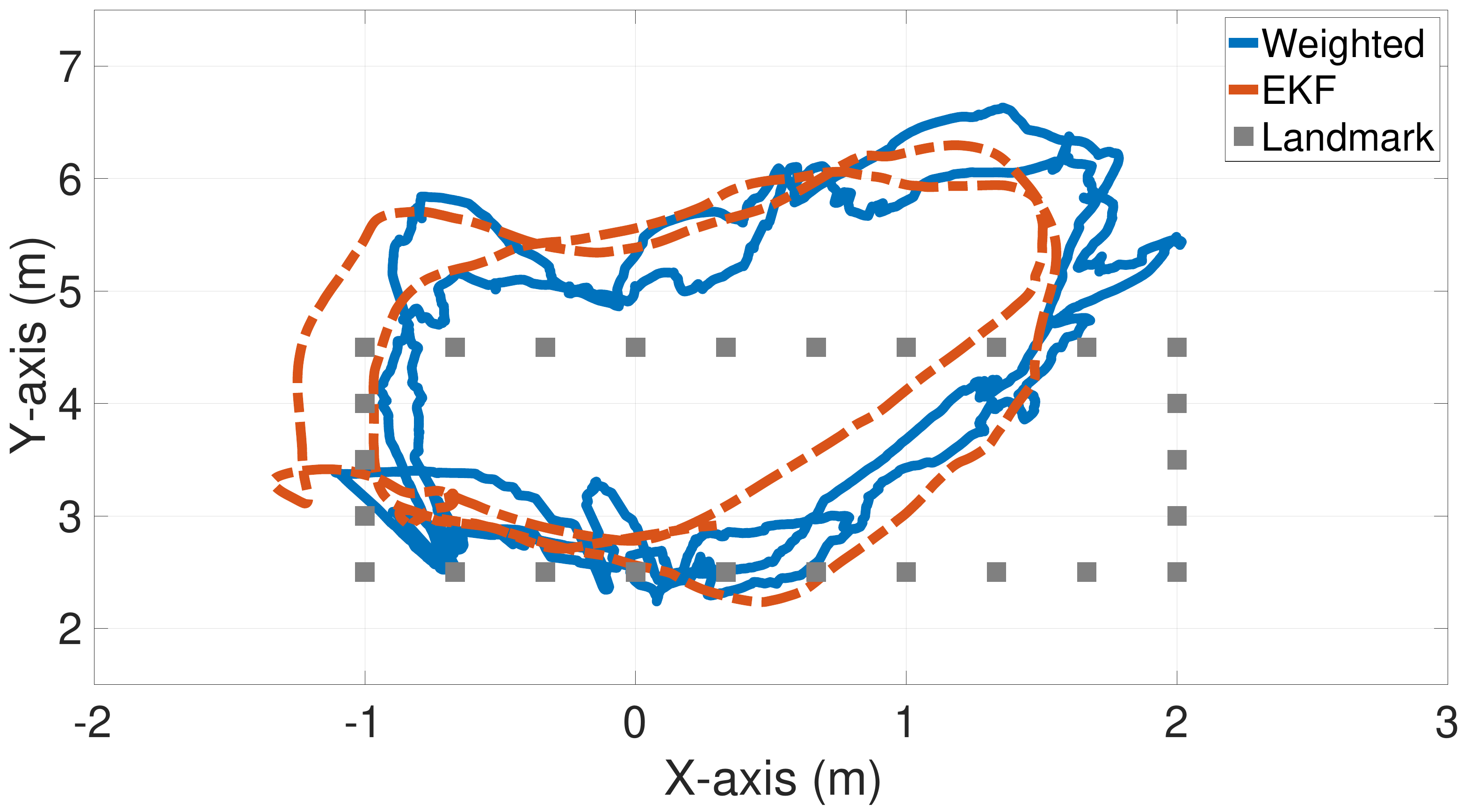}\\
    \includegraphics[width=\textwidth]{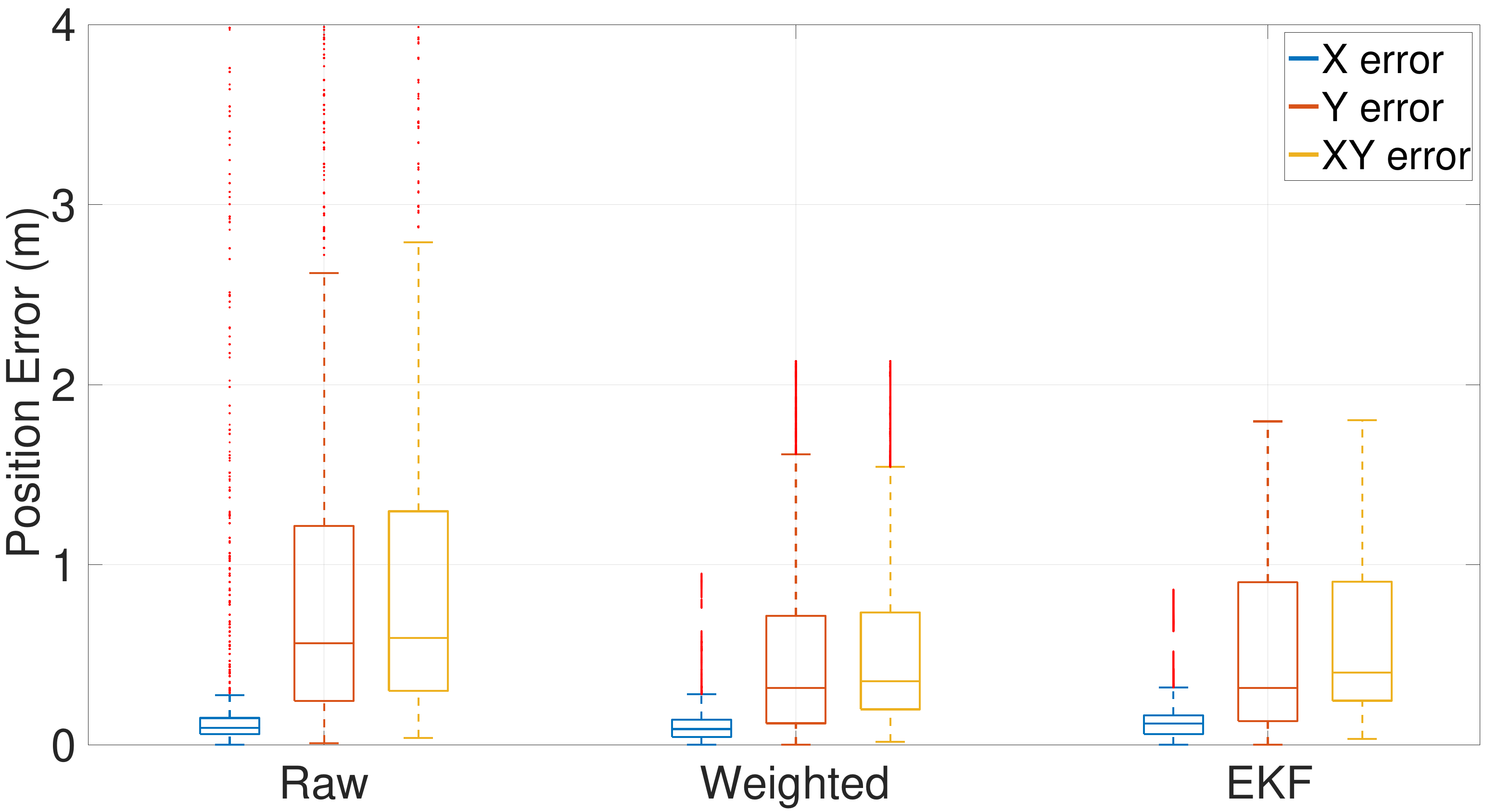}
    \subcaption{Rectangular trajectory.}
    \label{fig:traj_all:rec}
\end{subfigure}
\caption{Tracking results at 3.1~GHz for three walking trajectories. Each column shows (top to bottom) the extracted range (delay), AoA, and Doppler features, followed by the recovered trajectory and the corresponding position error.}
\label{fig:traj_all}
\vspace{-0.5em}
\end{figure*}

\begin{figure*}[t]
    \centering
    \begin{subfigure}[t]{0.329\linewidth}
        \centering
        \includegraphics[width=0.93\textwidth]{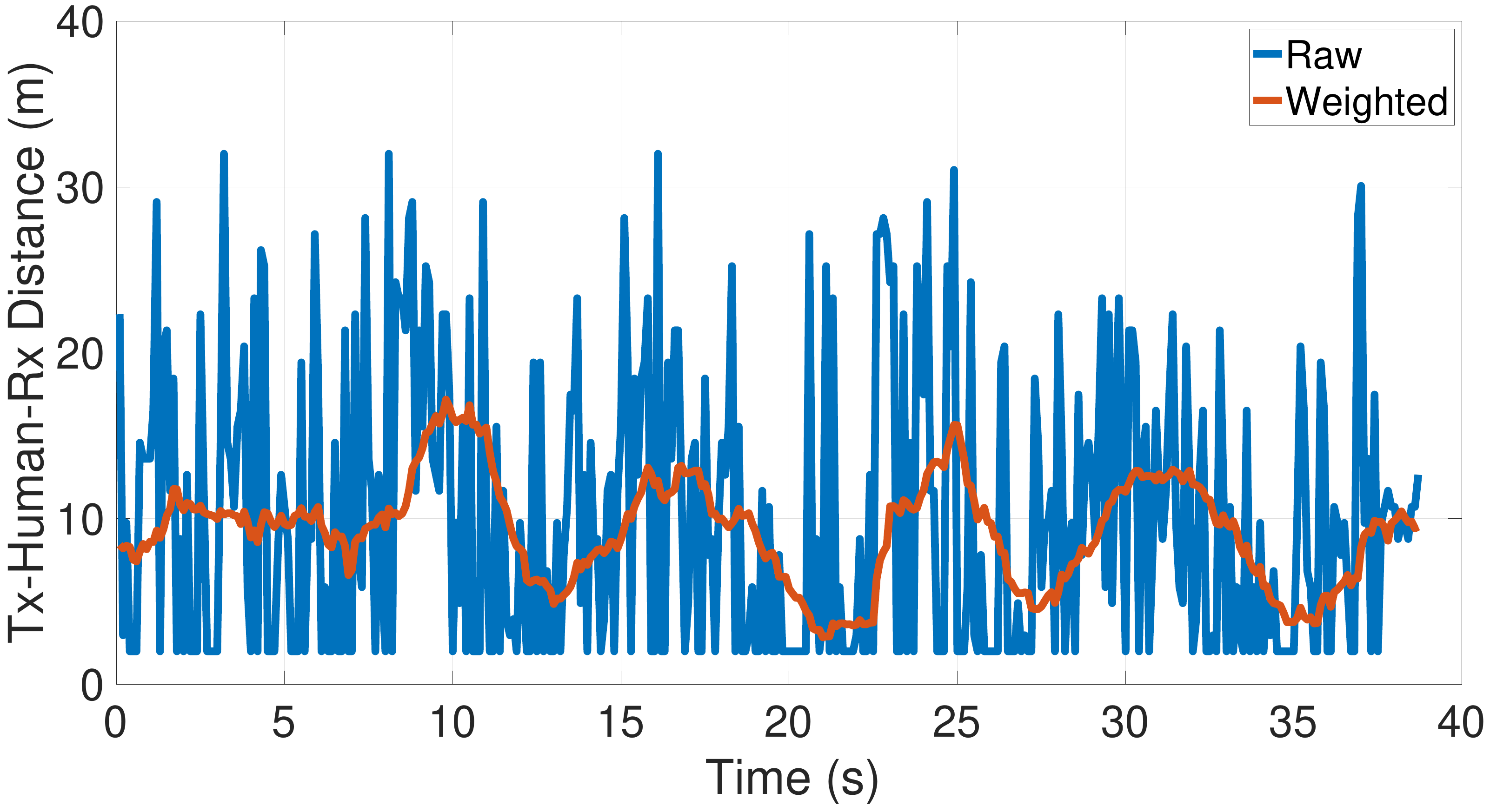}
        \subcaption{Range}
        \label{fig:traj2d:range}
    \end{subfigure}
    \hfill
    \begin{subfigure}[t]{0.329\linewidth}
        \centering
        \includegraphics[width=\textwidth]{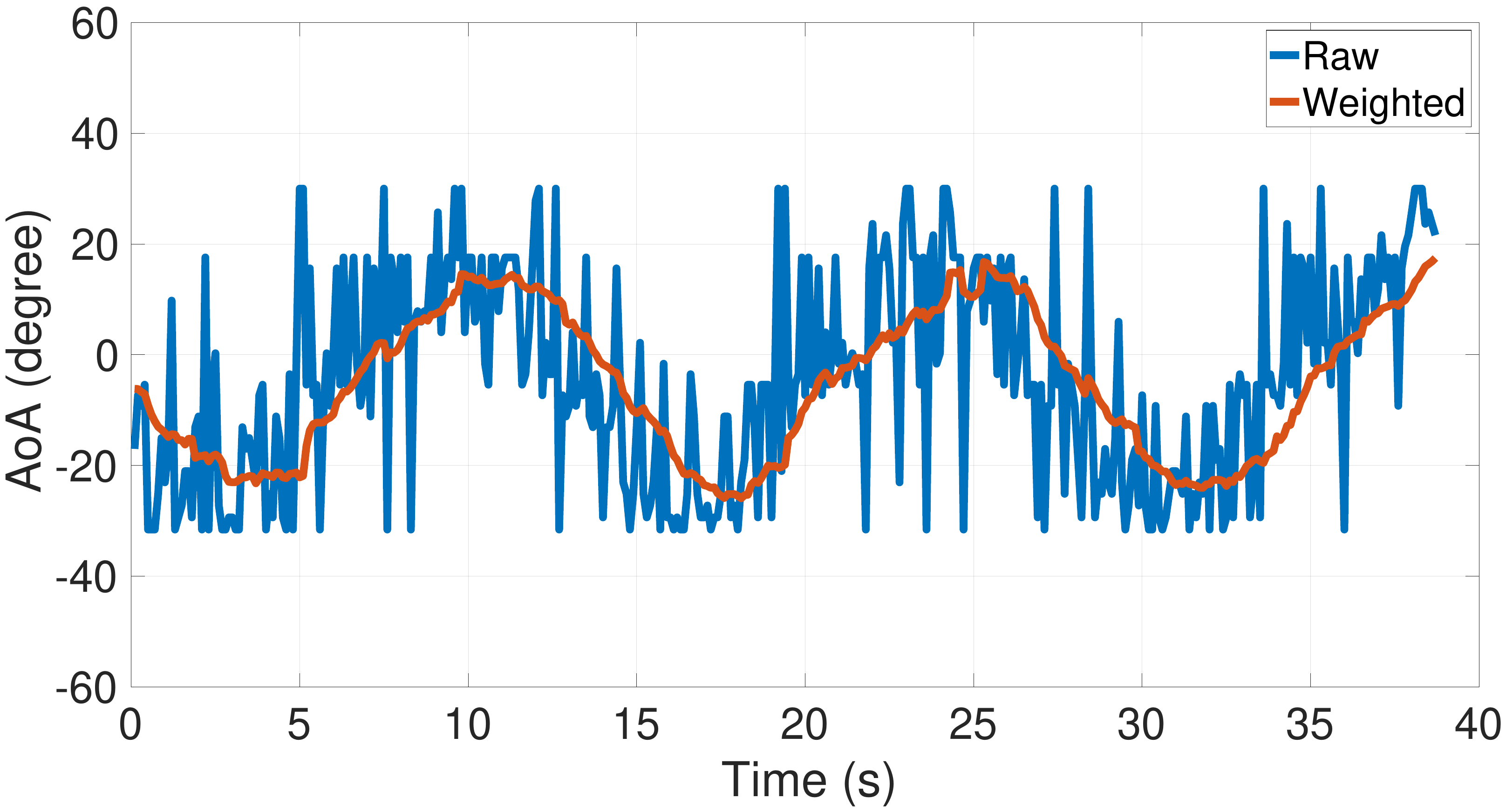}
        \subcaption{AoA}
        \label{fig:traj2d:aoa}
    \end{subfigure}
    \hfill
    \begin{subfigure}[t]{0.329\linewidth}
        \centering
        \includegraphics[width=\textwidth]{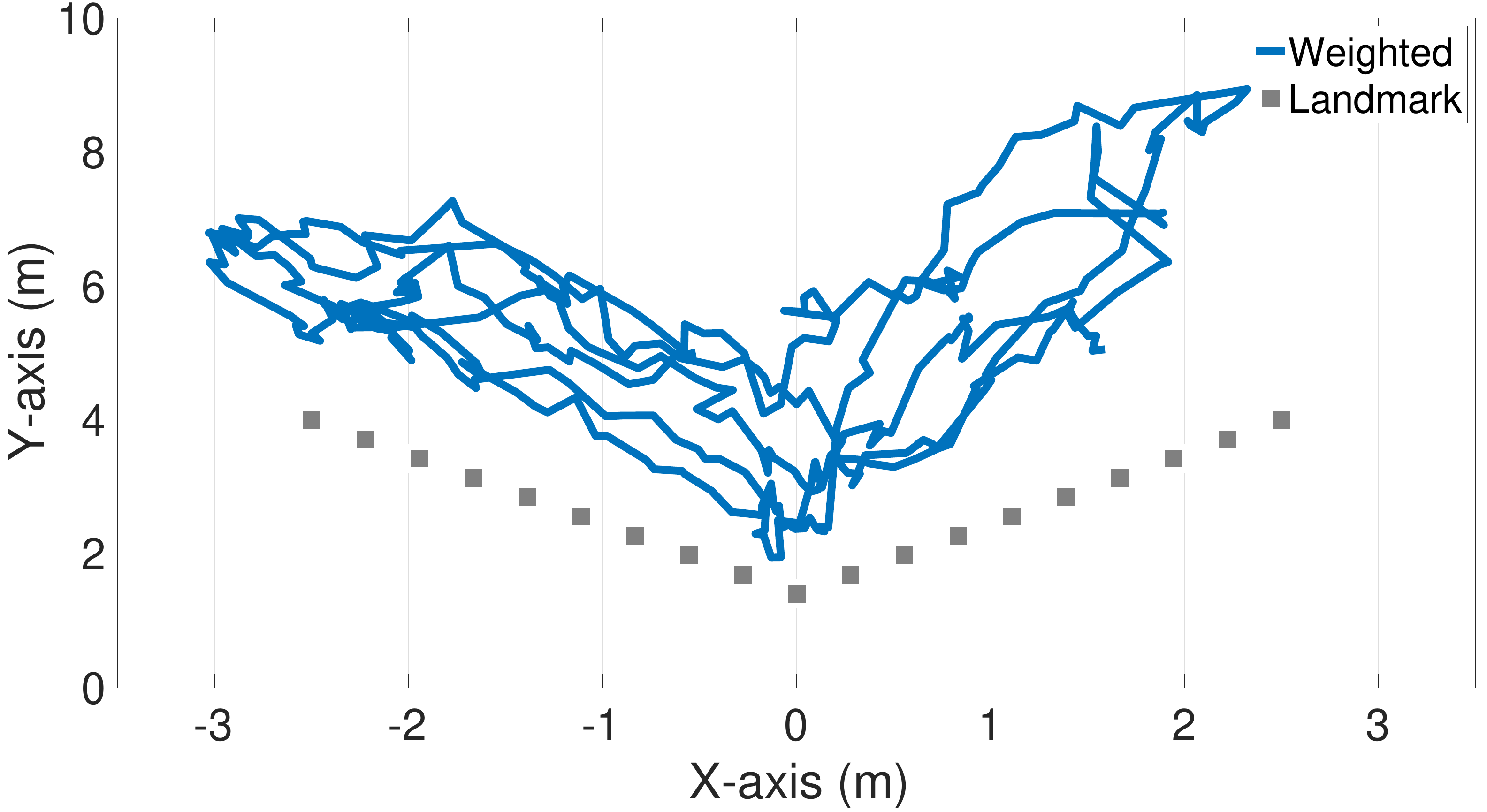}
        \subcaption{Weight-optimized trajectory}
        \label{fig:traj2d:traj}
    \end{subfigure}
    \caption{Tracking result using range and AoA only (no Doppler) under sparse sampling.}
    \label{fig:traj2d}
    \vspace{-1em}
\end{figure*}

\begin{figure*}[t]
    \centering
    \begin{subfigure}[t]{0.329\linewidth}
        \centering
        \includegraphics[width=0.93\textwidth]{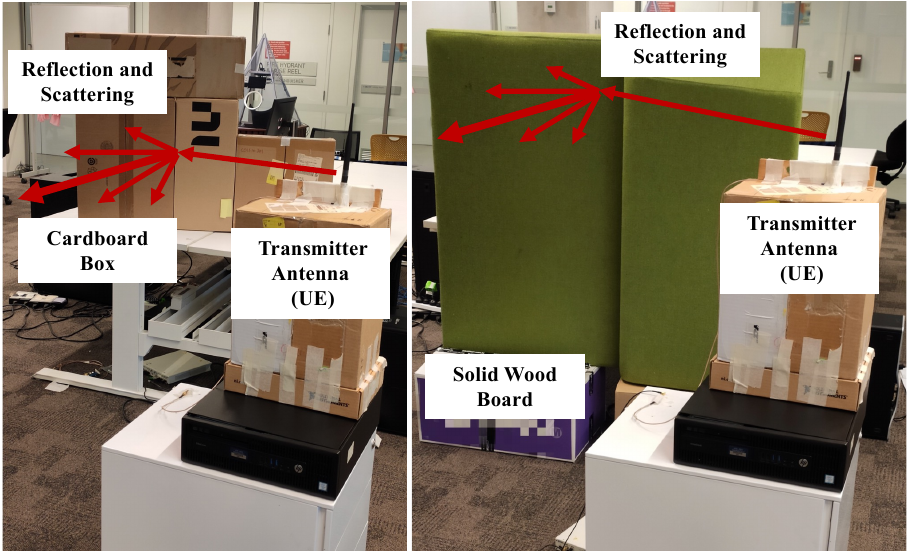}
        \subcaption{NLOS setups.}
        \label{fig:nlos:setup}
    \end{subfigure}
    \hfill
    \begin{subfigure}[t]{0.329\linewidth}
        \centering
        \includegraphics[width=\textwidth]{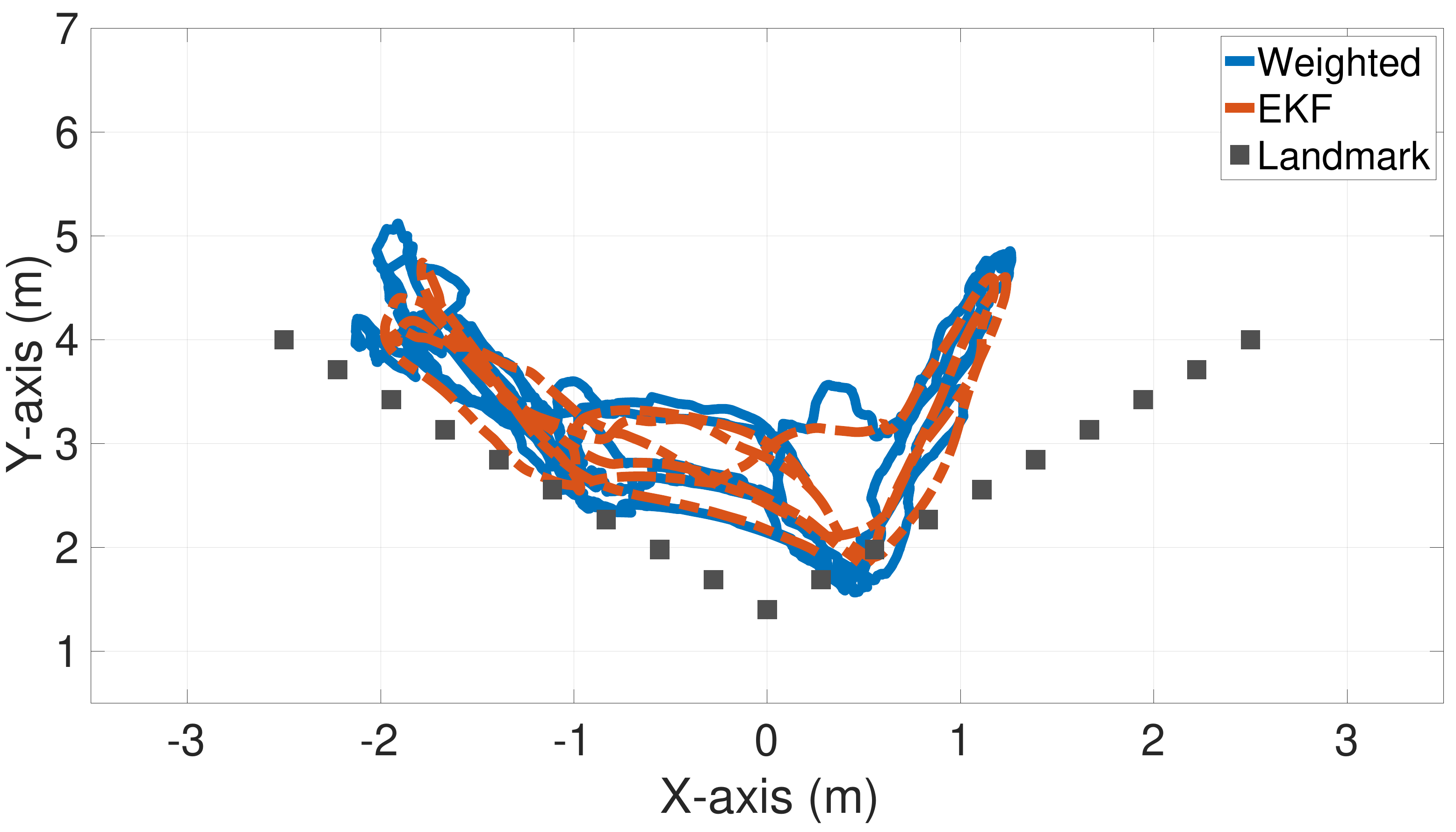}
        \subcaption{Hard cardboard box.}
        \label{fig:nlos:cardboard}
    \end{subfigure}
    \hfill
    \begin{subfigure}[t]{0.329\linewidth}
        \centering
        \includegraphics[width=\textwidth]{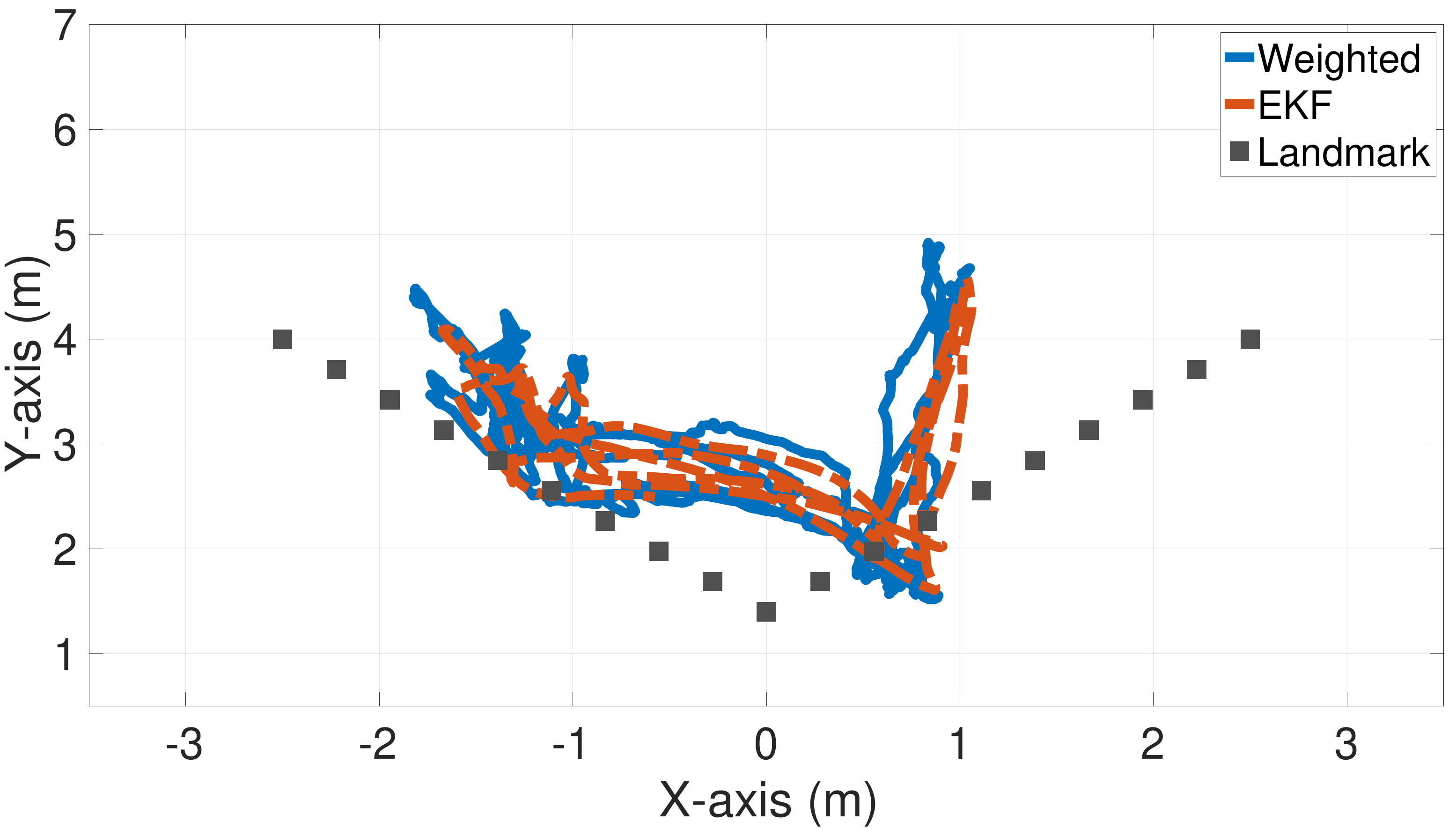}
        \subcaption{Solid wood board.}
        \label{fig:nlos:wood}
    \end{subfigure}
    \caption{Impact of NLOS conditions.}
    \label{fig:nlos}
    \vspace{-1em}
\end{figure*}

\begin{figure*}[t]
    \centering
    \begin{subfigure}[t]{0.329\linewidth}
        \centering
        \includegraphics[width=\textwidth]{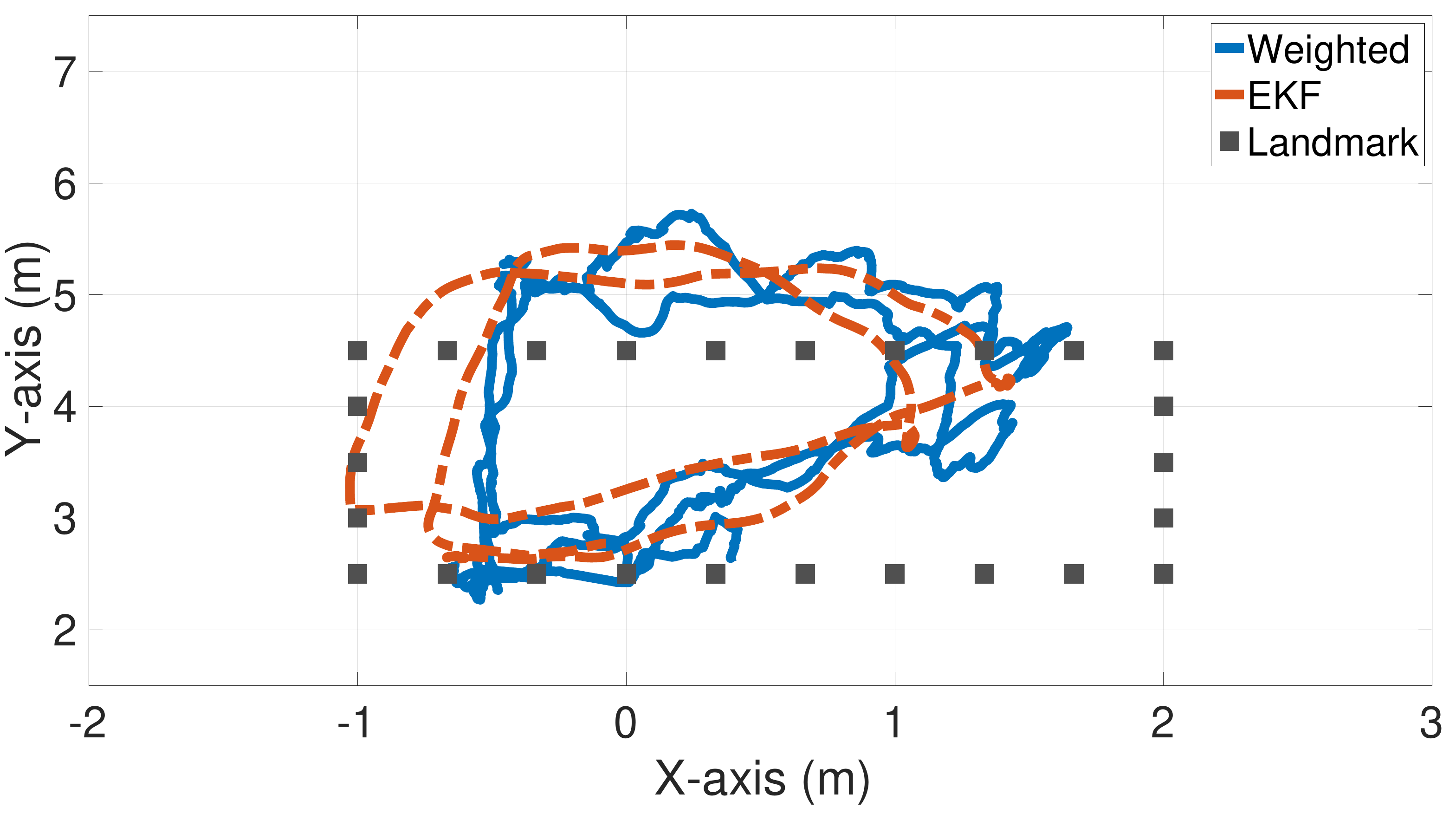}
        \subcaption{High-speed movement (1.2~m/s).}
        \label{fig:speed:fast}
    \end{subfigure}
    \hfill
    \begin{subfigure}[t]{0.329\linewidth}
        \centering
        \includegraphics[width=\textwidth]{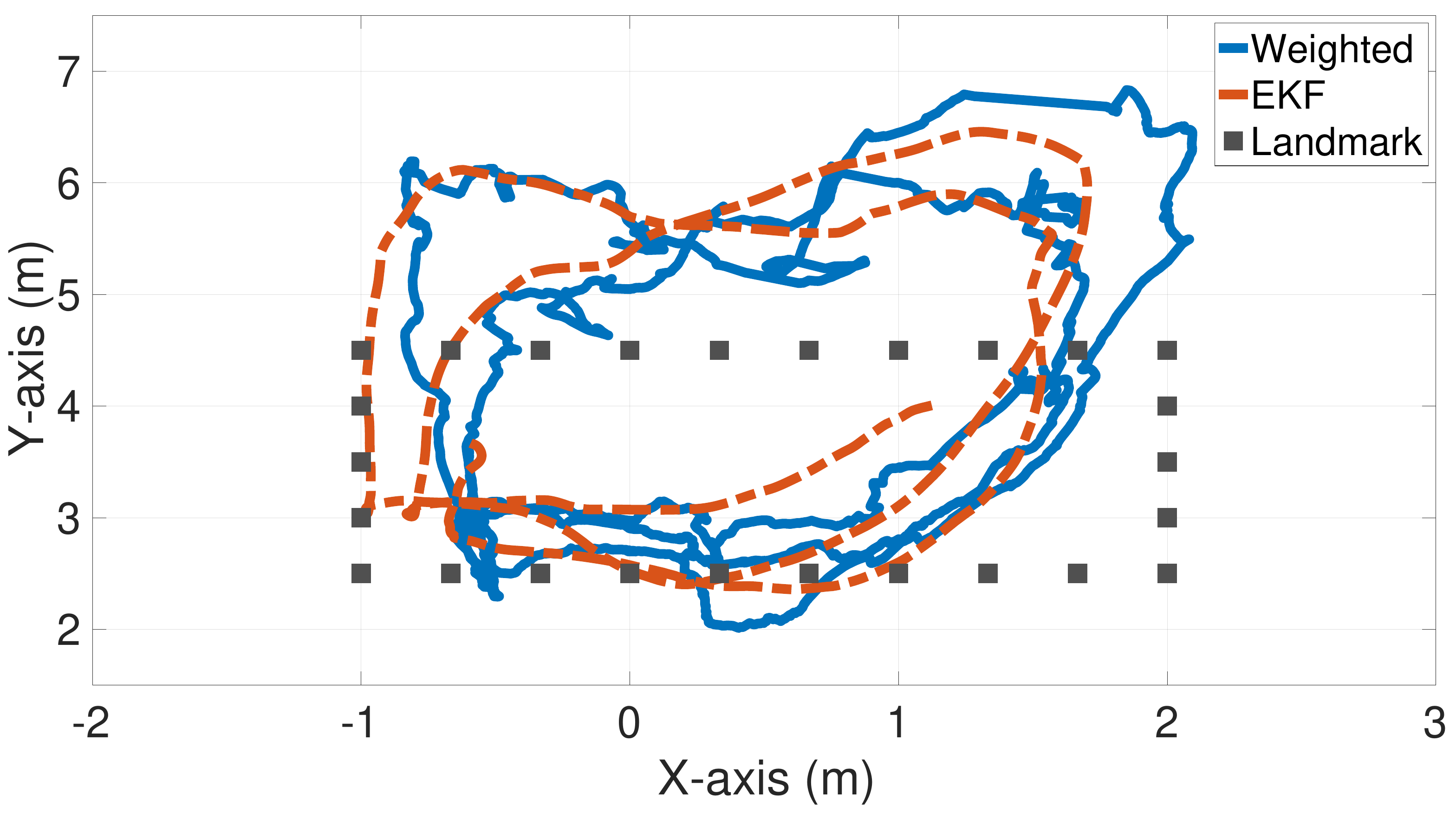}
        \subcaption{Medium-speed movement (1.0~m/s).}
        \label{fig:speed:medium}
    \end{subfigure}
    \hfill
    \begin{subfigure}[t]{0.329\linewidth}
        \centering
        \includegraphics[width=\textwidth]{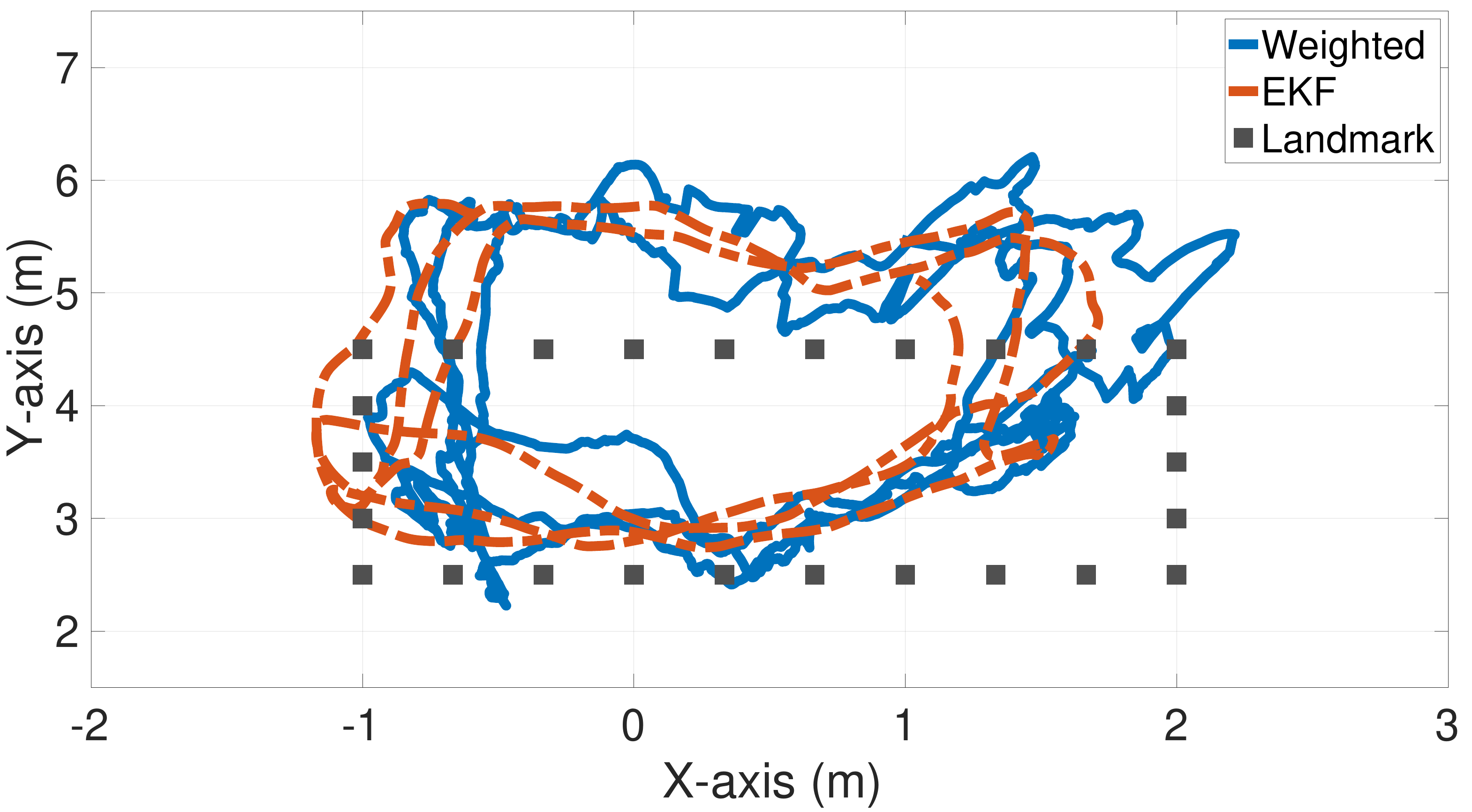}
        \subcaption{Low-speed movement (0.7~m/s).}
        \label{fig:speed:low}
    \end{subfigure}
    \caption{Impact of target speed.}
    \label{fig:speed}
    \vspace{-0.5em}
\end{figure*}

\begin{figure}[t]
\centering
\begin{minipage}{0.49\linewidth}
    \centering
    \includegraphics[width=\linewidth]{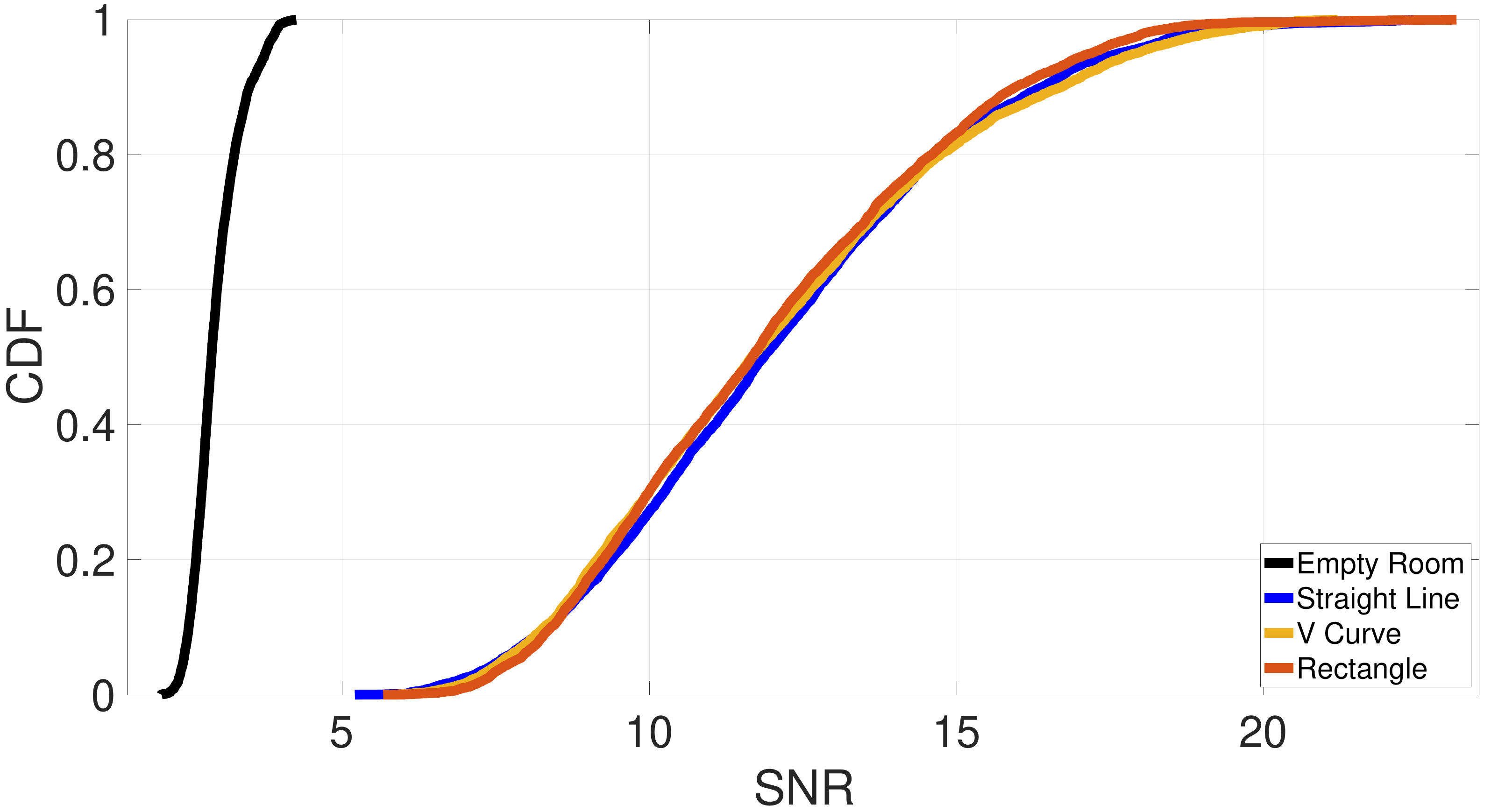}
    \caption{Detection threshold.}
    \label{fig:det_threshold}
\end{minipage}
\hfill
\begin{minipage}{0.49\linewidth}
    \centering
    \includegraphics[width=\linewidth]{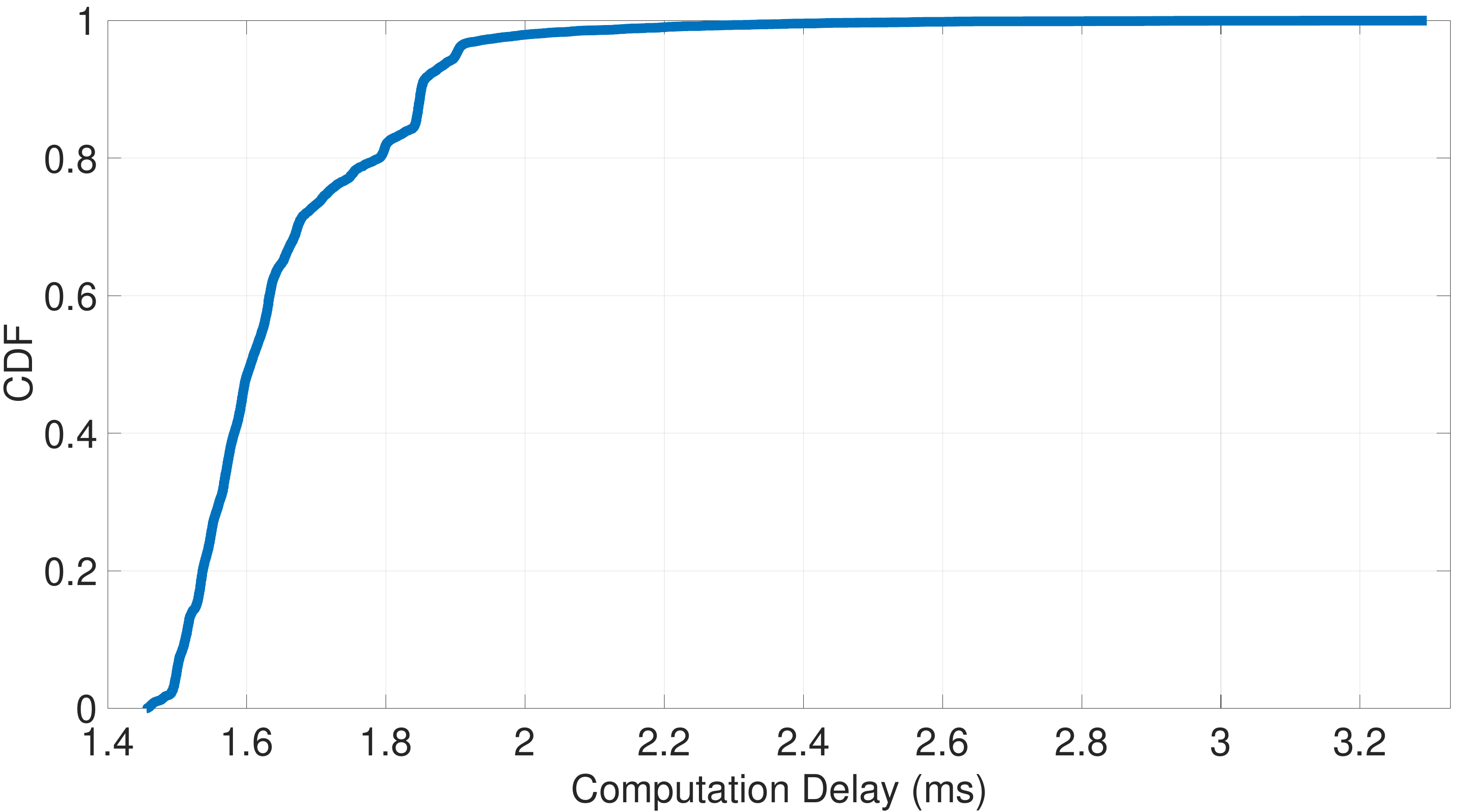}
    \caption{Computation latency.}
    \label{fig:runtime}
\end{minipage}
\vspace{-1em}
\end{figure}

\begin{figure*}[t]
    \centering
    \begin{subfigure}[t]{0.329\linewidth}
        \centering
        \includegraphics[width=\textwidth]{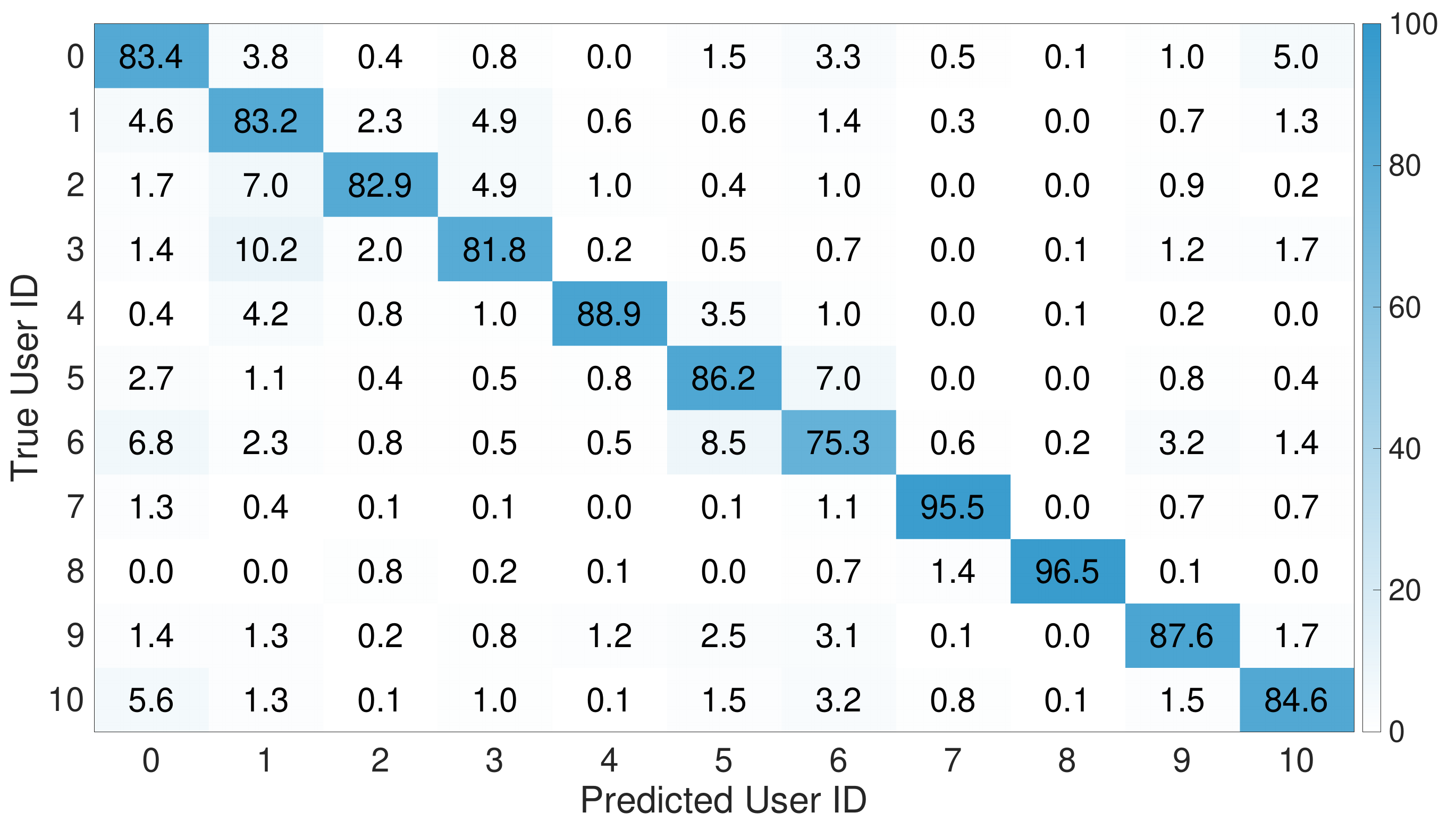}
        \subcaption{Confusion matrix (CASR)}
        \label{fig:gait_confusion:casr}
    \end{subfigure}
    \hfill
    \begin{subfigure}[t]{0.329\linewidth}
        \centering
        \includegraphics[width=\textwidth]{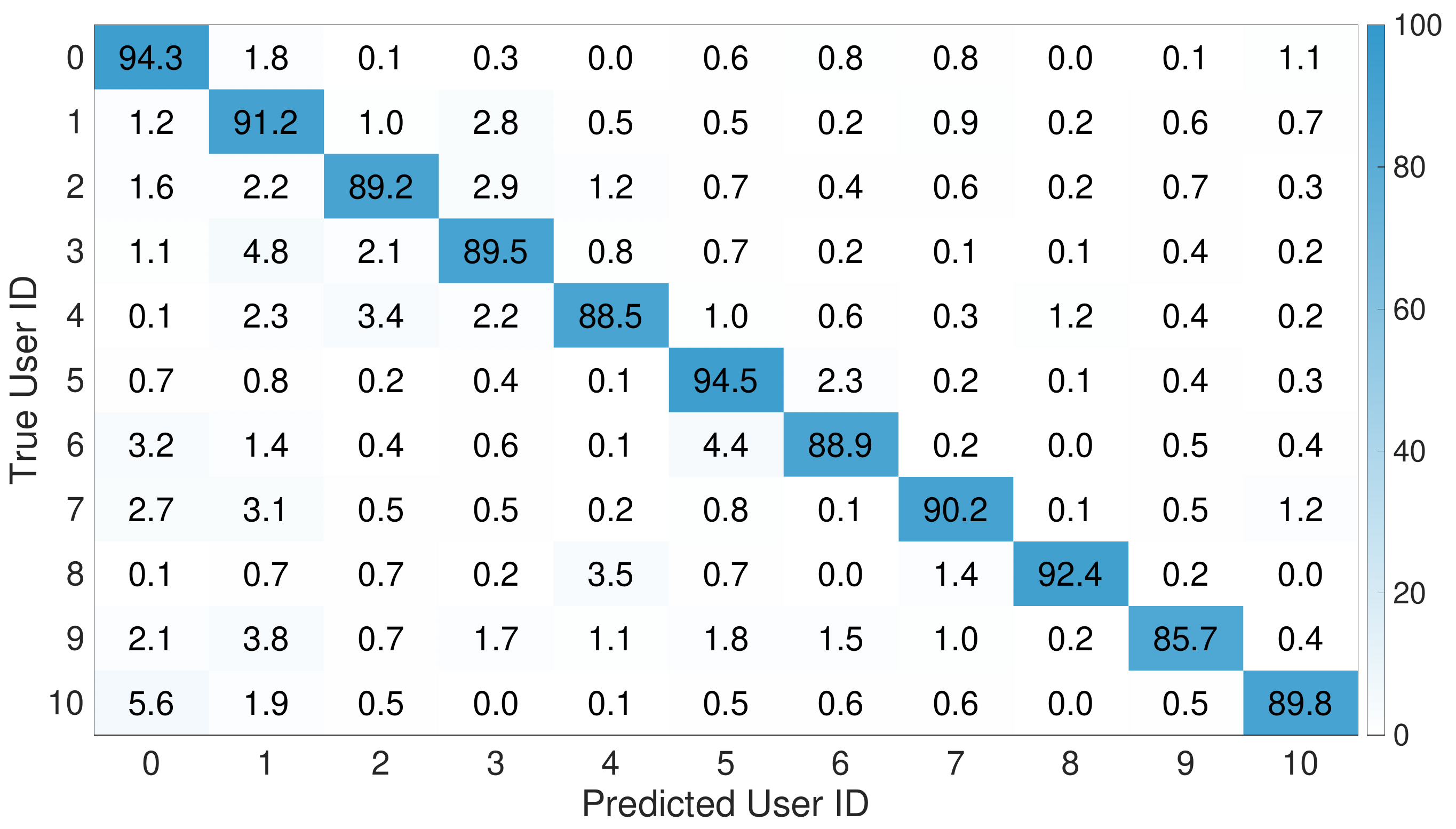}
        \subcaption{Confusion matrix (position-refined)}
        \label{fig:gait_confusion:posref}
    \end{subfigure}
    \hfill
    \begin{subfigure}[t]{0.329\linewidth}
        \centering
        \includegraphics[width=\textwidth]{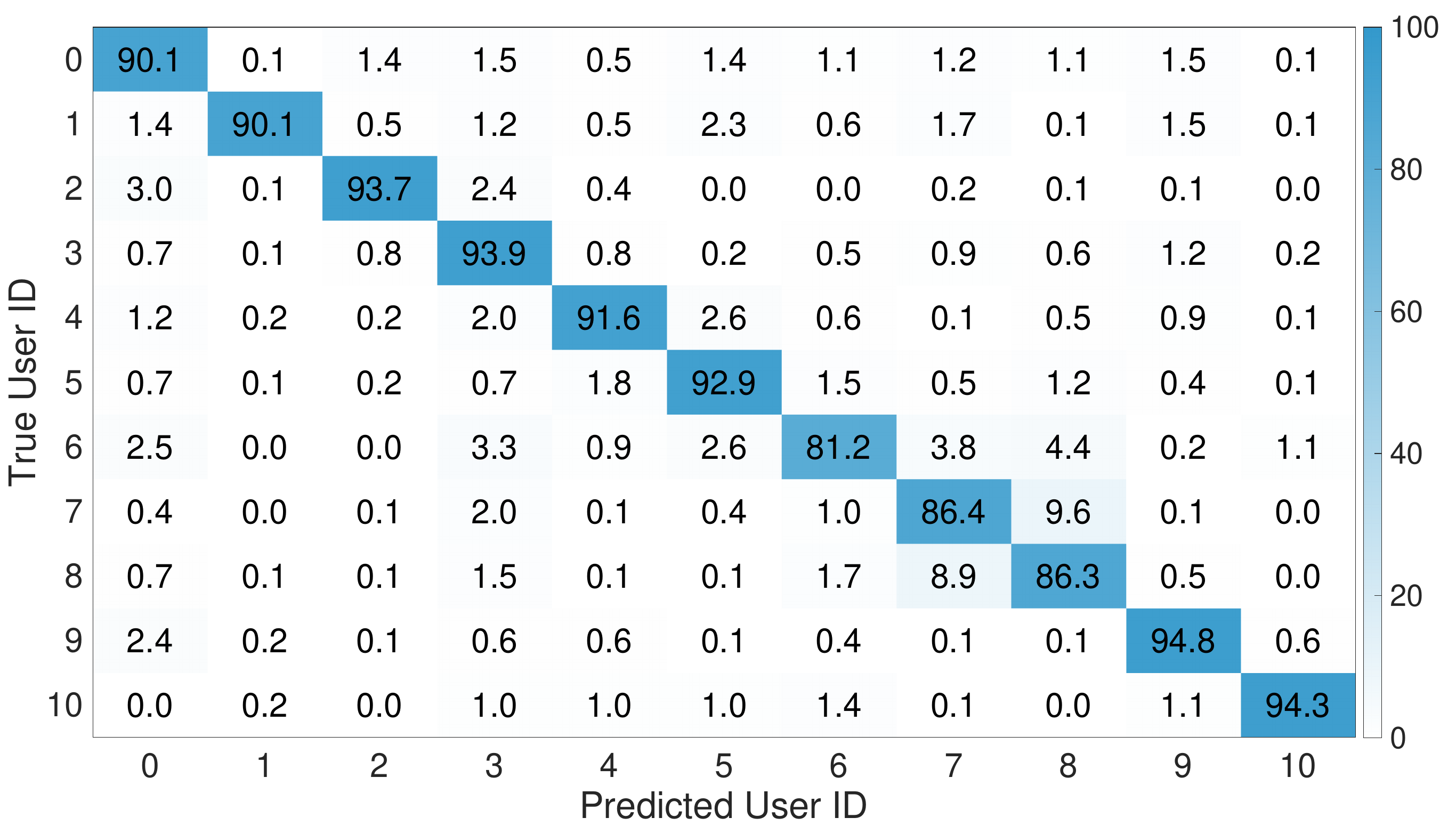}
        \subcaption{Confusion matrix (range-refined)}
        \label{fig:gait_confusion:rangeref}
    \end{subfigure}
    \caption{Gait identification confusion matrices for three micro-Doppler features: CASR baseline, position-refined, and range-refined.}
    \label{fig:gait_confusion}
    \vspace{-1.5em}
\end{figure*}

\begin{figure}[t]
\centering
    \includegraphics[width=0.75\linewidth]{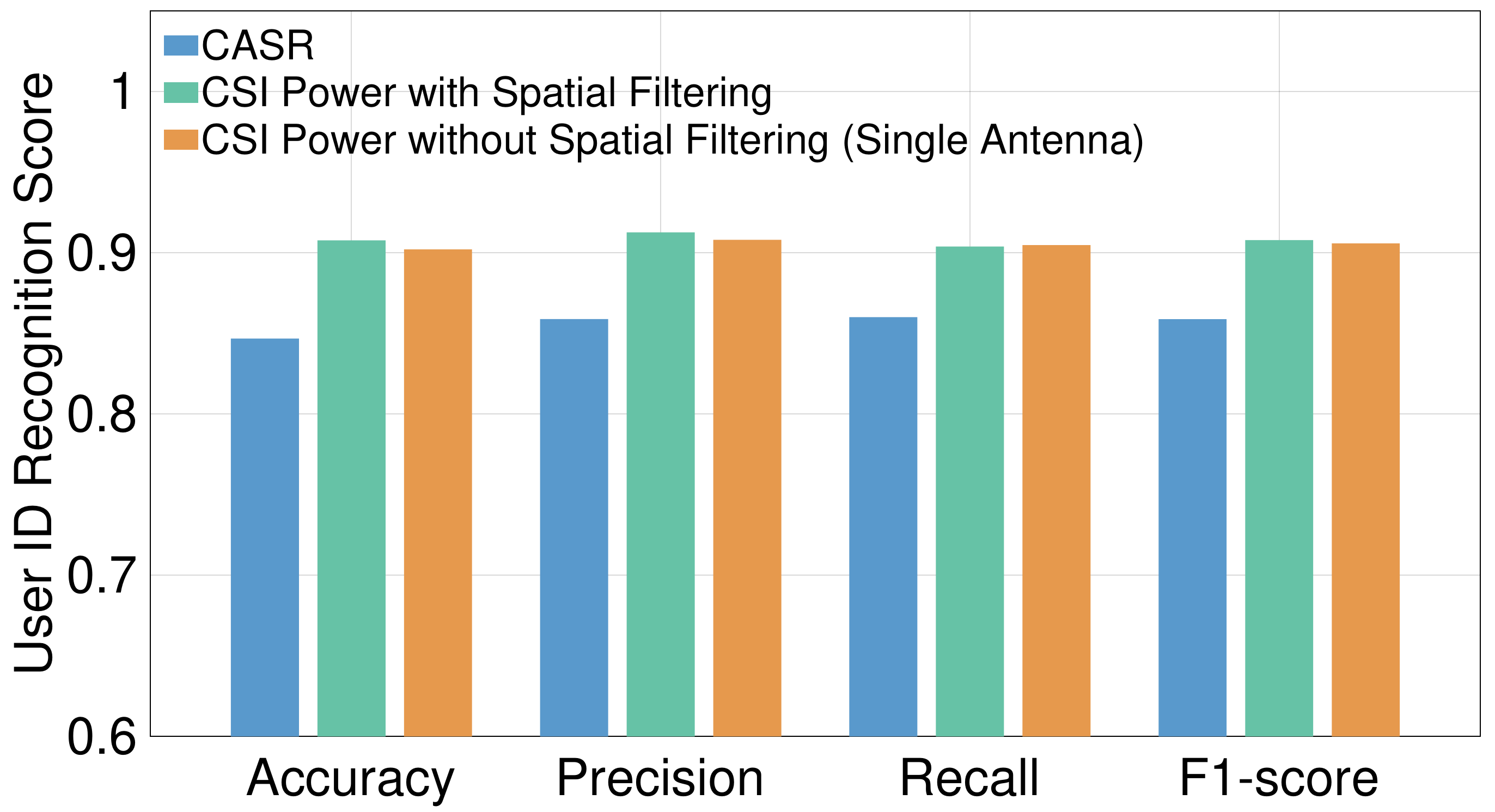}
    \caption{Gait identification performance.}
    \label{fig:userid_f1}
\vspace{-1.5em}
\end{figure}

\subsection{Real-World Results}
\label{subsec:real_world}
\subsubsection{Overall tracking performance}
We evaluate \textit{WiDFS2.5} on three indoor walking trajectories. Fig.~\ref{fig:traj_all:lin}, Fig.~\ref{fig:traj_all:v}, and Fig. \ref{fig:traj_all:rec} show the extracted range (delay), AoA, and Doppler features, together with the weight-optimized trajectories and the EKF-refined tracks. At first, with the limited bandwidth (20~MHz), range (delay) estimates are noisier than AoA and Doppler. The weight-optimization stage aggregates measurements over a 1.5~s window to suppress outliers, at the cost of a fixed latency. Using these fused features, the EKF enforces temporal consistency and yields smoother tracks than the weight-optimized trajectories alone. Second, the CSI-power 3D DFT processing also mitigates mirrored artifacts, most evidently in Doppler, reducing the sign ambiguity of real-valued spectra and improving peak stability. Third, the average EKF position errors are 0.28~m, 0.68~m, and 0.60~m for the linear, V-shaped, and rectangular trajectories, respectively (0.52~m overall). Errors along $x$ are typically smaller than along $y$, consistent with the Tx-Rx geometry providing stronger constraints along the baseline direction.

\subsubsection{Tracking performance under sparse sampling}
\label{subsubsec:sparse_sampling}
When Doppler processing is unavailable (e.g., due to sparse slow-time sampling), we fall back to a 2D range-AoA mode and estimate the trajectory using delay and AoA only. In this experiment, we downsample the V-shaped trajectory to 10~Hz (10 CSI samples per second), where the Doppler dimension is no longer reliable. As shown in Fig.~\ref{fig:traj2d}, compared with the full delay-AoA-Doppler pipeline, the per-CPI range and AoA estimates become noticeably noisier and exhibit more frequent abnormal jumps. After applying the proposed temporal smoothing and outlier suppression (weighted fusion), both features are substantially stabilized. Although the 2D result is less accurate and more sensitive to range noise than the full 3D pipeline, it still captures the coarse motion trend reliably.

\subsubsection{Impact of NLOS scenarios}
To assess robustness under NLOS conditions, we introduce controlled obstructions near the transmitter using a hard cardboard box and a solid wood board, as illustrated in Fig.~\ref{fig:nlos:setup}. The resulting trajectories in Fig.~\ref{fig:nlos:cardboard} and Fig.~\ref{fig:nlos:wood} show larger deviations than the LOS case. The wood board causes noticeably stronger degradation than the cardboard box, which is consistent with its stronger reflection and the more pronounced multipath. These results indicate that severe NLOS multipath can bias delay/AoA estimation and consequently degrade tracking accuracy.

\subsubsection{Impact of movement velocity}
Fig.~\ref{fig:speed} compares tracking at three walking speeds: 1.2~m/s, 1.0~m/s, and 0.7~m/s. Faster motion does not necessarily improve accuracy, consistent with~\cite{wang2023single}: although larger Doppler shifts can enhance separability, rapid non-rigid motion may introduce unstable scattering and fluctuations that offset this benefit.

\subsubsection{Motion detection threshold}
\label{subsubsec:motion_detection}
We evaluate the motion detector using the CDF of the detection statistic in Fig.~\ref{fig:det_threshold}. In static scenes (e.g., an empty room), the statistic stays below the decision threshold $\gamma_{\mathrm{cfar}}=5$ for nearly all samples, indicating a low false-alarm rate. Under human motion, the statistic is above the threshold across the tested trajectories, enabling reliable detection. Overall, the CFAR-style detector admits a unified threshold across scenes while sensitive to motion.

\subsubsection{Real-time performance}
Fig.~\ref{fig:runtime} reports the CDF of per-CPI processing latency. The average runtime is 1.65~ms per CPI (128 CSI samples), and over 98\% of CPIs are processed within 1.93~ms. The efficiency mainly comes from DFT-based feature extraction and the lightweight EKF recursion. This computational margin allows dense temporal updates and supports larger temporal windows for outlier suppression.

\subsection{Sensing Results}
\label{subsec:sensing_results}
We evaluate gait identification performance using the extracted micro-motion features. For user ID recognition, we employ a lightweight vision-transformer backbone, MobileViT-XXS~\cite{mehta2021mobilevit}, which is suitable for deployment on embedded platforms. Each sample is formatted as a micro-Doppler feature map and normalized on a per-sample basis. We use a 30\%/70\% train/test split, batch size 128, learning rate 0.001, and train for 384 epochs.We also apply simple data augmentations to improve robustness: reversing the map along the time axis or Doppler axis, small temporal shifts, additive noise , mild warping of the time axis  or Doppler axis, and a smooth non-rigid deformation of the 2D map.

Fig.\ref{fig:gait_confusion} visualizes the gait-identification confusion matrices for three feature representations. Compared with the CASR baseline in Fig.\ref{fig:gait_confusion:casr} (macro-F1 =0.859), both CSI-power based features yield clearer diagonals and fewer cross-subject confusions. In particular, the position-refined representation in Fig.\ref{fig:gait_confusion:posref} achieves the best overall performance (macro-F1 =0.908), while the range-refined variant in Fig.\ref{fig:gait_confusion:rangeref} attains a comparable macro-F1 of 0.906. Fig.~\ref{fig:userid_f1} further summarizes the corresponding accuracy/precision/recall/F1, confirming a consistent improvement over CASR.

\section{Conclusion}
We propose WiDFS2.5, a real-time bistatic passive tracking and sensing framework that operates in the CSI-power domain. The power-domain representation largely suppresses TO/CFO and inter-chain phase offsets without explicit calibration, while preserving target information in the static-dynamic cross term. We develop a lightweight 3D DFT pipeline to extract delay-AoA-Doppler features, and use physics-constrained gating (with mirror-aware selection) to resolve the mirror ambiguity of real-valued power spectra for robust peak picking. Building on these features, WiDFS2.5 integrates motion detection, multi-CPI fusion for outlier suppression, and EKF tracking with deterministic initialization to achieve accurate single-receiver tracking. Finally, position-refined micro-Doppler improves micro-motion sensing and boosts gait identification. Extensive simulations, 3.1~GHz LTE over-the-air experiments, and evaluations on the WiFi GaitID dataset validate the effectiveness and generality of WiDFS2.5.


\bibliographystyle{IEEEtran}
\bibliography{refs.bib}



\end{document}